\documentclass[12pt]{article}
\usepackage{lscape}
\usepackage{geometry}
\usepackage[onehalfspacing]{setspace}
\usepackage{amssymb}
\usepackage{amsfonts}
\usepackage{graphicx}
\usepackage{amsmath}
\usepackage{bbm}
\usepackage{epstopdf}
\usepackage{natbib}
\usepackage{bibunits}
\usepackage{morefloats}
\usepackage{float}
\usepackage{comment}
\usepackage{pdflscape}
\usepackage{adjustbox}
\usepackage{subcaption}
\usepackage{longtable}
\usepackage{apalike}
\usepackage{xfrac}
\usepackage[dvipsnames,table]{xcolor}
\usepackage{neuralnetwork}
\usepackage{listings}
\usepackage{lipsum}

\usepackage{qtree}
\usepackage{multicol}
\usepackage{booktabs}
\usepackage{threeparttable}
\usepackage{tabularx}
\usepackage{footnote}
\usepackage{wrapfig}

\usepackage{dcolumn}
\newcolumntype{d}[1]{D..{#1}} 

\usepackage{siunitx}
\usepackage{mathpazo} 
\usepackage[scaled]{helvet} 
\usepackage{courier} 
\normalfont
\usepackage[T1]{fontenc}
\usepackage{listings}
\usepackage{epigraph}
\def\sym#1{\ifmmode^{#1}\else\(^{#1}\)\fi}

\definecolor{DarkerPineGreen}{RGB}{0, 90, 80} 

\usepackage[pdftex,bookmarks,colorlinks]{hyperref}
\hypersetup{
  colorlinks,
  citecolor=DarkerPineGreen,
  linkcolor=red,
  urlcolor=blue}
\usepackage{cleveref}

\usepackage{tikz}

\usepackage{amsmath,bm}
\usepackage{xstring}

\usetikzlibrary{fit,positioning}

\newcommand\denselyConnectNodes[2]{
  \foreach \n [count=\lyrIdx, remember=\lyrIdx as \previdx, remember=\n as \prevn] in #2 {
    \foreach \y in {1,...,\n} {
      \ifnum \lyrIdx > 1
        \foreach \x in {1,...,\prevn}
          \draw[-] (#1-\previdx-\x) -- (#1-\lyrIdx-\y);
      \fi
    }
  }
}

\usepackage{etoolbox}

\makeatletter
\newcommand\halftiny{\@setfontsize\halftiny\@vipt\@viipt}
\newcommand\notsotiny{\@setfontsize\notsotiny{8.95415}{11.2828}}
\makeatother

\definecolor{bred}{RGB}{122, 0, 0}
\definecolor{bredd}{RGB}{127, 0, 0}

\definecolor{dpd}{rgb}{0.0, 0.05, 0.5}

\renewenvironment{abstract}
 {\small
  \begin{center}
  \bfseries \abstractname\vspace{-.5em}\vspace{0pt}
  \end{center}
  \list{}{
    \setlength{\leftmargin}{1.2cm}    \setlength{\rightmargin}{\leftmargin}  }  \item\relax}
 {\endlist}
\begin{document}
\sloppy
\title{\vspace*{-1.4cm} \LARGE {\textbf{\color{bred} From Reactive to Proactive Volatility Modeling \\ with Hemisphere Neural Networks}} \vspace{0.9cm}}
\small
\small 
\author{ \hspace{-1.15em} Philippe Goulet
Coulombe\thanks{%
Contact:  \href{mailto:p.gouletcoulombe@gmail.com}{\texttt{goulet\_coulombe.philippe@uqam.ca}}.  For helpful comments, we thank Frank Diebold,  Maximilian G\"obel,  Alain Guay,  Nicolas Harvie,  Michael Pfarrhofer,  Aubrey Poon,  Dalibor Stevanovic,  and Boyuan Zhang as well as participants at the IIF MacroFor, the AMLEDS seminar, the FinEML Conference 2023, the 6th Annual Workshop on Financial Econometrics and the CFE 2023.  The views expressed in this paper do not necessarily reflect those of the Oesterreichische Nationalbank or the Eurosystem.  This research was enabled in part by support provided by Calcul Québec and the Digital Research Alliance of Canada. This draft: February 15,  2024. The Python package is available \href{https://github.com/TheAionxGit/Aionx}{here}. }  \\
\textbf{\color{dpd} \texttt{\fontfamily{phv}\selectfont \footnotesize\normalsize \quad \quad \quad Universit\'{e} du Qu\'{e}bec \`{a} Montr\'{e}al} \quad \quad \quad}  \newline \medskip
\and Mikael Frenette \\
\textbf{\color{dpd} \texttt{\fontfamily{phv}\selectfont \footnotesize \normalsize \quad \quad \quad Universit\'{e} du Qu\'{e}bec \`{a} Montr\'{e}al} \quad \quad \quad}  \newline   \medskip
\and Karin Klieber \\
 \textbf{\color{dpd} \texttt{\fontfamily{phv}\selectfont \footnotesize \normalsize \quad \quad \quad Oesterreichische Nationalbank \enskip \quad \quad \quad}}
}

\date{}

\setstretch{0.98}

\maketitle
\vspace{-0.99cm}
\textbf{
\large
\center 
 \smallskip
  \bigskip
 }
 \setstretch{1.2}

\begin{abstract}

\noindent We reinvigorate maximum likelihood estimation (MLE) for macroeconomic density forecasting through a novel neural network architecture with dedicated mean and variance hemispheres.  Our architecture features several key ingredients making MLE work in this context.  First, the hemispheres share a common core at the entrance of the network which accommodates for various forms of time variation in the error variance.  Second, we introduce a volatility emphasis constraint that breaks mean/variance indeterminacy in this class of overparametrized nonlinear models. Third, we conduct a blocked out-of-bag reality check to curb overfitting in both conditional moments.  Fourth, the algorithm utilizes standard deep learning software and thus handles large data sets --  both computationally and statistically.  Ergo, our Hemisphere Neural Network (HNN) provides proactive volatility forecasts based on leading indicators when it can, and reactive volatility based on the magnitude of previous prediction errors when it must. We evaluate point and density forecasts with an extensive out-of-sample experiment and benchmark against a suite of models ranging from classics to more modern machine learning-based offerings. In all cases, HNN fares well by consistently providing accurate mean/variance forecasts for all targets and horizons.  Studying the resulting volatility paths reveals its versatility, while probabilistic forecasting evaluation metrics showcase its enviable reliability.  Finally, we also demonstrate how this machinery can be merged with other structured deep learning models by revisiting \cite{HNN}’s Neural Phillips Curve.

\end{abstract}

\thispagestyle{empty}





\clearpage


\clearpage 
\setcounter{page}{1}

\newgeometry{left=1.7 cm, right= 1.7 cm, top=2.3 cm, bottom=2.3 cm}

\section{Introduction}

Unlike traditional deep learning strongholds such as speech recognition and computer vision, applications in social sciences are typically nowhere near perfect prediction accuracy. In other words, signal-to-noise ratio is low for most economic applications, and in the vicinity of 0 for finance applications. Still, the recent literature shows that deep learning methods can do surprising yet informative predictions in economics \citep[see, e.g., ][]{smalter2017macroeconomic,medeiros2019,andreini2020deep,hauzenberger2020real,barbaglia2022testing,HNN}.  Thus, it is particularly pertinent to estimate heterogeneous prediction uncertainty -- in order to determine when to trust or distrust a neural network's forecast.   

In this paper,  we provide a principled and effective way to do so,  which comes in the form of a novel standalone density forecasting tool.  Its design also allows for it to be a building block that can be merged with elements of other macroeconometric deep learning models.  This is of independent interest given that deep neural networks (NN) and their associated software environments are fertile ground to build more structured models (either for the sake of interpretability,  increased performance,  or both; see \citealt{farrell2021deep,HNN}),  or to incorporate the ever-growing sources of non-traditional data.   

\vskip 0.2cm

{\noindent \sc \textbf{A {Hemisphere} Neural Network (Redux).}}  \cite{HNN} introduces the concept of a Hemisphere Neural Network (HNN) where a NN  is restricted so that its prediction is the sum of latent time series corresponding to the outputs of subnetworks. Those are constructed from groups of predictors separated at the entrance of the network into different hemispheres.  The  structure allows the understanding of the final layer's cells output as  latent states in a linear equation.  There,  the motivation was interpretability of the conditional mean through separability.  Here,  the point is to go beyond the conditional mean.  

This paper treats the mean and the variance of a predictive regression as two separate hemispheres in one neural network where the loss function is the negative log-likelihood.  The model features a common core at the entrance of the network which accommodates for various interactions between the conditional mean and variance structures. This resembles the autoregressive conditional heteroskedasticity (ARCH) behavior where mean parameters enter the volatility equation or volatility-in-means with the reverse operation.  But going straight for maximum likelihood estimation of the new architecture will fail,  for old and new reasons.  The most prominent of those is that the double descent phenomenon --  the modus operandi of modern deep learning -- will result  in the usual benign overfitting of the conditional mean \citep{belkin2019reconciling,hastie2019surprises,bartlett2020benign} \textit{and} malign underfitting of the conditional variance.   A key observation is that,  in vastly overparameterized models aiming for the first two moments,  in-sample overfitting of the first leads to underfitting of the second,  and vice versa. {\color{black} Then,  what will happen out-of-sample is anybody's guess. }  Accordingly,  left unchecked,   HNN could completely overfit the training data with either a perfect conditional mean path or an equally perfect conditional variance process -- the allocation between the two very disparate models left to random initialization choices. 

We overcome this particularly daunting roadblock by designing three main algorithmic modifications: a volatility emphasis constraint in estimation, a blocked out-of-bag recalibration, and blocked subsampling.  The resulting HNN will prove highly competitive in our (point and density) forecasting exercise and provide more reliable coverage than currently available machine learning (ML) based  alternatives.   This desirable consistency in performance is a direct byproduct of the three aforementioned "modifications" bringing what could be called "conformal restrictions" in estimation and prediction.  As the name suggests,  such operations are related to the rapidly growing ML literature on conformal prediction where a pseudo-out-of-sample metric is used as raw material to construct prediction intervals with coverage guarantees \citep{vovk2005algorithmic}.



\vskip 0.2cm

{\noindent \sc \textbf{Proactivity,  Reactivity,  and Related Literature.}} We neither restrict the mean nor the variance to follow a specific law of motion.  They are both neural (sub)networks taking a large panel of macroeconomic series as common input.  Neural networks successfully deal with high-dimensional input spaces and are implemented in highly optimized software environments providing fast computations.  We refer to proactive volatility forecasts as those leveraging leading indicators to predict heightened volatility before the model delivers a large forecast error.   Conversely,   reactive forecasts propagate shocks that already occurred, resulting in increased expected variance in the following periods--after the occurrence of an initial major shock.  HNN provides proactive volatility forecasts based on observed indicators when it can, and reactive volatility based on the magnitude of previous prediction errors when it must.

The "reactive" class of models has a very long and distinguished history in econometrics,  with (G)ARCH  \citep{engle1982autoregressive,bollerslev1986generalized} and stochastic volatility (SV) models \citep{Taylor1982stochvol,hull1987pricing,jacquier2002bayesian}.  The popularity of SV for macroeconomic forecasting is mostly unrivaled.  It is  the workhorse volatility process to close a Bayesian model and accounts for (slow) structural change in innovations' variance  \citep{stock2007has,AAG2013,clark2015macroeconomic,carriero2019large}.  SV and GARCH models can be augmented with indicators that may have proactive qualities \citep{guidolin2021boosting},  but this faces various important challenges,  like that of high-dimensionality,  and therefore,  traditional reactive specifications have nearly always dominated the landscape.      We find neural network adaptions for SV and GARCH to model time-varying volatility in financial time series in, e.g., \cite{luo2018neural,yin2022neuralgarch, yin2022varvol}.  However, estimating the predictive variance when applying deep learning models to estimate the predictive mean has turned out to be a very challenging task. Neural networks tend to be overconfident in making predictions \citep{guo2017calibration,amodei2016concrete} and deliver residuals close to 0 \citep{belkin2019reconciling} that are a rather elusive target in a secondary conditional variance regression.  Furthermore,  implementing GARCH or SV-like methods in the highly nonlinear structure of deep learning models implies a significant deviation from the very software environments making their computations feasible and efficient.  Recent contributions apply SV in nonlinear or nonparametric models such as Bayesian additive regression trees \citep{huber2022inference} or Bayesian neural networks \citep{hauzenberger2022bnn}.  However, these models rely on Bayesian estimation which often turns out to be computationally costly,  and the volatility prediction remains solely reactive by construction.  


Quantile and distributional regressions enjoy increasing popularity in the macroeconomic literature and have seldom been found to have proactive qualities \citep{adrian2019,adams2021forecasting,caldara2021understanding,delle2021modeling,guidolin2021boosting}.  Early propositions to overcome the normality assumption when modeling densities include the seminonparametric (SNP) model of \cite{gallant1987semi}. Recent contributions extend the concepts of quantile and density regressions to nonlinear nonparametric models.  \cite{clark2022tail} do so in the context of Bayesian additive regression trees (BART) whereas \cite{barunik2022learning} and \citet{chronopoulos2023forecasting} do related things with neural networks. 

From the deep learning literature side of the aisle,  we find extensions of traditional methods which allow for the estimation of high-order moments, mixtures of distributions,  as well as quantile regressions.  \citet{nix1994estimating} and \citet{bishop1994mixture} proposed estimating the first and second moments of the predictive distribution with two \textit{separate} neural networks.   Building on this idea, the recent literature proposes different variants of mean-variance neural networks \citep{dybowski2001pinn,khosravi2014optimized} as well as mixture density networks \citep{graves2013generating,lakshminarayanan2017simple}.   In that vein,  the DeepAR model of \citet{salinas2020deepar} is getting increasing attention.  Amazon's DeepAR is a sequence-to-sequence probabilistic forecasting model which estimates the parameters of a distribution with Recurrent Neural Networks (RNNs) based on maximum likelihood.  However, as documented in \citet{gasthaus2019probabilistic},  DeepAR tends to underestimate variance,  likely for the aforementioned double descent reasons.  We will also find in our experiments that the quality of DeepAR's density forecasts is erratic.  Lastly,  the estimation of quantile regressions using neural networks dates back to \citet{taylor2000qrnn} and has since been the subject of a copious amount of research  \citep{feng2010robust,wen2017multi,cannon2018mcqrnn,moon2021learning}.  

\vskip 0.2cm

{\noindent \sc \textbf{Intended Use.}} A relevant question is where HNN stands in this deluge of works.  It is an economical yet not any less sophisticated solution to quantify time-varying uncertainty surrounding deep learning-based macroeconomic forecasts.  It is fast,  malleable,  and easily understood -- through the use of only two (nonlinear) conditional moments.  We will see that it works well for many targets without any particular tuning, and that both point and density forecasts are highly competitive and reliable.   Lastly,  it will be easily merged with more structured models,  like that  of \cite{HNN}, giving a "complete" model of inflation \textcolor{black}{based on a nonlinear Phillips curve specification}. 

\vskip 0.2cm

{\sc \noindent \textbf{Summary of Forecasting Results}.}  In a thorough forecasting exercise using macroeconomic data for the US,  we find that HNN has a great capacity for adaptation in the face of a heterogeneous pool of series.  Adaptability is a recurring finding when applying machine learning tools to macroeconometric problems \citep{MRF} and can be linked back to the carefully crafted semi-nonparametric structure of the model.  Specifically,   it captures the Great Moderation pattern in real activity variables (i.e.,  long-run change) and yet,  without changing the specification nor hyperparameters,  can deliver a more "spiky" volatility process for the S\&P 500.  The estimated volatility path for longer-run forecasts of macroeconomic targets ($s=4$ quarters ahead) displays a behavior that at times resembles more that of a (smoothly) switching process,  in contrast to the slowly evolving SV process which dominates the literature.  Those higher volatility regions are proactive in the sense that they begin before the advent of a major prediction error, a behavior that is observed both in-sample (with out-of-bag estimates) and out-of-sample for many targets (e.g.,  GDP growth,  Unemployment Rate,  Inflation).

In terms of performance,  HNN  always ranks among the top models in terms of RMSE, log score, 
coverage rates,  and other metrics of calibration and probabilistic forecast evaluation.  More interestingly, it never suffers "catastrophic failures" (like massive undercoverage) that seldom occur on some targets for the other sophisticated competing models.  For instance,  it is not infrequent to see BART  and DeepAR substantially undercovering -- a phenomenon we delve into and attribute to harmless overfitting of the conditional mean leading to quite harmful underestimation {\color{black} and underfitting} of the conditional variance.  This implies that while conditional means can be used as per the model’s estimation (and perform well as such),  conditional variances cannot, and often fail in ways that basic manual quality control is unable to flag nor fix ex-ante.   HNN,  in contrast,  appears to have a level of reliability mostly on par with that of AR competitors.   The use of out-of-bag (and presumably non-overfitted) errors to calibrate or estimate the volatility process helps HNN a lot in being reliable out-of-the-box.  Our much simpler NN$_\text{\halftiny SV}$,  a reduction of HNN which forfeits proactive volatility but keeps a sophisticated conditional mean function and fits a SV model on OOB residuals in a second step,  is equally reliable in terms of coverage.    

The forecasting section also includes a series of vignettes.  First, we compare our approach to quantile regressions of various kinds.  A striking observation is that for real activity targets -- which asymmetry and non-normality have been heavily documented following \cite{adrian2019} -- the normal likelihood-based HNN  usually performs better or as well as the best (linear or nonlinear) quantile model.  This is true for both tails of the distribution and short and long forecasting horizons.  Second,  we investigate whether the use of Long Short-Term Memory Networks \citep[LSTM, ][]{hochreiter1997lstm} could further improve HNN results in any material way---they do not.  \textcolor{black}{In addition, we present results on monthly data for the US and a euro area forecasting exercise in the appendix. We find that HNN fares well with time series that are noisier and shorter.  However, as one could expect from this more hostile terrain, gains with respect to autoregressive SV models are more punctual and modest in size.}

\vskip 0.2cm

{\sc \noindent \textbf{Fusion With a Structured Deep Learning Model}.}  HNN and the overall apparatus developed in this paper can also be joined in a modular fashion with more structured deep learning models to obtain interpretable forecasts with reliable uncertainty quantification.  We construct such a model for inflation by embedding \cite{HNN}'s Neural Phillips Curve (NPC) within this paper's arsenal.  We find that the customized (yet restricted) HNN-NPC improves point and density forecasts over the plain density HNN.  We also compare with a simpler Bayesian model also providing interpretability and uncertainty estimates via SV \citep{chan2016}.   In line with \cite{HNN}'s findings, we see that the neural model better captures inflation dynamics during important episodes like in the outset of the Great Recession and the post-Pandemic inflation surge.  This is attributable to a very different reading of the contribution of real activity and expectations in less quiet economic times.  We enrich such results by showing that HNN-NPC was rather confident when predicting high inflation in early 2020, and nearly dismisses its own deflationary forecast in 2020.  Moreover,  its proactive volatility qualities are also apparent in 2008 where the volatility estimate climbs out of its bed well before SV does so (following the 2008 crash in oil prices).  All in all,  HNN-NPC demonstrates how deep learning can improve over classic approaches while retaining essential qualities of the latter  -- interpretability \textit{and} uncertainty quantification.

\vskip 0.2cm

{\sc \noindent \textbf{Outline}.} Section \ref{sec:HNN} introduces the mean-variance HNN by describing and motivating the network architecture,  and presenting the key algorithmic modifications to plain MLE.  Section \ref{sec:emp} conducts an extensive empirical analysis using macroeconomic data for the US. \textcolor{black}{In Section \ref{sec:npc} we extend HNN to a nonlinear Phillips curve model for inflation.} Finally, Section \ref{sec:con} concludes.

\section{The Architecture}\label{sec:HNN}

\begin{figure}[ht!]
\begin{center} 
\hspace*{-0.1cm}\includegraphics[scale=.23]{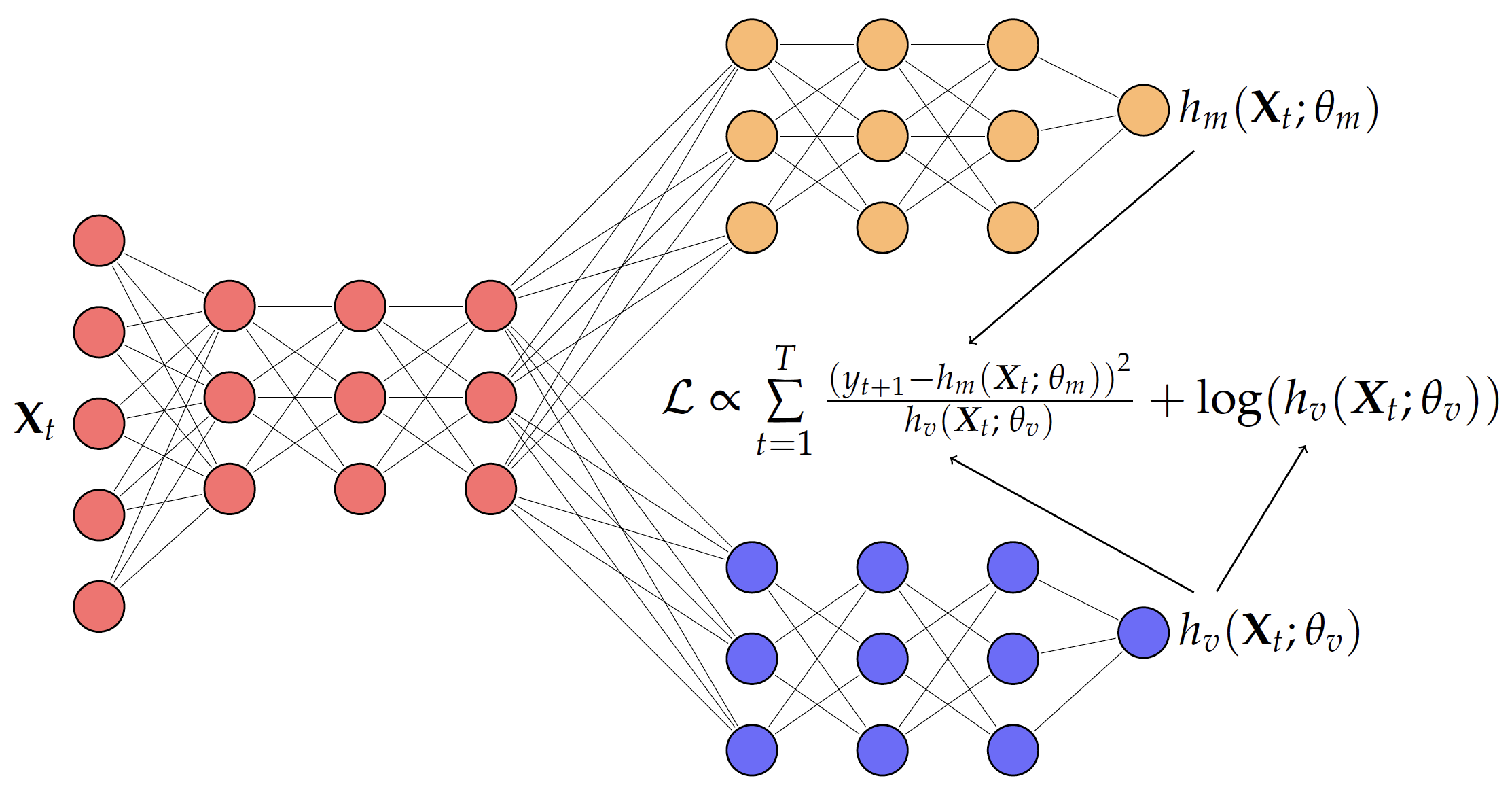}
\caption{Hemisphere Neural Network's Architecture for Mean/Variance Forecasting}\label{fig:hnnarchi}
\end{center}
\end{figure} 


This section describes our proposed neural network architecture to estimate the mean and variance of the predictive distribution of our target variable $y_{t+1}$ (or $y_{t+s}$ in the case of direct $s$ steps ahead forecasts).  We assume that $y_{t+1}$ follows a Gaussian distribution and depends on a (potentially very large) number of $K$ covariates denoted by $\boldsymbol{X}_t$:
\begin{equation}
y_{t+1} \sim \mathcal{N}(f(\bm X_t),g(\bm X_t))
\end{equation}
In this very general setup, the functions $f$ and $g$ are unknown and may be highly nonlinear.  To remain agnostic on the functional form of $f$ and $g$,  both will be approximated through a neural network (NN) structure \citep{hornik1989multilayer}.  Rather than estimating a standard deep NN model and hoping to make something out of the residuals,  we design a specific architecture derived from  \cite{HNN}'s original HNN that will obtain $g$ and $f$ jointly.  In our application,  hemisphere 1 ($h_m = f$) is the conditional mean and hemisphere 2 ($h_v = g$) is the conditional variance.  Both hemispheres are fully nonlinear, nonparametric functions of the input space $\boldsymbol{X}_t$ which ultimately output two time series: conditional mean ($\hat{y}_t$) and conditional variance ($\hat{\sigma}^2_t$).  Importantly,  they get their assigned roles from how they enter the loss function,  which is now proportional to the log-likelihood.  Accordingly,  the first building block of our approach is to have a neural network with objective function
\begin{equation}\label{eq:ouou}
\min_{\theta_{m}, \theta_{v}}  \sum_{t=1}^T  \frac{\left( y_{t+1} - h_m (\boldsymbol{X}_t;\theta_m)\right)^2 }{h_v(\boldsymbol{X}_t;\theta_v)}  + \log(h_v(\boldsymbol{X}_t;\theta_v))
\end{equation}
where $\theta_{m}$ and  $\theta_{v}$ are the network parameters consisting of the weights $w_j$ and the bias term $b_j$ (i.e., $\theta_m = (w_m, b_m)$ and $\theta_v = (w_v, b_v)$).  The next questions are (i) what is the structure of $h_m$ and $h_v$,  and (ii) how do we successfully solve \eqref{eq:ouou}.  The next paragraphs set out to answer those through a series of "ingredients" that reinvigorate what is otherwise a rather plain-looking MLE problem. 

\vskip 0.25cm

{\noindent \sc \textbf{Ingredient 1: Two Hemispheres and A Common Core.}}  Figure \ref{fig:hnnarchi} summarizes the network's architecture.  As can be seen, both hemispheres share the same input data as well as a few common layers before estimating the parameters of each hemisphere. The outcome of both hemispheres enter the loss function and thereby complete the model setup.  
Going backward from the loss towards the original inputs,  the "yellow" hemisphere is 
\begin{equation}\label{eq:hm}
\begin{aligned}
h_m (\boldsymbol{X}_t;\theta_m) &= \bm w_m^{(L_m)'}{\bm Z}_t^{(L_m-1)}  + \bm b_m^{(L_m)}, \quad \text{with} \\ 
{\bm Z}^{(l)}_t &= \phi^{(l)}( \bm w_m^{(l)'} {\bm Z}^{(l-1)}_t + \bm b_m^{(l)}), \quad \text{for } L_c \leq l \leq L_m-1,
\end{aligned}
\end{equation}
and the "blue" one is
\begin{equation}\label{eq:hv}
\begin{aligned}
h_v (\boldsymbol{X}_t;\theta_v) =& \text{log}(1 + \text{exp}(\bm w_v^{(L_v)'}{\bm Z}_t^{(L_v-1)}  + \bm b_v^{(L_v)})), \quad \text{with} \\ 
{\bm Z}^{(l)}_t =& \phi^{(l)}( \bm w_v^{(l)'} {\bm Z}^{(l-1)}_t + \bm b_v^{(l)}), \quad \text{for } L_c \leq l \leq L_v-1,
\end{aligned}
\end{equation}
where $\phi$ denotes a nonlinear activation function,  $L_m$ and $L_v$ are number of hidden layers for each hemisphere,  and $L_c$ is that of the common (red) core.  The \textit{Softplus} activation function in \eqref{eq:hv} constrains $\hat{h}_v(\mathbf{X}_t;\theta_v)$ to be positive at all times.  Clearly,   the definitions of $h_m (\boldsymbol{X}_t;\theta_m)$ and $h_v (\boldsymbol{X}_t;\theta_v)$ are still incomplete given that $\boldsymbol{X}_t$ has yet to make an appearance on the right-hand side of \eqref{eq:hm} and \eqref{eq:hv}.  The common core at the entrance of the networks brings such completion via 
\begin{equation}\label{eq:hc}
{\bm Z}^{(l)}_t = \phi^{(l)}( \bm w_c^{(l)'} {\bm Z}^{(l-1)}_t + \bm b_c^{(l)}), \quad \text{for } 1 \leq l \leq L_c
\end{equation}
with ${\bm Z}^{(0)}_t = \bm X_t$.  Thus, the first layer in each hemisphere uses the outputs of neurons from the last layer of the common block ${\bm Z}^{(L_c)}_t$.

Having dedicated mean and variance hemispheres requires little additional ink to motivate,  but the virtues of the common core, while numerous,  are more subtle.  Consider the following simple linear data generating process (DGP) with ARCH errors
\begin{align}
\left\{\begin{array}{l}
y_t=\mathbf{X}'_t \boldsymbol{\beta}  +\varepsilon_t, \quad \textcolor{black}{\varepsilon_t = \sigma_t^2 \epsilon_t, \quad \epsilon_t \sim \text{iid}}  \\
\sigma_t^2=c+a_1 \varepsilon_{t-1}^2+\ldots+a_p \varepsilon_{t-p}^2
\end{array}\right. 
\end{align}
{In this model,  the corresponding hemisphere outputs and parametrizations would be}  
\begin{align}
\left\{\begin{array}{l}
h_m(\boldsymbol{X}_t; \boldsymbol{\beta} ) =\mathbf{X}'_t \boldsymbol{\beta}  \\
h_v(\boldsymbol{X}_t;[ {\color{black}\boldsymbol{a}} \enskip \boldsymbol{\beta} ]) =c+a_1\left(y_{t-1}-\mathbf{X}'_{t-1} \boldsymbol{\beta}  \right)^2+\ldots+a_p\left(y_{t-p}-\mathbf{X}'_{t-p} \boldsymbol{\beta} \right)^2.
\end{array}\right.
\end{align}
As \cite{gourieroux1997arch} puts it "{Even in the simple case, we cannot estimate separately the parameters of the conditional mean and those appearing in the conditional variance.}" Obviously,  this does not mean all volatility models need to be estimated jointly.  What it suggests, however,  is that successful models of time series volatility often have some {cross-equation restrictions} between $h_m$ and $h_v$.  
 

Rather than introducing cross-equation restrictions,  which are likely both unfeasible and undesirable in a neural network setup, we discipline $h_m$ and $h_v$ with soft constraints, i.e., cross-equation regularization.  We achieve this by estimating common layers at the entrance of the network, which can be interpreted as hemispheres sharing weights. As emphasized in Figure \ref{fig:hnnarchi}, we estimate a few common layers for both hemispheres before separating the mean from the variance hemisphere, where hemisphere-specific neurons are presented in yellow for the former and in blue for the latter.  While this example details how the conditional mean parameters enter that of the variance,  the opposite sharing direction is also possible.  For instance,  latent structures driving volatility can flow in the mean hemispheres,  which conveniently allow for GARCH- or SV- or any volatility-in-means effect which have been popular in finance to study the time-varying risk premium \citep{EngleLilienRobins1987} and now in macroeconomics to quantify the real effects of uncertainty 
\citep[e.g., ][]{carriero2018measuring,shin2020new}.

By adding time trends to our set of covariates we approach a classical SV specification through a residuals trend-filtering perspective.  GARCH dynamics would suggest making $h_v$ a recurrent neural network.  As we will see,  HNN results will be quite competitive without this additional complication -- RNNs and LSTMs are notoriously harder and longer to train.  We nevertheless explore this possibility in Section \ref{sec:hrnn}.


\vskip 0.25cm

{\noindent \sc \textbf{Ingredient 2: Volatility Emphasis.}}  As we know from the \textit{double descent} phenomenon \citep{belkin2019reconciling,hastie2019surprises,bartlett2020benign}, a mildly deep and large network will yield a near perfect in-sample fit even in the presence of large amounts of noise \textit{and yet} produce stellar out-of-sample results.  When focusing on out-of-sample point forecasts,  we can safely embrace  double descent and its associated benefits.  However,  this creates trouble for the historical (in-sample) analysis of conditional mean estimates and double trouble for the conditional variance,  with the latter being unreliable both in- and out-of-sample.  Our volatility emphasis constraint, coupled with the next two ingredients,  will make MLE work in the context of densely parameterized models.

A first observation is that a model in the double descent region eradicates residuals,  yet MLE is supposed to obtain the parameters of their non-degenerated distribution.  A second  is that the reverse solution is also possible: a perfect volatility model with no conditional mean.  In other words,  when solving \eqref{eq:ouou} without further adjustments,   HNN can completely overfit the data with either $h_m$ or $h_v$,  giving rise to vastly different models.   This suggests that the overall prevalence of $h_m$ versus $h_v$,  when those are left completely unconstrained,  is not identified and cannot be obtained from in-sample estimation (in a similar spirit to the regularization parameter $\lambda$ in a ridge regression,  \citealt{GC2019}).  Note that early stopping can help in regularizing $h_m$ or $h_v$,  but will do symmetrically which is highly suboptimal in many applications where it is clear that one hemisphere should be more expressive than the other.    

As a solution, we bring in a constraint.  We fix the  average conditional predictive variance to a constant (i.e., $\texttt{mean}(h_v(\boldsymbol{X}_t;\theta_v))/ \texttt{var} (y_{t+1}) =\nu$) during estimation and let HNN learn deviations from it.  We refer to $\nu$ as the volatility emphasis parameter,  because it guides how much of the network fitting capacities \textcolor{black}{should go} to the volatility versus the mean.  Why does this work? First,  it serves as a solution to optimization cycling through near-perfect conditional mean versus conditional variance optima and the general indeterminate nature of the problem in overfitting situations.  Fixing the expressivity of $h_m$ allows us to let $h_v$ benignly overfit (if need be) the way it is typically done for the conditional mean estimation in plain squared error minimization.  As a result, the conditional mean is tied to deliver estimates that will look like a plausible OOB fit for every run (as set by $\nu$),    and conditional variance can be projected OOB and compared to OOB squared errors (Ingredient 3 \& 4) to obtain a non-overfitted volatility path in- and out-of-sample. 


While the final unconditional variance is readjusted and not imposed (see Ingredient 4 below),  we should not choose $\nu$ lightly, as it will influence the relative flexibility of $h_m$ and $h_v$ and estimated paths.  Clearly, from experience, we expect $\nu$ to be close to 1 for stock returns and lower for other macroeconomic targets,  especially those exhibiting persistence.   In theory,  $\nu$ could be cross-validated,  but to avoid the obvious practical cost of doing so,  we rather set $\nu$ through a very well-informed guess.  We estimate a standard NN with an analogous architecture,  calculate the mean of the squared blocked OOB residuals,  and set $\nu = \texttt{mean}(\hat{\varepsilon}_{t,\text{NN}}^2)/ \texttt{var} (y_{t+1})$ where $\hat{\varepsilon}_{t,\text{NN}}$ denotes the blocked OOB residuals.   In effect,  if one were to conduct basic conformal prediction-based inference for a plain NN in a macroeconomic time series context and assume homoscedasticity,  $\texttt{mean}(\hat{\varepsilon}_{t,\text{NN}}^2)$ and $\hat{\varepsilon}_{t,\text{NN}}$ in general would be natural inputs to obtain coverage-guaranteed (out-of-sample) prediction intervals \citep{chernozhukov2018exact}.  The presence of the denominator $\texttt{var} (y_{t+1})$ brings $\nu$ in universal units (i.e.,  between 0 and 1)\footnote{In practice,  the original estimate can go marginally above 1 since the inputs are OOB rather than training residuals,  and the plain NN model may do worse out-of-bag than simply taking the sample average when facing extremely low signal-to-noise ratios.  We enforce an upper bound at 0.99, effectively forcing $\nu$ to always deliver a $R^2> 1 \%$. } and is implicit in our calculations because all the data will be standardized at the entrance of the network and scaled back to original units at the exit.  A possibility for future work is to cross-validate $\nu$ in the neighborhood of the informed guess or update it through iterative HNN estimations,  but our current empirical results suggest this extra legwork may not be necessary.  

The inevitable failing of an unchecked HNN (and DeepAR later on) as well as the usefulness of the volatility emphasis constraint can also be intuitively understood from basic MLE econometrics for linear regression.  Even when fitting the simplest linear model without shrinkage,  the MLE estimate of the error variance is always biased downward: it yields $\frac{\sigma^2}{T}<\frac{\sigma^2}{T-K}$ where $K$ is the number of regressors and the second expression is the OLS version.  This potentially major discrepancy is straightforward to correct when the number of degrees of freedom is known, which it is not in the deep learning context.   The best course of action when the analytical calculation of degrees of freedom is impossible is the use of pseudo-out-of-sample metrics (of which cross-validation is the better known).  Thus,  curbing many problems at once,  we fix $\nu$ to a plausibly unbiased value ex-ante calculated from out-of-bag sampling and an approximated $h_m$.      

Two outstanding issues remain.  If the originally imposed $\nu$  is not exactly in tune with $h_m$'s final performance,  we may want to adjust the average conditional variance accordingly.   Another observation is that the volatility emphasis constraint fixes the expressivity of the conditional mean,  but not that of the variance.  Thus,  it does not prevent $h_v$ from overfitting what is left free in the likelihood,  and may offer implausibly accurate conditional variance forecasts in-sample that will not be matched out-of-sample.   We will get back to this when covering the fourth and final ingredient.



\vskip 0.25cm

{\noindent \sc \textbf{Ingredient 3: Blocked Subsampling.}}  We now turn to the important ingredient that has been implicit throughout.  Bagging in our context entails two major benefits.  First, the use of quantities which are immune to extreme overfitting.  Second,  it helps with optimization itself.  There is no guarantee that a single run of stochastic gradient descent initiated randomly will yield the "true parameters". In that sense,  our approach does not aim to succeed where traditional maximum likelihood estimation (MLE) would likely fail.  This is less concerning when considering an ensemble of multiple runs, similar to what is common practice for point prediction with NNs \citep{anatomy}.  In this case, we employ \textcolor{black}{$B=1000$ runs},  which may seem excessive for out-of-sample predictions but is suitable for OOB "time series" that utilize an average of $(1-\texttt{subsampling.rate}) \times 1000$ runs at each point in time. \footnote{Considering 300-something runs can be sufficient but results may change in a very marginal way depending on the seed.}

More precisely,  the calculations proceed as follows.  Assume we have a sample of size 100 \textcolor{black}{and choose a subsampling rate of 0.80}. We estimate HNN using data points from 1 to 85, and project it "out-of-bag" on the 20 observations not used in training. This gives us $h_j(\boldsymbol{X}_{80:100};\hat{\theta}_{j,b})$ for a single allocation $b$ (for $b = 1, \dots, B$) while $h_j(\boldsymbol{X}_{1:80};\hat{\theta}_{j,b})$ are still \texttt{NA}s.  By considering many such random (non-overlapping blocked) allocations where "bag" and "out-of-bag" roles are interchanged,  we obtain the final $h_{t,m}$ and intermediary (see next ingredient) $h_{t,v}$ paths  by averaging over $B$ at each $t$ such that 
\begin{align}
{h}_j(\boldsymbol{X}_t;\hat{\theta}_{j}) = \frac{1}{(1-0.80) \times B} \sum_{b=1}^B I({h}_j(\boldsymbol{X}_t;\hat{\theta}_{j,b})\neq  \texttt{NA}){h}_j(\boldsymbol{X}_t;\hat{\theta}_{j,b})  \quad \text{for } j \in \{m, v \}.
\end{align}
Interestingly,  this procedure fits within the framework of \cite{newton1994approximate}'s Weighted Bayesian Bootstrap \textcolor{black}{and, in particular, of} \cite{newton2021weighted}'s extension of it for generic ML losses. In short,  randomly weighted optimization of the loss provides an approximate Bayesian posterior.

\vskip 0.25cm

{\noindent \sc \textbf{Ingredient 4: Blocked Out-of-Bag Reality Check.}} To obtain a proper estimate of the unconditional \textit{variance} we introduce a recalibration step based on the blocked out-of-bag residuals. This is done by our reality check, which scales back the $h_v(\boldsymbol{X}_t;\theta_v)$ path making use of the OOB residuals.\footnote{In a sense, this step takes the concept of conformal prediction -- a method to form prediction intervals without making distributional assumptions \citep{vovk2005algorithmic,lei2013distribution,linusson2020efficient} -- to a conditionally heteroskedastic environment.  Uncertainty in future predictions is based on the residuals of a held-out validation set, which is used to recalibrate the prediction intervals.  \cite{chernozhukov2018exact} extends the applicability of such methods to dependent data using a block approach. } {\color{black} In our setup, the initial guess for $\nu$ -- coming from a plain NN and not the dual estimation of $h_v$ and $h_m$ -- might not exactly match the resulting average volatility of HNN's OOB residuals.}  Moreover,  even after early stopping,  the raw $h_v$ may be overly wiggly and reflect conditional variance overfitting.  Hence, we recalibrate $h_v$ using HNN's blocked OOB residuals by running 
\begin{equation}
\log \left(\hat{\varepsilon}_{t, \text{HNN}}^2\right)=\underbrace{\zeta_0 +\zeta_1 \log \left({h}_v(\mathbf{X}_t;\hat{\theta}_v)\right)}_{\Large {\delta}_t}+\xi_t
\end{equation}
and then update the in-sample volatility such that
\begin{equation}
\hat{h}_v(\mathbf{X}_t;\left[ \hat{\theta}_v,  \hat{\zeta_0},\hat{\zeta_1}, \hat{\varsigma} \right]) \leftarrow \exp(\hat{\delta}_t) \times \hat{\varsigma},  
\end{equation}
where $\hat{\varsigma}$ is the estimate of the scaling object $\varsigma= \mathrm{E}[\exp(\xi_t)]$.  If there is a mismatch between $\nu$ and the new OOB residuals coming from HNN,  $\zeta_0$ can adjust for it.  $\zeta_1$'s role is to move accordingly \textit{and} damper the overall variation in the final $\hat{h}_v$ in the event that the raw ${h}_v$ overfits.  This is because $\hat{\varepsilon}_{t, \text{HNN}}^2$,  coming from blocked subsampling,   is a suitable approximation to the kind of prediction errors HNN will encounter in the real out-of-sample.  Thus,  if necessary,  $\zeta_1$ acts as a raccord between $h_v$ and "reality".

The above operations can be seen as a direct neural translation of \cite{wooldridge2015introductory}'s Section 8.4  on weighted least squares (WLS). Note that the constant $\hat{\varsigma}$ is not part of Wooldridge's textbook because only relative (observation) weights are needed for the WLS application.  In our case,  we need an absolute metric and $\mathrm{E}[\exp(\xi_t)]$ is not merely equal to 1 as a result of $\exp()$ being a nonlinear function and $\xi_t$ likely being non-normal.  We sample with replacement from the vector of $\hat{\xi}_t = \log \left(\hat{\varepsilon}_{t, \text{HNN}}^2\right)-\hat{\delta}_t$ to estimate the expectation.  For out-of-sample volatility predictions,  we thus use $\hat{h}_v(\mathbf{X}_t^{\text{test}}; \left[ \hat{\theta}_v,  \hat{\zeta_0},\hat{\zeta_1}, \hat{\varsigma} \right])$.

\vskip 0.15cm

{\noindent \sc \textbf{Hyperparameters.}} For each hemisphere we estimate a standard feed-forward fully connected network, which features two hidden layers ($\texttt{layers} = 2$). The same holds for the common block at the entrance of the network. Moreover, each layer (common or not) is given $\texttt{neurons} = 400$. We choose the \textit{ReLU} activation function ($\operatorname{ReLU}(x)=\max \{0, x\}$) throughout the hidden layers and define a linear activation function for the output of $h_m$.  To prevent the error variance from being negative, a natural choice for the output activation function for $h_v$ is the \textit{Softplus} function ($\operatorname{Softplus}(x)=\log \left(1+\exp \left(x\right)\right)$), which imposes these bounds (${h}_v(\mathbf{X}_t;\theta_v) > 0 \enskip \forall t$) and is, in effect,  a soft \textit{ReLU}.

Hyperparameters for the optimization of the algorithm are set as follows.  The maximum number of epochs is 100 and the learning rate is 0.001. Similar to \citet{HNN}, we perform early stopping by using only a subset (80\%) of the training sample for the estimation of the parameters and determine with the remaining set (i.e., 20\%) when to stop the optimization.  We set $B=1000$.  The patience parameter in early stopping is 15 epochs.  As a form of ridge regularization on network weights,  early stopping may improve the efficiency of the algorithm and prevents the network from overfitting \citep{raskutti2014early}. In addition, we apply dropout with a \texttt{dropout rate} of 0.2. We use the Adam optimizer and choose the whole sample for the batch size.  Network weights $w_m$ and $w_v$ are initialized using $\mathcal{N}\left(0, {\small \sfrac{3}{100}} \right)$.  Those choices are common to all target variables.

\section{Macroeconomic Point and Density Forecasting}\label{sec:emp}                    


We test our proposed approach by modeling and forecasting key macroeconomic and financial variables of the US economy. We base our analysis on the FRED-QD database of \citet{mccracken2020fred}, which is available on a quarterly frequency and features 248 US macroeconomic and financial aggregates. Our sample ranges from 1960Q1 to 2022Q4. All variables but prices are transformed according to \citet{mccracken2020fred} to achieve approximate stationarity.\footnote{\color{black} NONBORRES  (Reserves of  Depository institutions (Nonborrowed)),  TOTRESNS (Reserves of  Depository institutions (total)),  GFDEBTNx (total public debt),  and BOGMBASEREALx (real monetary base) have been dropped because of their large shift in scale between in- and out-of-sample.  While estimation and predictions were robust to their inclusion (by putting a very small weight on those variable),  out-of-sample variable importance metrics were affected (see \cite{anatomy} for further discussion on this issue in the context of Shapley Values). }  Prices are in log first differences (inflation rate) rather than second differences (acceleration rate).  All predictors are standardized to have zero mean and unit variance which is necessary for NN-based models and redundant for the others.  We include two lags for each variable $X_{t,k}$ and add 100 linear trends to the set of covariates allowing for exogenous slow time variation in the parameters, and approximate trend filtering of the residuals   à la stochastic volatility if the DGP requires so.  Missing values at the beginning of the training sample are imputed using the EM algorithm of \cite{stock1999forecasting}.  

The target variables are GDP growth, change in the unemployment rate, headline CPI inflation, housing starts growth as well as S\&P 500 stock returns. For each of them, we compute the one-step and four steps ahead predictive mean and variance for our hold-out sample starting in 2007Q1 and ending 2022Q4.   NN-based models are re-estimated every two years whereas standard models are updated every quarter, all on an expanding window basis. Our forecasting exercise is based on a pseudo-out-of-sample analysis, which does not account for ragged edges or revisions in the underlying data set. Since we deal with large, dense models and none of the NN-based models put a disproportionate weight on a few indicators not available in real time, extending to a real-time exercise will not entail significant deviations from the results presented.


We explore the performance of the HNN by comparing the results to a set of competing models. This set is comprised of simple linear benchmarks including AR processes with SV and GARCH (AR$_\text{SV}$ and AR$_\text{G}$) as well as a high-dimensional Bayesian linear regression endowed with shrinkage and SV (BLR). In terms of nonlinear modeling choices, we consider standard neural network specifications (NN$_\text{SV}$ and NN$_\text{G}$), Bayesian additive regression trees (BART, see \citet{BART}) as well as Amazon's DeepAR \citep{salinas2020deepar}. Details on the implementation of the benchmark models can be found in Appendix \ref{app:bench}.  NN$_\text{SV}$ and NN$_\text{G}$ use the same architecture as HNN's conditional mean, but are trained by minimizing the usual squared errors and the volatility processes are fitted in a second step on the resulting out-of-bag residuals.  Those plain NNs allow to directly assess the relevance of a data-rich and densely parameterized nonlinear volatility function, and document HNN's proactivity versus standard approaches for often similar conditional means.  Lastly,  those two NN benchmarks allow to quantify the various merits of modeling jointly the first two conditional moments.  As discussed in Section \ref{sec:HNN}, it is not difficult to think of DGPs where this could make a sizable difference,  but knowing in what terrain we are standing is inevitably an empirical question.  

The rest of this rather rich set of competitors allows us to span the space of relevant one-shot deviations from our framework.  First, comparing the results to linear models sheds light on whether modeling nonlinearities pays off for macroeconomic point \textit{and} density forecasting. 
Second,  BART and DeepAR are the natural go-to nonlinear, nonparametric ML tools providing density forecasts.  Tree ensembles are always very stubborn benchmarks for learning tasks with tabular data, and BART provides a probabilistic extension of boosting that produces natively density forecasts.  DeepAR's architecture resembles that of a very crude HNN where there is only a common (LSTM) core,  no hemispheres,  and all remaining ingredients of Section \ref{sec:HNN} have been dropped.  Accordingly,  performance differentials with HNN will procure a rough estimate of the (non-)marginal benefits of those propositions.



For each of our six target variables, we evaluate compactly the resulting point forecasts using the root mean square error (RMSE), the probabilistic forecasting accuracy by means of the log score ($\mathcal{L}$) and the share of variation explained in residuals' magnitude via the $R^2_{|\varepsilon_t|}$ of absolute residuals.  For the out-of-sample (OOS) forecasted values at time $t$ for $s \in \{1,4\}$ we compute:  
\begin{align}
\text{RMSE}_{s} &= \sqrt{ \frac{1}{\#\text{OOS}}\sum_{t \in \text{OOS}} (y_{t+s}-\hat{y}_{t,s})^2},  \\[15pt]
\mathcal{L}_{s} &= - \frac{1}{\#\text{OOS}}\sum_{t \in \text{OOS}} \log \left( \varphi (\varepsilon_{t,s}; \hat{\sigma}_{t,s})\right),  \\[18.5pt]
R^2_{|\varepsilon_t|,s} &= 1 - \frac{\sum_{t \in \text{OOS}} (\lvert \varepsilon_{t,s} \rvert - \hat{\sigma}_{t,s})^2}{\sum_{t \in \text{OOS}} (\lvert \varepsilon_{t,s} \rvert - \eta)^2},
\end{align}
where $\varepsilon_{t,s} = y_{t+s}-\hat{y}_{t,s}$,  $\eta$ is the standard deviation of the in-sample residuals,  and $\varphi (. \phantom{.}; \hat{\sigma}_{t,s})$ is a normal density with zero mean and standard deviation $\hat{\sigma}_{t,s}$.  While exotic in appearance,  this last metric is only the out-of-sample goodness of fit \textcolor{black}{ of what would be the second stage regression in a weighted least squares problem.}  Surely,  it does not have all the qualities of other scoring rules and $|\varepsilon_t|$ is not exactly realized volatility,  yet it arguably  provides a metric that is much easier to interpret \textit{quantitatively}.   Lastly, it must be interpreted with care: a model can reach a high $R^2_{|\varepsilon_t|}$ because it has a failing conditional mean and the unexploited predictability flows in $|\varepsilon_t|$.   Thus, a sufficient condition for safely gazing at the $R^2_{|\varepsilon_t|}$ of a particular model is for it to also have a low RMSE.  Intuitively,  a high-performing model with a fine $\mathcal{L}$ will have both a low RMSE and a high $R^2_{|\varepsilon_t|}$.  Moreover, in Section \ref{sec:calibration} we present additional density forecasting measures, which are the continuous ranked probabilitiy score (CRPS) and the coverage rate (68\%), and assess model calibration using probability integral transforms (PITs).


We report evaluation metrics for the full test sample,  as well as a subsample that ends prior to the Pandemic Recession.  Given the unpredictable and unprecedented wild swings of 2020,  those observations are always discarded for the real activity targets.   In the interest of space and since  AR$_\text{\halftiny SV}$ and AR$_\text{\halftiny G}$ often yield very similar results out-of-sample, we only report the best of the two according to $\mathcal{L}$ for each target/out-of-sample pair.  All reported RMSEs are ratios with respect to that of the OLS-based AR(2). \textcolor{black}{Preferred models are those with low values in terms of RMSE and $\mathcal{L}$ and high values for $R^2_{|\varepsilon_t|}$.}


\subsection{Results}\label{sec:fcast_results}

\begin{figure}[!t]
\begin{minipage}{\linewidth}

\begin{minipage}{\linewidth}
\centering
\captionof{figure}{\normalsize \textbf{GDP} ($s=1$) \vspace*{-0.3cm}} \label{tab:GDPC1_h1_l2}
\includegraphics[width=1.02\textwidth]{{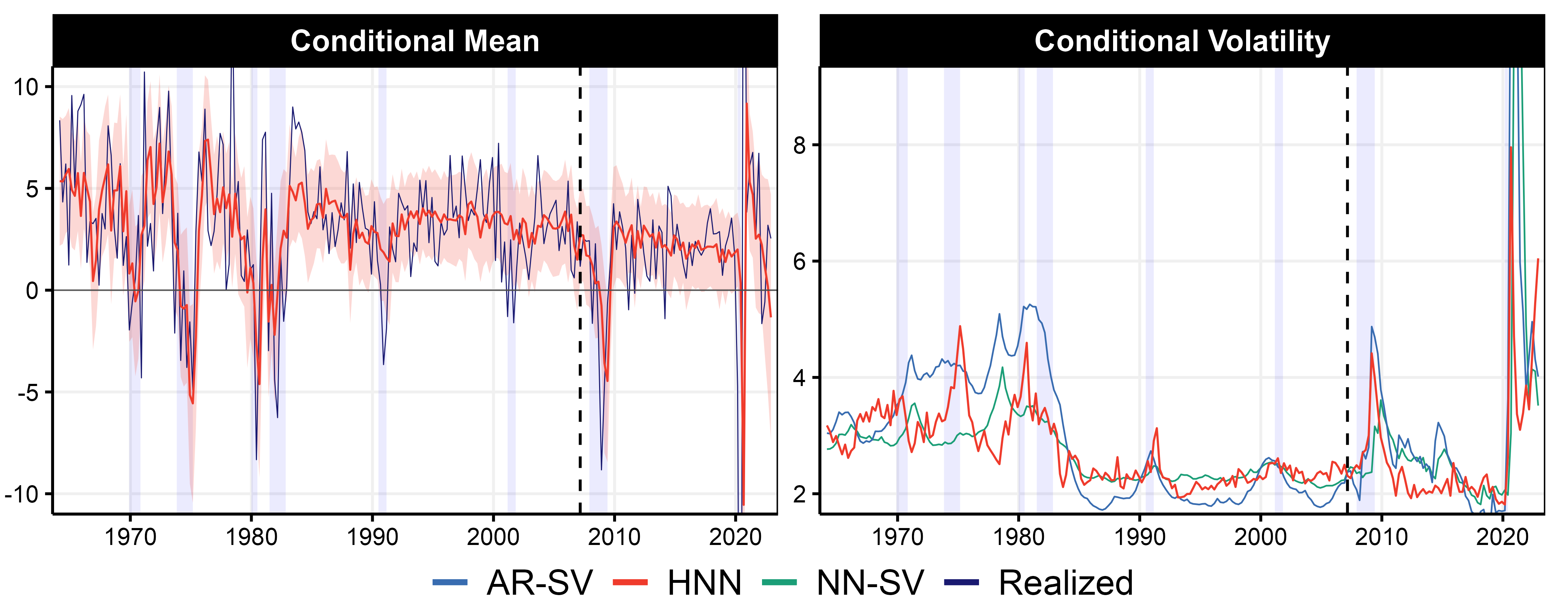}}
\end{minipage}

\hspace*{0.15cm}
\begin{minipage}{\linewidth}
\begin{threeparttable}
\center
\scriptsize
\footnotesize
\setlength{\tabcolsep}{0.225em}
 \setlength\extrarowheight{2.5pt}
 \begin{tabular}{l rrrrrrrrrrrrrrr | rrrrrrrrrrrrrrr} 
\toprule \toprule
\addlinespace[2pt]
& & \multicolumn{14}{c}{2007Q1 - 2019Q4} & &  \multicolumn{13}{c}{2007Q1 - 2022Q4, Excluding 2020} \\
\cmidrule(lr){3-15} \cmidrule(lr){18-30} \addlinespace[2pt]
& &  HNN &  & NN$_\text{\halftiny SV}$ &  & NN$_\text{\halftiny G}$ &  & {\notsotiny DeepAR} & & BART && AR$_\text{\halftiny SV}$  &  & BLR & & &
 HNN &  & NN$_\text{\halftiny SV}$ &  & NN$_\text{\halftiny G}$ &  & {\notsotiny DeepAR} & & BART && AR$_\text{\halftiny SV}$  &  & BLR  &\\
\midrule
\addlinespace[5pt] 
{\notsotiny RMSE} &   & \textbf{0.83} &   & 0.93 &   & 0.93 &   & 0.92 &   & 0.86 &   & 1.01 &   & 0.89 &   &   & \textbf{0.85} &   & 0.96 &   & 0.96 &   & 0.93 &   & 0.92 &   & 1.00 &   & 0.94 &   \\
$\mathcal{L}$ &   & \textbf{-3.93} &   & -3.82 &   & -3.80 &   & -3.18 &   & -3.88 &   & -3.75 &   & -3.69 &   &   & \textbf{-3.87} &   & -3.70 &   & -3.63 &   & -3.23 &   & -3.71 &   & -3.69 &   & -3.63 &   \\
$R^2_{|\varepsilon_t|}$ &   & \textbf{0.30} &   & 0.18 &   & 0.21 &   & 0.04 &   & 0.07 &   & -0.23 &   & -1.23 &   &   & \textbf{0.18} &   & -1.57 &   & -3.82 &   & 0.08 &   & -22.20 &   & -0.68 &   & -1.19 &   \\

\bottomrule \bottomrule
\end{tabular}
\begin{tablenotes}[para,flushleft]
  \scriptsize 
    \textit{Notes}: The upper panels show the conditional mean and the conditional variance for HNN and selected benchmarks. Up to 2006Q4 we show the in-sample results of the respective model followed by the out-of-sample results (from 2007Q1 to 2022Q4), indicated by the dotted line. The table presents the root mean square error (RMSE) relative to the AR model with constant variance, the log score ($\mathcal{L}$), and the $R^2_{|\varepsilon_t|}$ of absolute residuals.
  \end{tablenotes}
\end{threeparttable}
\end{minipage}\hfill
 
\end{minipage}
\end{figure}

\begin{figure}[!t]
\begin{minipage}{\linewidth}

\begin{minipage}{\linewidth}
\centering
\captionof{figure}{\normalsize \textbf{GDP} ($s=4$) \vspace*{-0.3cm}} \label{tab:GDPC1_h4_l2}
\includegraphics[width=1.02\textwidth]{{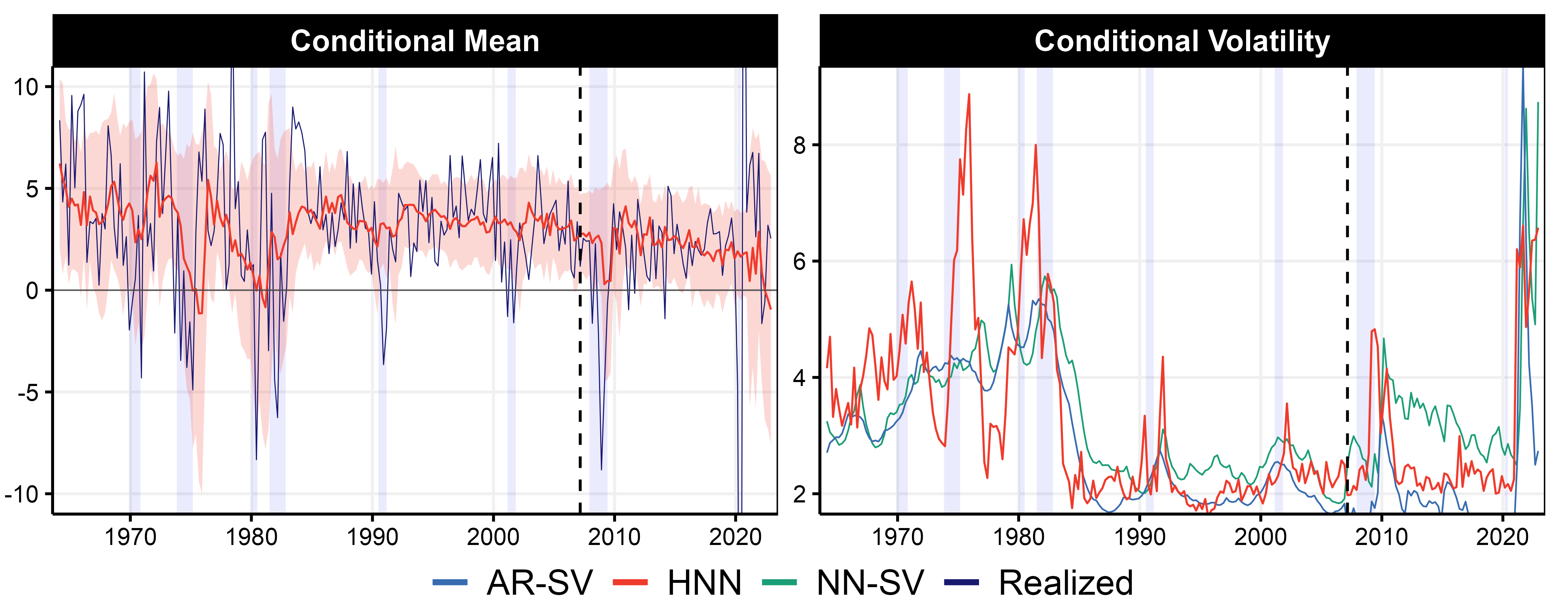}}
\end{minipage}

\hspace*{0.15cm}
\begin{minipage}{\linewidth}
\begin{threeparttable}

\center
\scriptsize
\footnotesize
\setlength{\tabcolsep}{0.225em}
 \setlength\extrarowheight{2.5pt}
 \begin{tabular}{l rrrrrrrrrrrrrrr | rrrrrrrrrrrrrrr} 
\toprule \toprule
\addlinespace[2pt]
& & \multicolumn{14}{c}{2007Q1 - 2019Q4} & &  \multicolumn{13}{c}{2007Q1 - 2022Q4, Excluding 2020} \\
\cmidrule(lr){3-15} \cmidrule(lr){18-30} \addlinespace[2pt]
& &  HNN &  & NN$_\text{\halftiny SV}$ &  & NN$_\text{\halftiny G}$ &  & {\notsotiny DeepAR} & & BART && AR$_\text{\halftiny SV}$  &  & BLR  & & &
 HNN &  & NN$_\text{\halftiny SV}$ &  & NN$_\text{\halftiny G}$ &  & {\notsotiny DeepAR} & & BART && AR$_\text{\halftiny SV}$  &  & BLR   &\\
\midrule
\addlinespace[5pt] 
{\notsotiny RMSE} &   & 0.90 &   & 0.88 &   & 0.88 &   & 1.07 &   & \textbf{0.88} &   & 0.99 &   & 0.91 &   &   & 0.85 &   & 1.46 &   & 1.46 &   & 0.99 &   & \textbf{0.81} &   & 0.98 &   & 0.95 &   \\
$\mathcal{L}$ &   & -3.70 &   & -3.54 &   & -3.52 &   & -2.83 &   & \textbf{-3.70} &   & -3.04 &   & -3.59 &   &   & \textbf{-3.61} &   & -3.33 &   & -3.36 &   & 1.27 &   & -3.55 &   & -3.05 &   & -3.51 &   \\
$R^2_{|\varepsilon_t|}$ &   & \textbf{0.28} &   & 0.12 &   & 0.27 &   & 0.09 &   & -0.03 &   & 0.07 &   & -0.67 &   &   & 0.06 &   & 0.07 &   & -0.08 &   & -0.03 &   & -9.86 &   & \textbf{0.07} &   & -0.41 &   \\
\bottomrule \bottomrule
\end{tabular}
\begin{tablenotes}[para,flushleft]
  \scriptsize 
    \textit{Notes}: For more details we refer to Figure \ref{tab:GDPC1_h1_l2}.
  \end{tablenotes}
\end{threeparttable}
\end{minipage}\hfill
 
\end{minipage}
\end{figure}

\begin{figure}[!t]
\begin{minipage}{\linewidth}

\begin{minipage}{\linewidth}
\centering
\captionof{figure}{\normalsize \textbf{Inflation}  ($s=1$)  \vspace*{-0.3cm}} \label{tab:CPIAUCSL_h1_l2}
\includegraphics[width=1.02\textwidth]{{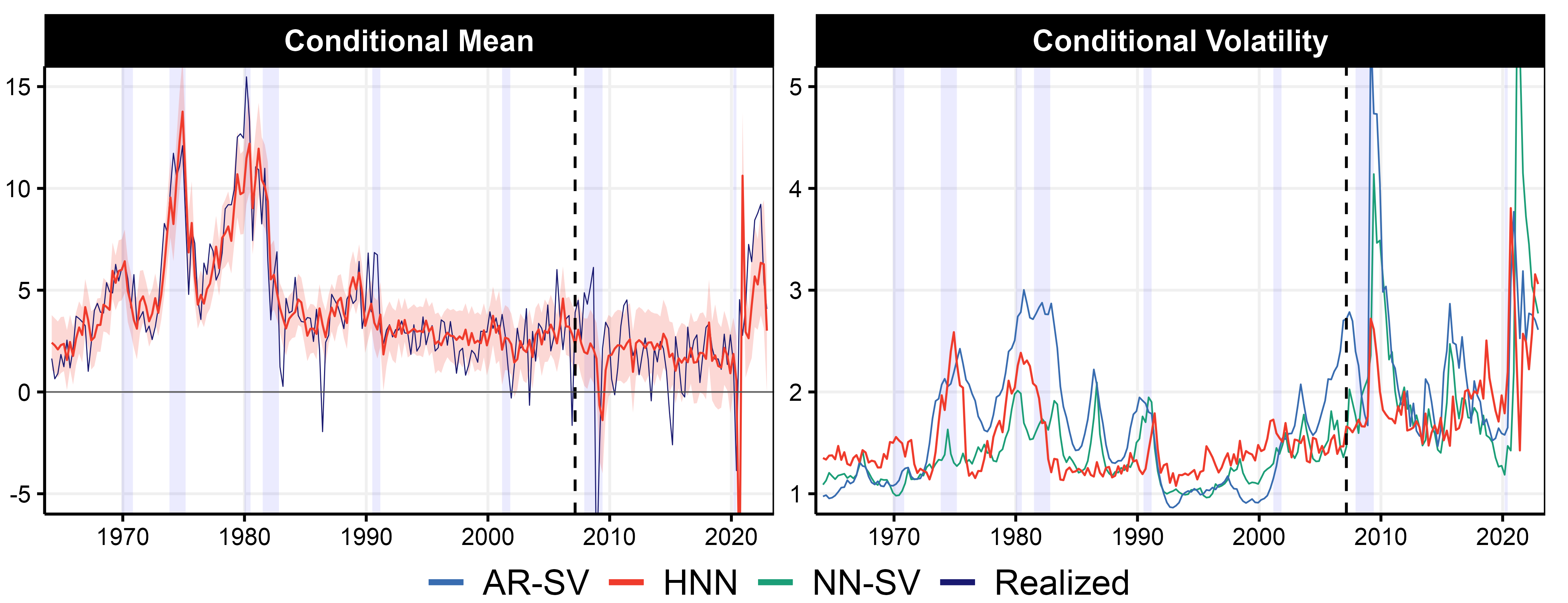}}
\end{minipage}

\hspace*{0.15cm}
\begin{minipage}{\linewidth}
\begin{threeparttable}
\center
\scriptsize
\footnotesize
\setlength{\tabcolsep}{0.225em}
 \setlength\extrarowheight{2.5pt}
 \begin{tabular}{l rrrrrrrrrrrrrrr | rrrrrrrrrrrrrrr} 
\toprule \toprule
\addlinespace[2pt]
& & \multicolumn{14}{c}{2007Q1 - 2019Q4} & &  \multicolumn{13}{c}{2007Q1 - 2022Q4} \\
\cmidrule(lr){3-15} \cmidrule(lr){18-30} \addlinespace[2pt]
& &  HNN &  & NN$_\text{\halftiny SV}$ &  & NN$_\text{\halftiny G}$ &  & {\notsotiny DeepAR} & & BART && AR$_\text{\halftiny SV}$  &  & BLR & & &
 HNN &  & NN$_\text{\halftiny SV}$ &  & NN$_\text{\halftiny G}$ &  & {\notsotiny DeepAR} & & BART && AR$_\text{\halftiny SV}$  &  & BLR  &\\
\midrule
\addlinespace[5pt] 

{\notsotiny RMSE} &   & \textbf{0.94} &   & 0.95 &   & 0.95 &   & 1.02 &   & 1.07 &   & 1.11 &   & 1.05 &   &   & 1.14 &   & 1.17 &   & 1.17 &   & \textbf{0.93} &   & 0.96 &   & 1.00 &   & 1.23 &   \\
$\mathcal{L}$ &   & -3.63 &   & -3.74 &   & \textbf{-3.82} &   & -3.57 &   & -2.91 &   & -3.26 &   & -3.60 &   &   & -3.41 &   & -2.72 &   & -3.47 &   & \textbf{-3.52} &   & -1.30 &   & -3.32 &   & -3.33 &   \\
$R^2_{|\varepsilon_t|}$ &   & -0.06 &   & -0.02 &   & -0.03 &   & \textbf{0.15} &   & -0.48 &   & 0.04 &   & -0.32 &   &   & \textbf{0.17} &   & -0.02 &   & 0.08 &   & 0.15 &   & -0.41 &   & 0.06 &   & -0.02 &   \\

\bottomrule \bottomrule
\end{tabular}
\begin{tablenotes}[para,flushleft]
  \scriptsize 
    \textit{Notes}: For more details we refer to Figure \ref{tab:GDPC1_h1_l2}.
  \end{tablenotes}
\end{threeparttable}
\end{minipage}\hfill
 
\end{minipage}
\end{figure}

\begin{figure}[!t]
\begin{minipage}{\linewidth}

\begin{minipage}{\linewidth}
\centering
\captionof{figure}{\normalsize \textbf{S\&P 500 } ($s=1$) \vspace*{-0.3cm}} \label{tab:SP500_h1_l2}
\includegraphics[width=1.02\textwidth]{{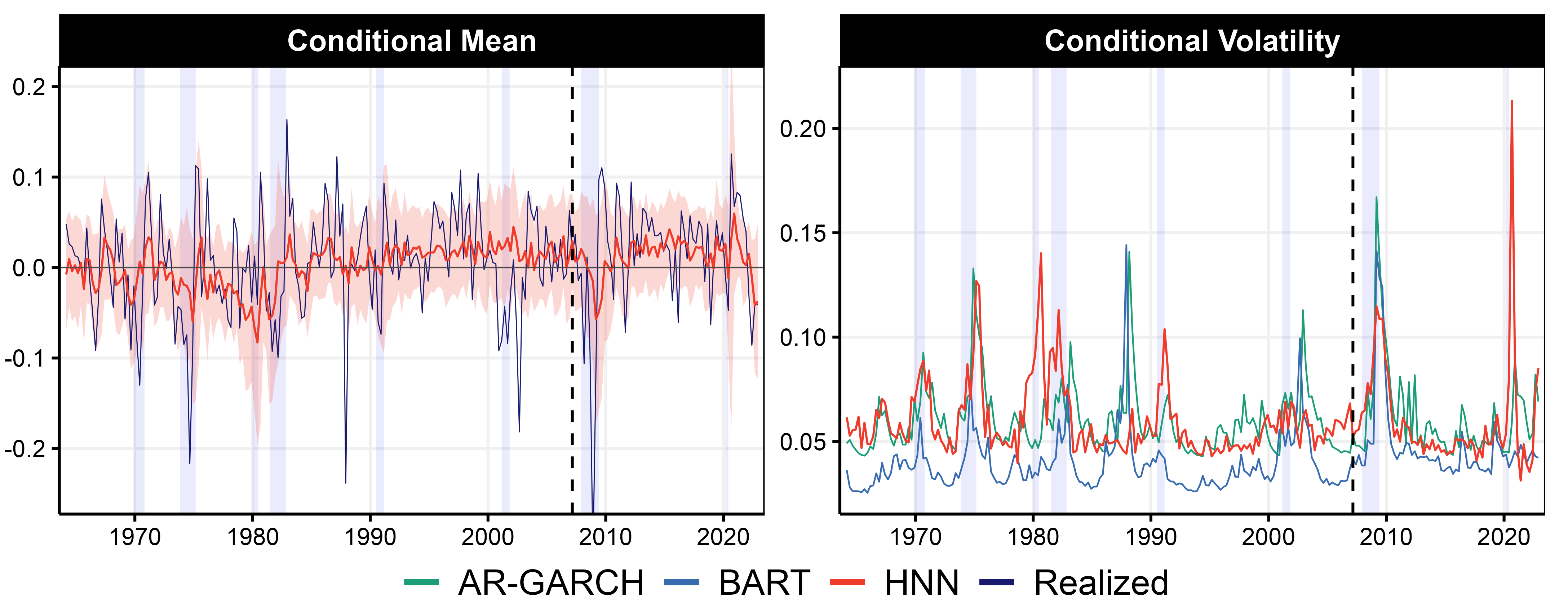}}
\end{minipage}

\hspace*{0.15cm}
\begin{minipage}{\linewidth}
\begin{threeparttable}
\center
\scriptsize
\footnotesize
\setlength{\tabcolsep}{0.225em}
 \setlength\extrarowheight{2.5pt}
 \begin{tabular}{l rrrrrrrrrrrrrrr | rrrrrrrrrrrrrrr} 
\toprule \toprule
\addlinespace[2pt]
& & \multicolumn{14}{c}{2007Q1 - 2019Q4}  & & \multicolumn{13}{c}{2007Q1 - 2022Q4} \\
\cmidrule(lr){3-15} \cmidrule(lr){18-30} \addlinespace[2pt]
& &  HNN &  & NN$_\text{\halftiny SV}$ &  & NN$_\text{\halftiny G}$ &  & {\notsotiny DeepAR} & & BART && AR$_\text{\halftiny G}$ &  &  BLR & & &
 HNN &  & NN$_\text{\halftiny SV}$ &  & NN$_\text{\halftiny G}$ &  & {\notsotiny DeepAR} & & BART && AR$_\text{\halftiny G}$  &  & BLR   &\\
\midrule
\addlinespace[5pt] 
{\notsotiny RMSE} &   & 0.96 &   & 1.09 &   & 1.09 &   & 1.02 &   & \textbf{0.92} &   & 0.94 &   & 0.98 &   &   & 0.93 &   & 1.04 &   & 1.04 &   & 1.06 &   & \textbf{0.89} &   & 0.92 &   & 0.96 &   \\
$\mathcal{L}$  &   & \textbf{-1.55} &   & -1.24 &   & -1.27 &   & -1.34 &   & -1.28 &   & -1.35 &   & -1.25 &   &   & \textbf{-1.52} &   & -1.30 &   & -1.32 &   & -1.13 &   & -1.34 &   & -1.39 &   & -1.29 &   \\
$R^2_{|\varepsilon_t|}$ &   & 0.26 &   & 0.04 &   & 0.06 &   & \textbf{0.30} &   & 0.11 &   & 0.24 &   & -0.13 &   &   & 0.07 &   & 0.04 &   & 0.05 &   & 0.22 &   & 0.12 &   & \textbf{0.25} &   & -0.14 &   \\
\bottomrule \bottomrule
\end{tabular}
\begin{tablenotes}[para,flushleft]
  \scriptsize 
    \textit{Notes}: For more details we refer to Figure \ref{tab:GDPC1_h1_l2}.
  \end{tablenotes}
\end{threeparttable}
\end{minipage}\hfill
 
\end{minipage}
\end{figure}

We report the main forecasting results through a series of dashboards for selected targets featuring essential statistics and visualizations. \textcolor{black}{This detailed analysis is conducted target by target to assess the individual performance and to provide some basic economic reasoning.}  We compare HNN's conditional mean and volatility paths to that of selected benchmarks in the upper left and upper right panel of Figure \ref{tab:GDPC1_h1_l2} to Figure \ref{tab:SP500_h1_l2}.  We plot in-sample estimates up to the start of our hold-out (i.e., up to 2007Q1) and the recursively re-estimated out-of-sample ones thereafter (i.e., from 2007Q1 to 2022Q4).  Our analysis is complemented by investigating the main drivers of the mean and the variance hemisphere.  For this purpose,  we measure Variable Importance (VI) as in \cite{MRF} and \cite{HNN}, which is itself inspired from what is traditionally done to interpret Random Forests \citep{breiman2001}. Details are given in Appendix \ref{app:VI}.  Additional results for leftover quarterly targets ($s=4$ cases  and unemployment) can be found in Appendix \ref{app:results}.

In general,  HNN exhibits remarkable adaptability when faced with a diverse range of series. It adeptly captures the Great Moderation pattern in real activity variables,  manages to produce a more "spiky" volatility pattern for the S\&P 500,  and sometimes exhibits behavior more akin to a smoothly switching process than to SV for predicting macroeconomic variables at longer horizons.  These higher volatility periods demonstrate a proactive nature by often preceding significant prediction errors. This behavior is observable both in-sample, with out-of-bag estimates, and out-of-sample. In terms of performance, HNN consistently ranks among the top models across all metrics.  Moreover,  it does not experience substantial undercoverage, which occasionally plague other sophisticated competing models.



\newpage

\hspace*{-0.7cm}

\vspace{-2.5em}
For the one-step ahead predictions of GDP growth HNN clearly outperforms all benchmarks. This finding holds for point and density forecasts as well as for both samples (see Figure \ref{tab:GDPC1_h1_l2}). Considering the sample ending in 2019Q4 we find that all nonlinear techniques yield high predictive power with HNN giving the lowest RMSE and the lowest log score ($\mathcal{L}$).  Moreover,  HNN gives the highest $R^2_{|\varepsilon_t|}$ at 30\%, distancing the nearest competitors NN$_\text{\halftiny SV}$ and NN$_\text{\halftiny G}$ by about 10 percentage points,  and BART/DeepAR by more than 20.   Similar findings are obtained from including Covid-19 pandemic observations. Coupled with the good forecasting performance, this implies that our model captures a substantial part of the "realized volatility". Extending the hold-out to the end of 2022 reveals that HNN also yields a good performance after the Covid-19 pandemic whereas other models (especially, BART and BLR) lose ground against the linear benchmark. 



The right panel of Figure \ref{tab:GDPC1_h1_l2} nicely demonstrates the reactive and proactive behavior of HNN's volatility hemisphere. The variance path increases at early stages of turmoil despite accurate predictions in previous periods,  and is thus better prepared to receive larger errors than the green line corresponding to the SV specification.  In fact,  HNN shoots up at about the same time as AR$_\text{\halftiny SV}$,  for which higher predictions errors have already been accumulating at that point.  In line with \cite{adrian2019}, we find that the conditional variance is affected by developments in financial markets whereas the predictive mean is driven by labor market variables as well as real activity measures such as imports, exports and manufacturers' new orders (see Figure \ref{fig:vi_gdps1} in Appendix \ref{app:VI}).



\hspace*{-0.7cm}

\vspace{-2em}
Figure \ref{tab:GDPC1_h4_l2} shows the results for the one-year ahead prediction of GDP growth. Again, high-dimensional models (except for DeepAR) yield high predictive accuracy when focusing on the hold-out ending in 2019. BART gives the best point forecasting performance, closely followed by BLR and NN specifications. In terms of log scores, HNN and BART clearly outperfom the other models. Moreover, $R^2_{|\varepsilon_t|}$ shows that HNN explains nearly a third of the realized volatility similar to NN$_\text{\halftiny G}$. When including the periods after 2020, HNN beats all its linear and nonlinear competitors with respect to density forecasting performance. While alternative NN specifications perform rather poorly at the end of 2021 and the beginning of 2022, HNN yields highly competitive predictions and acknowledges the elevated uncertainty until the end of the sample. 


Similar to the one-step ahead case, the volatility hemisphere shows proactive tendencies. The conditional variance picks up the uncertainty in the underlying data set early, resulting in superior density predictions. During the Great Moderation and in the periods after the Global Financial Crisis the variance is low and stable, narrowing the predictive distribution to rather certain estimates. Note that this also holds for the one-step ahead case. Again, variables measuring financial conditions are important drivers of the mean and the variance hemisphere. These include debt-to-income ratios of several sectors in the US economy as well as real disposable business income (see Figure \ref{fig:vi_gdps4} in the appendix). As shown by \cite{adrian2019}, the impact of financial conditions on GDP growth seems to be robust at multiple horizons. 
Similarly, \cite{de2017forecasting} and \cite{amburgey2023real}, amongst others, emphasize the importance of financial conditions on real activity, especially in the tails.

\hspace*{-0.7cm}

\vspace{-2em}
When interest centers on one-step ahead inflation predictions (see Figure \ref{tab:CPIAUCSL_h1_l2}) we find that our neural network models yield high forecasting performance for both point and density predictions as well as both samples. HNN outperforms all other models with respect to point forecasting performance in terms of RMSE and NN$_\text{\halftiny G}$ gives the lowest log score for density performance, closely followed by HNN. A similar pattern is observed when extending the sample to the end of 2020. 
Noteworthy, we see that the HNN overestimates the effects of the Covid-19 pandemic in its mean estimate which is, however, accompanied by a large variance, implying that the model acknowledges the unprecedentedly high uncertainty involved. This nearly completely discounts the dramatic deflation forecast and results in a highly competitive $\mathcal{L}$.  During this time, the variance hemisphere is mainly driven by employment variables and money stock which were heavily affected by the Covid-19 shock and exhibited major fluctuations in 2020 (see Figure \ref{fig:vi_cpis1} in the appendix). BART gives the worst log scores compared to the other models as it tends to underestimate the variance during most periods. 

Given the policy needs for interpretable inflation forecasts, we extend the HNN to a more structural approach proposed in \cite{HNN}. Relying on a nonlinear Phillips curve specification, the architecture of the neural network is designed to provide among other things a measurement of economic slack and inflation expectations.  As will be shown in Section \ref{sec:npc}, the Neural Phillips Curve (NPC) model equipped with a mean and a variance hemisphere predicts inflation reasonably well and substantially outperforms its competitors.


\hspace*{-0.7cm}

\vspace{-2em}
Results for the quarterly forecasts of the S\&P 500 presented in Figure \ref{tab:SP500_h1_l2} show that our HNN outperforms all competitors in terms of density predictions and yields highly competitive point forecasts following BART and AR$_\text{\halftiny G}$. Moreover, HNN explains almost a third of the variance of absolute residuals, similar to the DeepAR, but with allegedly much less "leftover conditional mean predictability" in it given DeepAR's higher RMSE.  

When comparing the conditional volatility estimated by the set of models, we see that the predictive variance of HNN follows a different pattern than those of the other models. Our proposed approach attaches higher weight to macro uncertainty and gives a countercyclical variance path (see the upper left panel of Figure \ref{tab:SP500_h1_l2}).  Since the architecture of HNN allows for proactive and reactive volatility,  the predictive variance takes into account signals from the input data set while at the same time accommodates for various forms of time variation. This way, it offers great flexibility going beyond the reactive structure of GARCH and SV processes and accounts for nonlinear relations between the covariates and the target. It relates to the strand of literature exploring the predictive power of exogenous variables for forecasting stock market volatility \citep[see, e.g., ][]{campbell2009stock,paye2012deja,engle2013stock,guidolin2021boosting,ma2022stock} and sheds light on the economic sources of the volatility process. We find that the variables shaping the variance hemisphere are related to both, financial and economic conditions. Main drivers are variables closely moving with the economic business cycle, such as new housing permits, imports and employment, but also variables measuring financial conditions and stock market variables (see Figure \ref{fig:vi_sp500s1} in the appendix).

\hspace*{-0.7cm}
\begin{figure}[!t]
\begin{minipage}{\linewidth}

\begin{minipage}{\linewidth}
\centering
\captionof{figure}{\normalsize \textbf{Housing Starts} ($s=1$) \vspace*{-0.3cm}} \label{tab:HOUST_h1_l2}
\includegraphics[width=1.02\textwidth]{{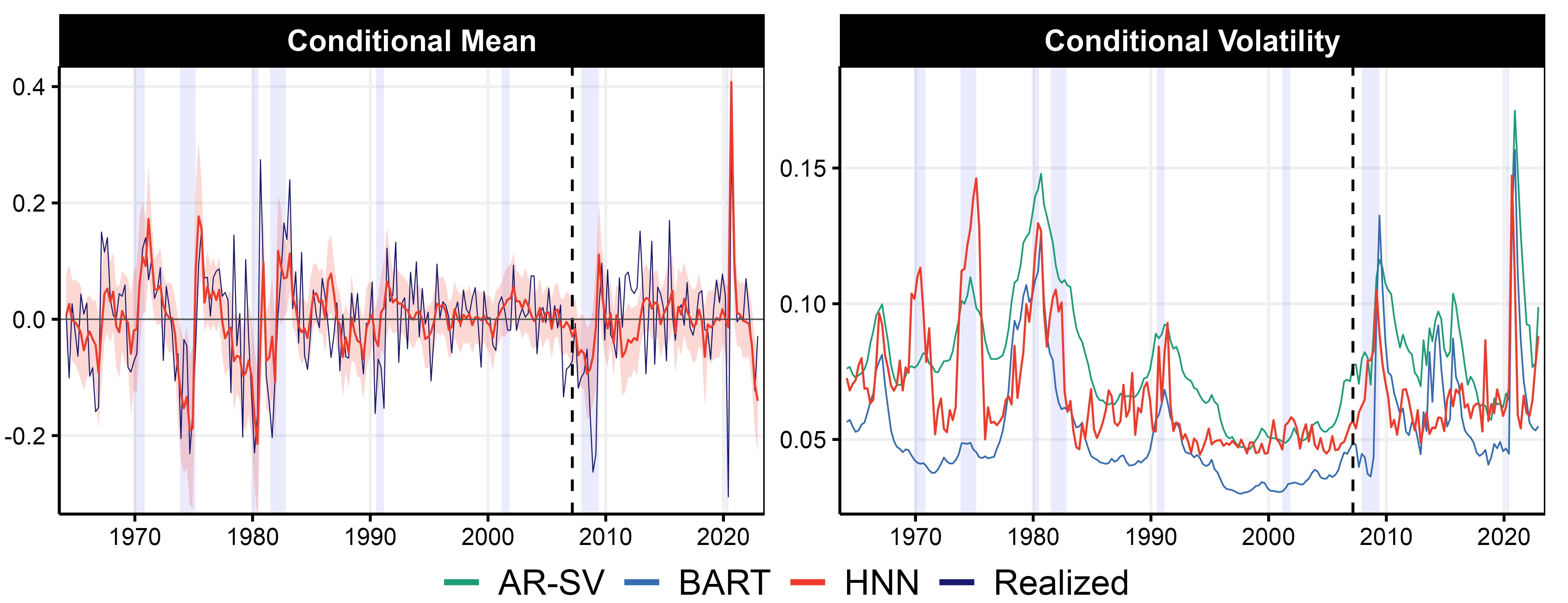}}
\end{minipage}

\hspace*{0.15cm}
\begin{minipage}{\linewidth}
\begin{threeparttable}
\center
\scriptsize
\footnotesize
\setlength{\tabcolsep}{0.225em}
 \setlength\extrarowheight{2.5pt}
 \begin{tabular}{l rrrrrrrrrrrrrrr | rrrrrrrrrrrrrrr} 
\toprule \toprule
\addlinespace[2pt]
& & \multicolumn{14}{c}{2007Q1 - 2019Q4} & &  \multicolumn{13}{c}{2007Q1 - 2022Q4} \\
\cmidrule(lr){3-15} \cmidrule(lr){18-30} \addlinespace[2pt]
& &  HNN &  & NN$_\text{\halftiny SV}$ &  & NN$_\text{\halftiny G}$ &  & {\notsotiny DeepAR} & & BART && AR$_\text{\halftiny SV}$  &  & BLR  & & &
 HNN &  & NN$_\text{\halftiny SV}$ &  & NN$_\text{\halftiny G}$ &  & {\notsotiny DeepAR} & & BART && AR$_\text{\halftiny SV}$  &  & BLR   &\\
\midrule
\addlinespace[5pt] 
{\notsotiny RMSE} &   & 0.99 &   & 1.01 &   & 1.01 &   & 1.06 &   & 0.96 &   & 0.99 &   & \textbf{0.96} &   &   & \textbf{0.86} &   & 0.87 &   & 0.87 &   & 0.97 &   & 0.93 &   & 1.00 &   & 0.99 &   \\
$\mathcal{L}$ &   & -1.14 &   & -1.08 &   & -1.08 &   & 0.07 &   & -0.98 &   & -1.15 &   & \textbf{-1.16} &   &   & -1.07 &   & -0.97 &   & -0.93 &   & -0.05 &   & -0.67 &   & \textbf{-1.15} &   & -0.92 &   \\
$R^2_{|\varepsilon_t|}$ &   & 0.14 &   & -0.06 &   & -0.01 &   & -0.14 &   & -0.03 &   & \textbf{0.36} &   & -0.27 &   &   & 0.09 &   & -0.09 &   & -0.12 &   & 0.02 &   & 0.03 &   & \textbf{0.15} &   & -0.03 &   \\
\bottomrule \bottomrule
\end{tabular}
\begin{tablenotes}[para,flushleft]
  \scriptsize 
    \textit{Notes}: For more details we refer to Figure \ref{tab:GDPC1_h1_l2}.
  \end{tablenotes}
\end{threeparttable}
\end{minipage}\hfill
 
\end{minipage}
\end{figure}  

\vspace{-2em}
Turning to the short-term predictions of housing starts, which is presented in Figure \ref{tab:HOUST_h1_l2}, we see a remarkable performance of the neural network models for the periods during and after the Covid-19 pandemic. In terms of density forecasts, HNN's predictive accuracy is only challenged by the highly competitive performance of AR$_\text{\halftiny SV}$. 
While BLR outperforms all competitors for point and density predictions before 2020, controlling for nonlinearities gains in importance thereafter. All nonlinear models catch up on the AR benchmark with HNN yielding the lowest RMSE and a $\mathcal{L}$, which is comparable to AR$_\text{\halftiny SV}$. Moreover, our proposed model explains about 10 \% of the realized volatility measured by the $R^2_{|\varepsilon_t|}$ of absolute residuals, following AR$_\text{\halftiny SV}$ which, however, gives a substantially higher RMSE.

Visual inspection of the conditional mean of the HNN (see the upper right panel of Figure \ref{tab:HOUST_h1_l2}) reveals some noteworthy patterns for the observations during the Covid-19 pandemic in 2020. Even though HNN underestimates the unprecedented downturn in the first quarter of 2020, it manages to take advantage of the signals provided by the unconventional behavior of various variables in the set of covariates for more accurate point and density forecasts than its competitors in the following periods. 
This raises the question: what drives this? For both hemispheres, we find that financial conditions have high predictive power. Moreover, new housing permits as well as commodity price developments (in particular, metals and fuels) play an important role (see Figure \ref{fig:vi_housts1} in the appendix).

\subsection{Calibration and Alternative Density Forecasts Evaluation Metrics}\label{sec:calibration}


Given the remarkably consistent performance of the HNN's density predictions,  we challenge these results by adding  evaluation metrics including the continuous ranked probability score (CRPS), the 68 \% coverage rate and a PIT-based test for auto-calibration \citep{knuppel2022score}.  First, we compute the CRPS introduced by \cite{gneiting2007strictly}, which is a proper scoring rule for predictive distributions and enjoys the advantage of being less sensitive to outliers. 
Let $F$ denote the cumulative distribution function and $\mathfrak{f}$ the predictive density with $\hat{y}_{t,s}$ and $\hat{y}'_{t,s}$ being independent random draws from the predictive density. The CRPS is then defined as
$$\text{CRPS}_{t,s}(y_{t,s}) = \int_{-\infty}^{\infty} (F(z) - \mathbbm{1}\{y_{t,s} \leq z \})^2 dz = E_{\mathfrak{f}} \lvert \hat{y}_{t,s} - y_{t,s} \rvert - 0.5 E_{\mathfrak{f}} \lvert \hat{y}_{t,s} - \hat{y}'_{t,s} \rvert,$$
where $\mathbbm{1}\{y_{t,s} \leq z \}$ defines an indicator function, which returns the value 1 if $y_{t,s} \leq z$ and 0 otherwise. In the figure below we report the CRPS averaged across the hold-out relative to our AR benchmark. 

Second, we consider the nominal coverage rate, which measures the frequency of the forecasts falling inside a specific interval. The predictive density is considered too wide (narrow), if the realized frequency exceeds (drops below) the nominal level chosen for the interval.  Formally, this boils down to
$$I_{t,s}^{\gamma} = 
\begin{cases} 
1 \text{ if } \hat{y}_{t,s} \in [L_{t,s}^{\gamma}, H_{t,s}^{\gamma}] \\
0 \text{ if } \hat{y}_{t,s} \notin [L_{t,s}^{\gamma}, H_{t,s}^{\gamma}],
\end{cases}$$
where $L_{t,s}^{\gamma}$ and $H_{t,s}^{\gamma}$ are the lower and upper limits of the interval. We compare the relative frequency of interval hits, $\hat{\gamma}_{s} = \frac{1}{\#\text{OOS}}\sum_{t \in \text{OOS}} I_{t,s}^{\gamma}$, to the pre-specified coverage rate $\gamma$, which we set to 68 \%. 

The last metrics tests for auto-calibration based on probability integral transforms (PITs). Following \cite{knuppel2022score} we use the PIT of the implied forecast distribution based on the energy score, which is given by
$${U}_{ES,t,s} =  \mathcal{P}_{\mathfrak{f}} \left( E_{\mathfrak{f}} || \hat{y}_{t,s}-\hat{y}'_{t,s} || \leq E_{\mathfrak{f}} || \hat{y}_{t,s}-y_{t,s} || \right),$$ 
where $|| \cdot ||$ gives the Eucledian distance and $\mathcal{P}_{\mathfrak{f}}$ and $E_{\mathfrak{f}}$ the probability and the expected value under the forecast distribution, respectively. We then test for standard uniformity of $\{U_{ES,t}\}_{t=1}^T$. A model is said to be well calibrated if we do not reject the null hypothesis of auto-calibration. This implies that regardless of any further transformations the forecast distribution will not improve. 

\begin{figure}
\vspace*{0.45em}
\caption{Alternative Evaluation Metrics}\label{fig:modelcalib}
\begin{minipage}{0.5\textwidth}
\centering \hspace*{0.5cm} 2007Q1 - 2019Q4
\end{minipage}
\begin{minipage}{0.5\textwidth}
\centering \hspace*{0.5cm} 2007Q1 - 2022Q4
\end{minipage}

\begin{minipage}{\textwidth}
\centering \hspace*{0.7cm} \textbf{Relative CRPS}
\end{minipage}

\begin{minipage}{0.5\textwidth}
\includegraphics[scale = 0.58]{{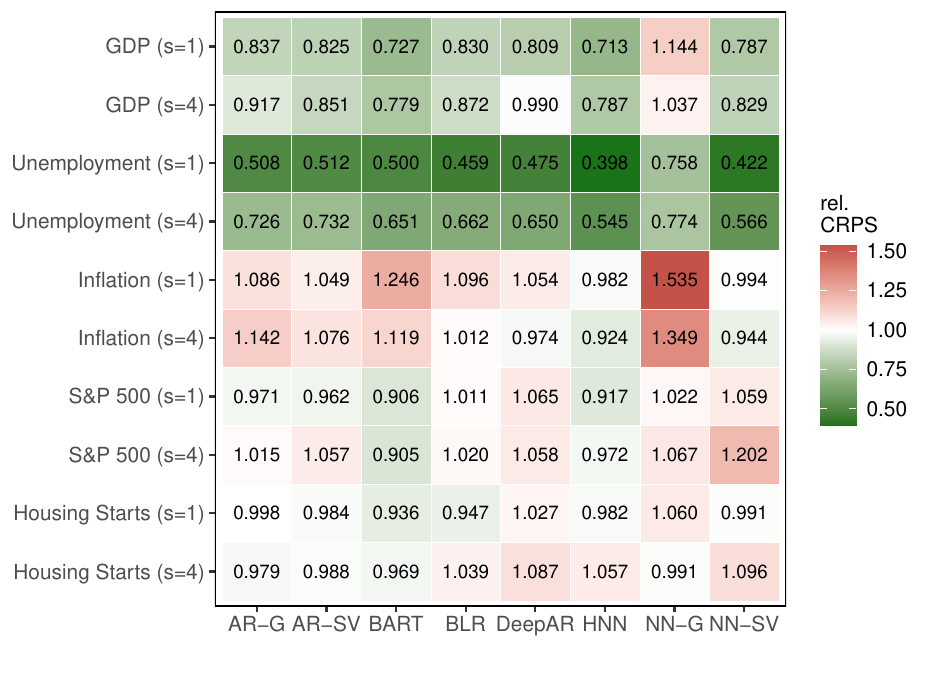}}
\end{minipage}
\begin{minipage}{0.5\textwidth}
\includegraphics[scale = 0.58]{{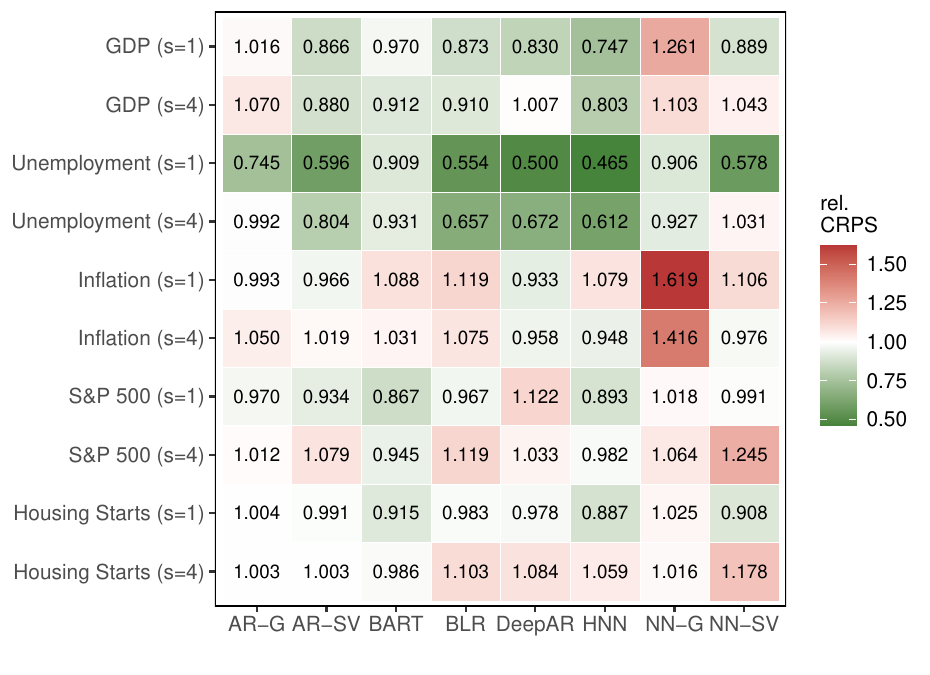}}
\end{minipage}

\begin{minipage}{\textwidth}
\centering \hspace*{0.7cm} \textbf{Coverage 68\%}
\end{minipage}

\begin{minipage}{0.5\textwidth}
\includegraphics[scale = 0.58]{{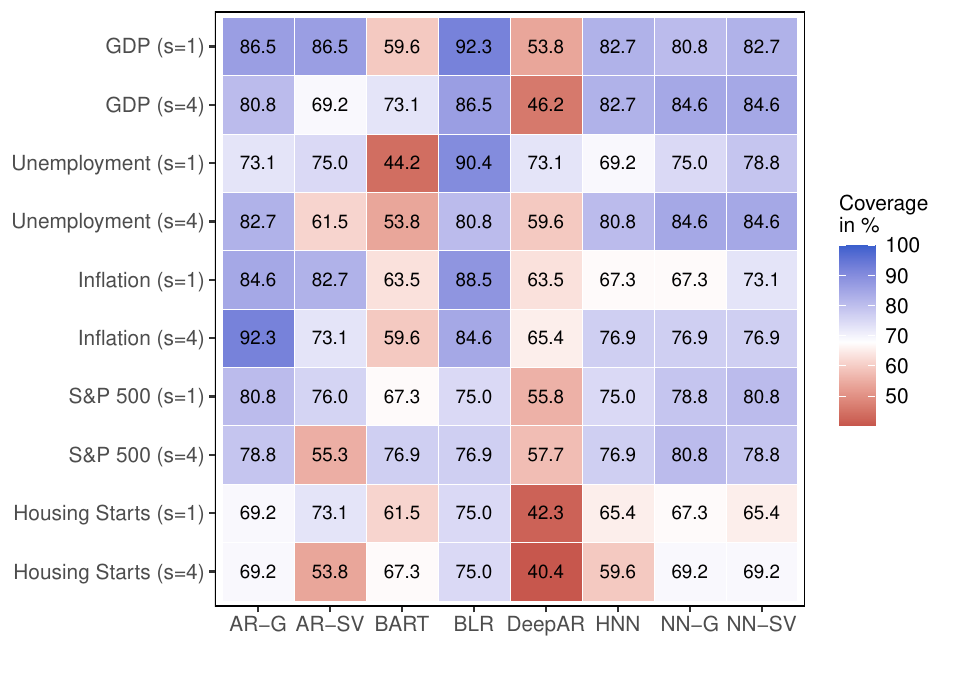}}
\end{minipage}
\begin{minipage}{0.5\textwidth}
\includegraphics[scale = 0.58]{{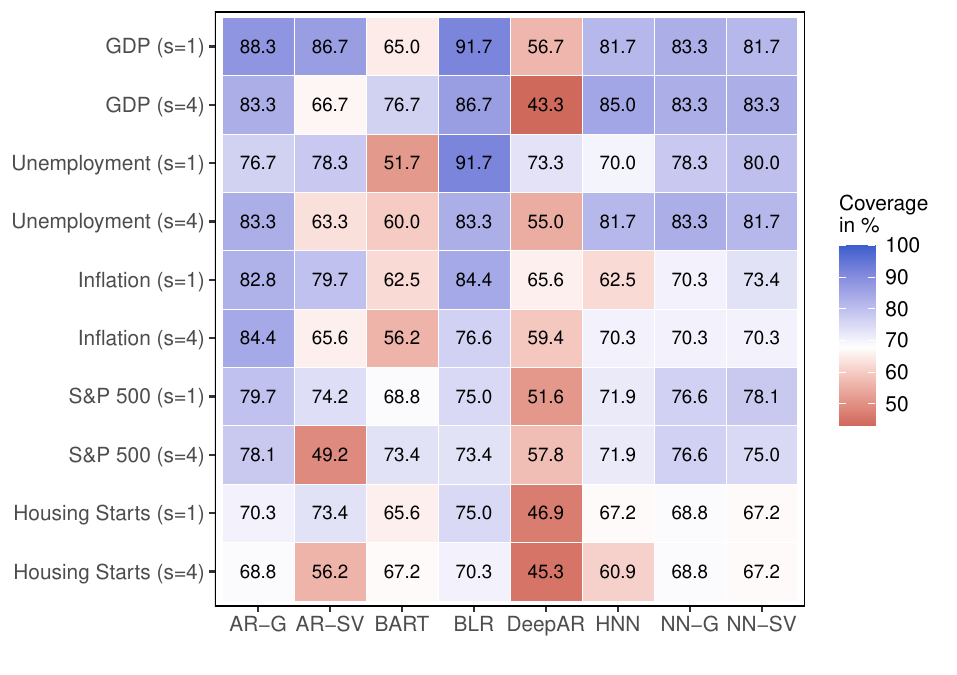}}
\end{minipage}

\begin{minipage}{\textwidth}
\centering \hspace*{0.7cm} \textbf{Auto-Calibration Test p-values (PIT-based)}
\end{minipage}

\begin{minipage}{0.5\textwidth}
\includegraphics[scale = 0.58]{{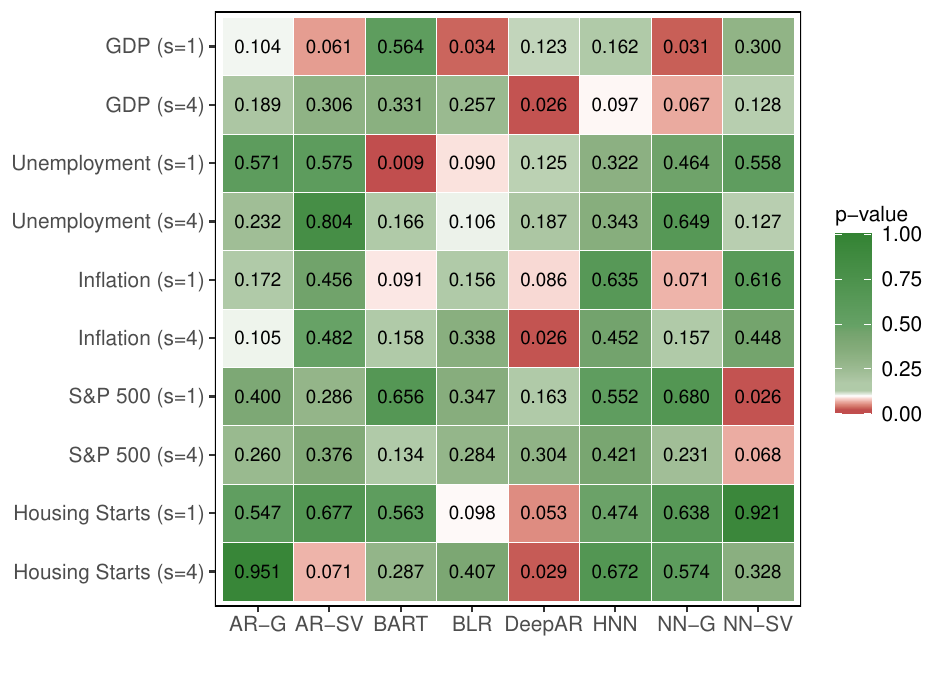}}
\end{minipage}
\begin{minipage}{0.5\textwidth}
\includegraphics[scale = 0.58]{{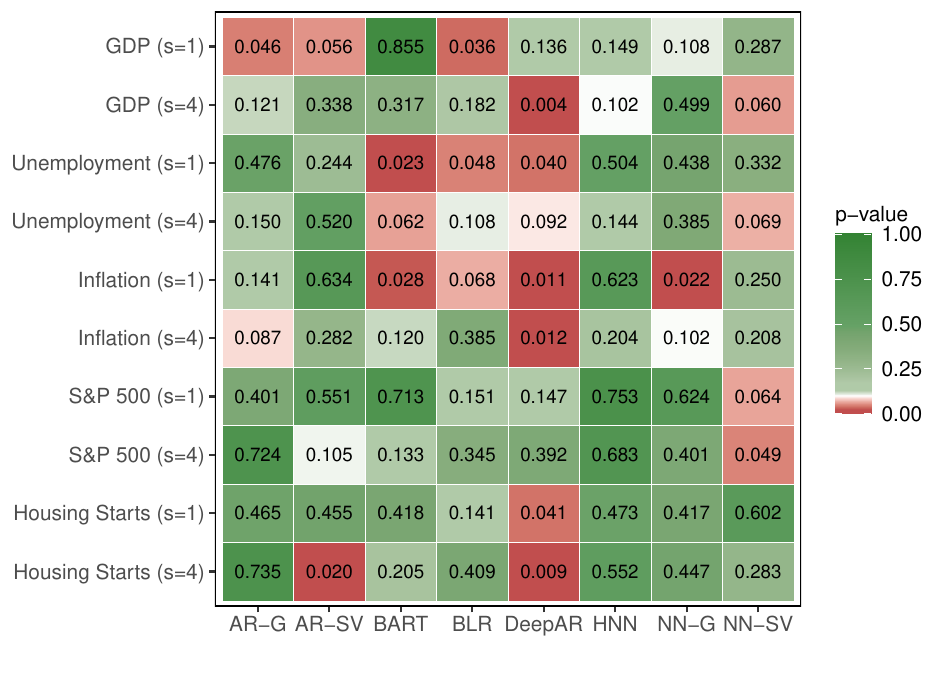}}
\end{minipage}

\begin{minipage}{\textwidth}
\scriptsize
\textit{Notes:} The upper panel shows continuous ranked probability score (CRPS) relative to the AR model with constant variance. The middle panel gives the 68 \% coverage rate. The lower panel tests for auto-calibration based on probability integral transforms (PITs) as proposed in \cite{knuppel2022score}. For the real activity variables (i.e., GDP and unemployment) we exclude the year 2020 from the evaluation sample.
\end{minipage}

\end{figure}

Figure \ref{fig:modelcalib} presents the additional scores for both evaluation samples. The left panels focus on the sample ending before the Covid-19 pandemic (i.e., 2007Q1 to 2019Q4) whereas the right panels present the full sample results (i.e., 2007Q1 to 2022Q4). 
Overall, the results confirm the promising performance of HNN. The relative CRPS shows substantial improvements of our approach against the AR model for most targets. The coverage rate is close to the selected level (which is 68 \%) and shows no tendency of structurally underestimating the variance. Unlike other models, the HNN shows no evidence against auto-calibration. 

Analyzing each metric in more details gives some interesting insights. Comparing the relative CRPS of the HNN with other nonlinear models underlines its remarkable density forecasting performance. HNN outperforms the linear benchmark in all cases except for high-order forecasts of housing starts and one-step ahead inflation forecasts when considering the full sample. Largest gains compared to the AR benchmark can be found for the unemployment rate.  This holds for all models.  In the case of HNN,  this is attributable to excellent point and density forecasts (almost 30\% reduction in RMSE for both horizons combined with a very high $R^2_{|\varepsilon_t|}$ for $h=4$ in Table \ref{tab:add_targets}).   BART often yields competitive results except for inflation.  DeepAR and BLR often show similar density forecasting performance to the AR benchmark, yielding scores close to 1. The NN models with time-varying volatility are often worse and only sometimes better (e.g., for unemployment) than linear models.

The coverage rate shows substantial underestimation of the variance for BART and DeepAR. The predictive densities of BART and DeepAR are too narrow for most targets and both evaluation samples. AR$_\text{\halftiny SV}$ gives a rather mixed picture with density predictions of GPD growth and inflation one-step ahead being too wide while higher-order densities being too narrow. HNN, on the other hand, gives ratios close to the 68\% level. For two target variables (i.e., GDP growth and four steps ahead unemployment), we get densities that are rather wide whereas the predictive distributions for housing starts and short-term inflation show slight underestimation of the variance. AR$_\text{\halftiny G}$ as well as BLR tend to give wide predictive distributions with coverage ratios close to 80 \% or above. Similarly, the NNs with SV or GARCH yield rather wide densities. 

The test results for auto-calibration, presented in the last two panels in Figure \ref{fig:modelcalib}, suggest that the models are well calibrated for most targets with the exception of DeepAR. In this case, the test clearly rejects auto-calibration for inflation and housing starts regardless of the forecast horizon and GDP growth for four steps ahead predictions. When considering the full sample, this also holds for the unemployment rate. The other competitors give low p-values for at least one or two of the estimated target variables and even more when focusing on the full sample. HNN, on the other hand, shows auto-calibrated results for all targets when evaluating the full sample and all but one target (i.e., GDP four steps ahead at the 10\% level) when considering the periods before the Covid-19 pandemic.

\subsection{An Understanding of BART and DeepAR's Difficulties }\label{sec:deepmarde}


Some results from the previous section warrant a digression from our main thread of investigation.  As is particularly apparent from coverage results in Section \ref{sec:calibration},  but also from log scores throughout  Section \ref{sec:fcast_results} and additional results in Table \ref{app:results},  the quality of BART and DeepAR's \textit{probabilistic} forecasts is rather uneven.   In both cases,  point forecasts often rank very highly but $\mathcal{L}$ and other density evaluation metrics show clear signs of distress.  While BART's problems are frequently contained to in-sample historical estimates (quite visible in Figures \ref{tab:SP500_h1_l2} and \ref{tab:HOUST_h1_l2}),  that of DeepAR are rather generalized.  This section first describes the facts,  and then provides suggestive explanations for the phenomena. Finally, we discuss what can be done in both cases to alleviate such substantial problems so that,  hopefully,  the operation produces some wisdom to draw from for future applications of such methods.


First, let us stress that it is certainly not excluded that an extensive amount of tuning for both could non-trivially improve the probabilistic performance of both approaches,  but this is not what is typically seen in the literature \citep{BART,green2012modeling,linero2018bayesian}.  There are practical reasons, of course,  but also statistical ones,  like the instability of cross-validation in such environments,  or that tuning hyperparameters is not exactly Bayesian.   Lastly,  we typically expect  proper calibration (whether the conditional mean is very proficient or not) to be obtained independently of tuning.  For instance,  this is what we get from AR$_\text{\halftiny SV}$, HNN,  and NN$_\text{\halftiny SV}$.  

A particularly telling example is the following.  HNN turns in similar outperformance for both GDP and unemployment at the two horizons under study---as one would rightfully expect from two strongly cross-correlated targets.   While HNN is decisively superior for both targets at $s=1$,  BART and HNN yield a nearly identical (and stellar) log score for GDP ($s=4$).   Yet,  BART ranks last among all models in terms of log score for Unemployment Rate ($s=4$) whereas HNN reaches gains similar to GDP ($s=4$).  Rarely does it hurt to look at the data underlying summary statistics,   and we use the case of unemployment at $s=4$ to guide the following discussion.  Moreover, it is a target for which the inherent "true" uncertainty is manifest.  

Figure \ref{bartdeepartrouble} reports times series corresponding to the second panel of Table \ref{tab:add_targets} in Appendix \ref{app:results} where BART and DeepAR are reported to have fine RMSEs with dismal log scores. We compare conditional means of HNN, BART and DeepAR to the realized value in the left panel and show conditional volatility in the right panel. In-sample, BART's and DeepAR's fitted values nearly perfectly overlap with the realized ones, and are suggestive of an unrealistically good predictive ability. Accordingly, BART estimates a very low volatility path with little fluctuations throughout these periods.  DeepAR inconsistently estimates the general volatility level to be at roughly the same level as HNN (for this specific case) and thus reports massive over-coverage in-sample. 
While BART's volatility estimates seem to be reasonable for the hold-out sample, those of DeepAR follow a strange path.  As mentioned above, this behavior is not exceptional to this case,  but is recurrent. For instance,  a similar behavior is observed for S\&P 500 (Figure \ref{tab:SP500_h1_l2}) as well as housing starts (Figure \ref{tab:HOUST_h1_l2}) where BART reports noticeably lower estimates of average volatility than either AR$_\text{\halftiny SV}$,  AR$_\text{\halftiny G}$ or HNN.  This suggests probabilistic forecasts of these models often lack appropriate levels of uncertainty and historical analysis based on in-sample results, which are frequently conducted in macroeconometrics, may provoke misleading implications and, hence, should be interpreted with care.

\begin{figure}[h!]
\vspace*{0.75em}
\caption{\textbf{Unemployment Rate} ($s=4$)}\label{bartdeepartrouble}  
\vspace*{-0.35cm}
\begin{center} 
\hspace*{-0.1cm}\includegraphics[width=1.02\textwidth]{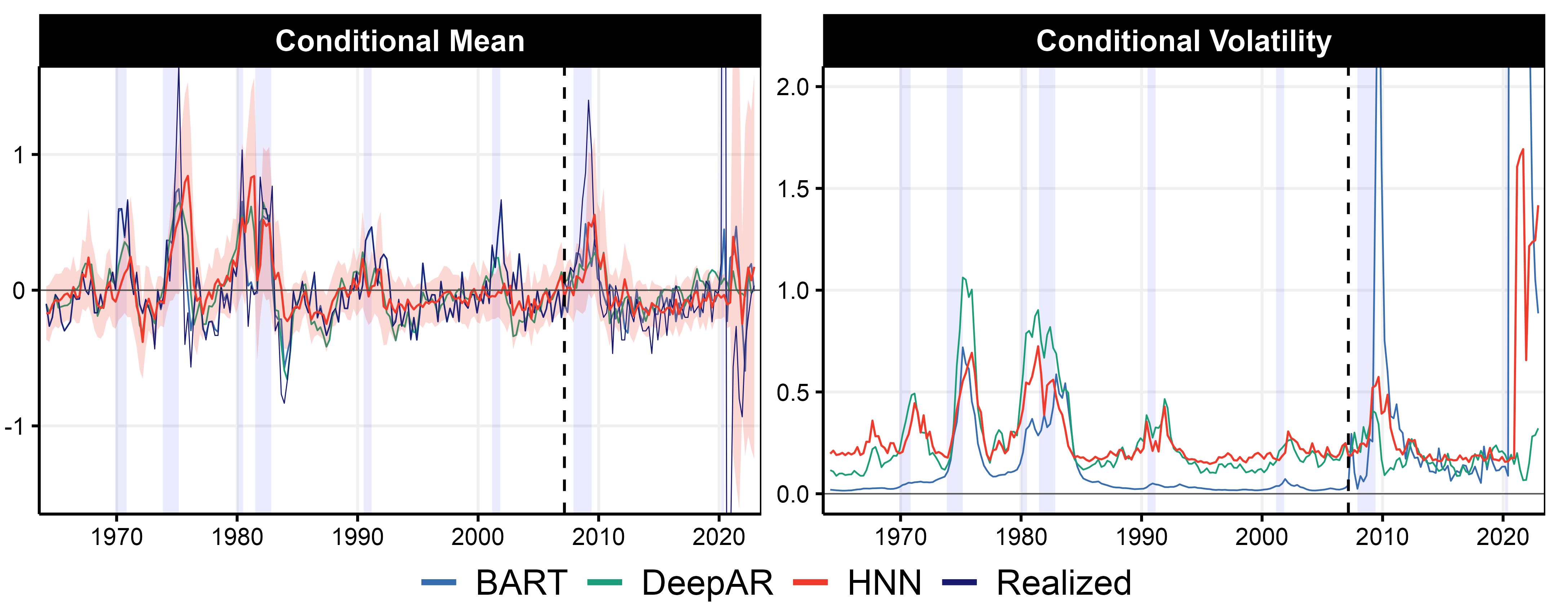}
\end{center}
\end{figure} 

What is causing this? In short, benign overfitting is,  as the name suggests,  benign for the conditional mean (out-of-sample).  In our results,  BART and DeepAR exhibit this phenomenon that has been described for neural networks \citep{belkin2019reconciling} and tree ensembles \citep{MSoRF}.  Without further precautions,  this overfitting is, however, malign for the conditional variance. The discussion of Ingredients  2 and 3  in Section \ref{sec:HNN} already alluded that such problems would arise if not addressed directly,  which we do for HNN by introducing the volatility emphasis parameter $\nu$ and using blocked subsampling to recalibrate the variance hemisphere.  In practical terms, if one is only interested in minimizing out-of-sample RMSEs, the near-perfect in-sample fits attained by BART and DeepAR can safely be ignored. However, when it comes to deeper investigations, such as uncertainty quantification or in-sample analysis, the best course of action is to tread lightly. At this juncture, it becomes preferable to inspect separately the two models.

The overfitting issues of BART are more blatant in-sample. Hence, our results suggest that one should be careful using BART estimates in-sample to draw any economic conclusion \textit{even if} BART yields the best (point or density) forecasts results out-of-sample. Note that the phenomenon is even more pronounced at the monthly frequency where pure noise is prevalent. Given that BART provides a reasonably convenient environment to go beyond mere conditional mean modeling and think about more structural objects (like some kind of time-varying parameters or any latent states), it is important to have reliable historical estimates. In the literature, we find possible solutions tailored to Random Forests, for which \cite{MSoRF} finds a similar pattern, such as using blocked out-of-bag quantities \citep{HNN,MACE}. However, out-of-bag sampling appears computationally unfeasible and would change the meaning of the Bayesian setup. 
More promising is some extensive case-by-case empirical tuning of the level of volatility prior. Again, some careful thinking is necessary about how such tuning should be conducted as BART will also have a preference for overly parameterized models in a pseudo-out-of-sample setup similar to what we see for out-of-sample. A possibility is to use an auxiliary model immune to such complication, like an autoregressive process.  Or,  when a preferred BART specification is chosen,  to increase such priors as long as the out-of-sample fit is mostly intact -- analogous to reducing the unnecessary depth of trees in a Random Forest \citep{MSoRF}. Note that, while not directly addressed here, some of BART's in-sample overconfidence spills out on the test sample mostly for calibration metrics, making it not as reliable as HNN or AR$_\text{\halftiny SV}$.  Some of the aforementioned solutions could likely help in that regard.

Regarding DeepAR, one could legitimately presume that early stopping and dropout could help avoiding the perfect in-sample fit seen in our results. Yet, the double descent phenomenon and associated neural networks oddities often make it difficult to use traditional regularization intuition as guidance.  Moreover,   as we have noted in Section \ref{sec:HNN},  unconstrained fully nonparametric models of mean and variance can overfit using either moment and the allocation between the two will depend on a mostly unknown mapping between obscure architecture choices and final results.  In further (unreported) experiments,  adjusting the number of layers and neurons can sometimes help or hurt, and in a mostly unpredictable way, i.e.,  there is no clear mapping between the total number of neurons and density forecasting underperformance.  The regularities are rather that (i) RMSEs are only remotely affected by such choices,  (ii) in-sample nominal coverage is often extremely high,  and (iii) out-of-sample coverage varies greatly but primarily on the low end. The most promising way of proceeding seems to be cross-validation based on blocked pseudo-out-of-sample density forecasting evaluation metrics. However, this would entail an unfeasible computational burden and, given the small sample size, probably places excessively high expectations on the power of cross-validation.



\subsection{On the Costs and Benefits of Recurrence}\label{sec:hrnn}     

\noindent Recurrent neural networks are specifically designed to process sequential data \citep[see, ][]{rumelhart1986learning,qin2017dual}. By keeping an internal memory state, which is used as additional input at each time step, RNNs capture patterns and dependencies over time. That is, RNNs receive information from two sources: external shocks to the system and the internal state from previous periods and as such, mimic the structure of a (G)ARCH process.  

While RNNs have become popular in various domains such as natural language processing or speech recognition \citep[see, ][]{young2018rnn}, they also come with limitations. Due to their recurrent nature and sequential processing, training can get computationally expensive, especially for long time series. Even more troublesome, RNNs are susceptible to vanishing or exploding gradients. To address these challenges, we implement a LSTM network \citep{hochreiter1997lstm}, which uses a gating mechanism to filter pertinent information, and restrict each hemisphere to use only one recurrent layer, effectively reducing the depth by half compared to the original architecture. All other hyperparameter choices are unchanged (see Section \ref{sec:HNN}).

As is evident from Table \ref{tab:hrnn} in Appendix \ref{app:results}, there is no need for taking on the burden of endowing HNN with a recurrent (LSTM) structure. Across all targets we find that gains from HNN-LSTM are either small or nonexistent. 
For point forecasts we get very similar results from both model specifications. HNN yields lower RMSEs in most cases, regardless of the forecast horizon. In cases where HNN-LSTM beats our standard specification, it is by very small margins. The only exception is the one-step ahead prediction of inflation and housing starts when considering the full sample. This difference in performance for inflation, however, can be diminished when considering the structural approach presented in Section \ref{sec:npc}.
When focusing on density predictions, HNN yields better forecasting accuracy as its recurrent counterpart.  We conclude that we can easily extend our proposed model to more complex types of neural networks, which yield, however, very similar results at the cost of higher computational burden.


\subsection{A Comparison with Quantile Regression Approaches}\label{sec:qreg}

\noindent Since all benchmarks considered so far estimate the variance process reactively, we expand our set of competitors by quantile regressions, which feature proactivity. We include a linear Bayesian quantile regression (BQR) with shrinkage, a quantile version of BART (QBART) as well as of the AR(2) model (QAR).
By estimating different quantiles of the predictive distribution based on the input matrix $\bm X_t$, quantile regressions directly and proactively model the uncertainty surrounding the response variable. This makes them a fair but hard-to-beat benchmark. However, estimating multiple quantiles for each target and horizon adds complexity and computational burden to our exercise. Besides rather statistical phenomena such as quantile-crossing \citep{bassett1982quantile}, which describes the lack of monotonicity when estimating conditional quantile functions, results may not have a straightforward interpretation in some cases. Consider, for example, the Neural Phillips Curve model with proactive volatility. Its extension to quantile regression would entail the estimation of an output gap measure for each quantile and, thus, raise the question of how to interpret the meaning of the resulting slack variables.

We evaluate the (tail) forecasting accuracy of all models using log scores and quantile-weighted continuous ranked probability score (CRPS$_{\omega}$). The weights are set such that the metrics allows for analyzing downside risks via the left tail and upside risks via the right tail of the distribution. To complete our analysis we also check the forecasting performance with respect to the center of the distribution.\footnote{Unweighted CRPS for HNN and the main benchmark models can be found in Figure \ref{fig:modelcalib}.} For details on model specification and implementation of the quantile regression approaches as well as the additional evaluation metrics we refer to Appendix \ref{app:bench}. Results are presented in Table \ref{tab:qreg_s1} and Table \ref{tab:qreg_s4} in the appendix.


We find that HNN remains highly competitive when investigating the predictive distribution in more detail and comparing its performance to quantile regression approaches. For real activity targets HNN either ranks first or is very close to the best performing model for both horizons, both samples and in each part of the distribution. 
This is remarkable since non-normality and, in particular,  asymmetry  of conditional distribution have become a major focal point of applied macroeconometric research.    In a well-known article,  \cite{adrian2019} show that when it comes to estimating the conditional distribution of GDP, quantile regressions perform well because the resulting distributions are left-skewed in recessionary periods and closer to symmetry during expansion.  We find that,  even though HNN builds upon the usual normality assumption,  its sophisticated mean and variance functions provide the necessary flexibility to adjust and capture dynamics in the tails.  HNN yields results very close to the Bayesian quantile regression for one-step ahead GDP growth and even outperforms it when considering the full sample. For four steps ahead, QBART tops the list of competing models but is again closely followed by HNN, especially when including the observations of the Covid-19 pandemic.  Therefore,  HNN's results are always in the ballpark of the (ex-post) best quantile regression model.

Turning to the results for the remaining targets, HNN outperforms all competitors for the unemployment rate one-step ahead and for four steps ahead when evaluating the sample up to the Covid-19 pandemic. Here, we find substantial gains in the tails as well as the center of the distribution. For higher-order forecasts of the unemployment rate (when considering the full sample) HNN yields highly competitive results compared to QBART. Similarly, QBART turns out to be the main competitor for higher-order density predictions of inflation. 
For the one-step ahead case we often find strong performance of BQR (see, e.g., GDP growth, housing starts, S\&P 500), except for inflation where HNN yields the lowest CRPS. For S\&P 500, especially for higher-order forecasts, BART remains the best performing model. Even though the Bayesian models, either plain or as quantile extensions, turn out to be very difficult to beat, HNN follows closely and thus, captures upside and downside risks to a similar extent.

\section{A Neural Phillips Curve with Proactive Volatility}\label{sec:npc}
           
\begin{figure}[t!]
\begin{center} 
\caption{Architecture of the Neural Phillips Curve with Proactive Volatility}\label{fig:npc}
\hspace*{-0.2cm}\includegraphics[scale=.61]{{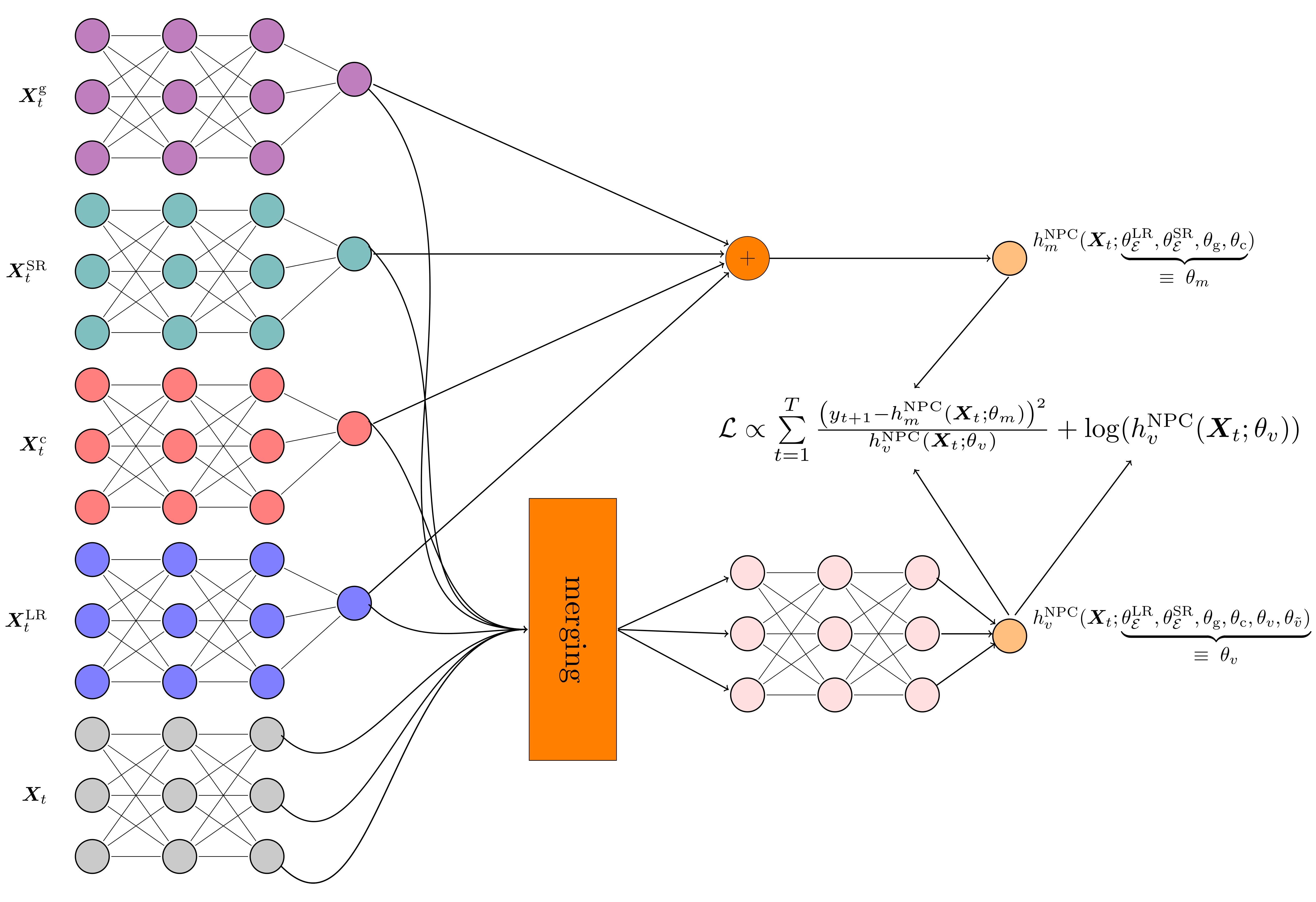}}
\end{center}
\vspace*{-1cm}
\end{figure}

Given their high importance for economic policy decisions in central banks and governmental institutions, inflation forecasts should be decent and preferably interpretable through some basic macroeconomic reasoning. No less important is to be aware of the level of uncertainty associated with a specific inflation forecast.   As we have seen, HNN is a promising tool  that manages to incorporate large amounts of data and captures nonlinearities via its sophisticated mean and variance specification.  However,  the anatomy of $h_m$ remains mostly unknown.  In this section, we achieve both goals by bringing back interpretability of the conditional mean for this particular target and, at the same time, modeling the time-varying level of decency with the variance hemisphere. 
The resulting model,  of appreciable architectural complexity, is \cite{HNN}'s Neural Phillips Curve embedded within this paper's probabilistic forecasting methodology (henceforth, HNN-NPC). Precisely, we impose a Phillips Curve structure on the fully nonparametric $h_m$ in \eqref{eq:ouou}, resulting in
\begin{align*} 
 h_m^{\text{NPC}}(\boldsymbol{X}_t;[\theta_{\mathcal{E}}^{\text{LR}},  \theta_{\mathcal{E}}^{\text{SR}},  \theta_g,  \theta_c]) = h_{\mathcal{E}}^{\text{LR}}(\boldsymbol{X}_t^{\mathcal{E}_\text{LR}};\theta_{\mathcal{E}}^{\text{LR}})+h_{\mathcal{E}}^{\text{SR}}(\boldsymbol{X}_t^{\mathcal{E}_\text{SR}};\theta_{\mathcal{E}}^{\text{SR}}) + h_{g}(\boldsymbol{X}_t^{g};\theta_g)+ h_{c}(\boldsymbol{X}_t^{c};\theta_c)
\end{align*}
where $\boldsymbol{X}_t^{i}$ with $i \in \{ \mathcal{E}_\text{LR}, \mathcal{E}_{\text{SR}},  g,  c \}$ being the subsets of columns that correspond,  respectively,  to long-run expectations, short-run expectations,  "output gap",  and commodity prices hemispheres.  Their definitions in terms of FRED-QD are identical to that of the original paper.  The exact list of mnemonics can be found in Appendix \ref{sec:mne}. The new hemispheric structure of the conditional mean and its goal to gain interpretability implies the need to drop the common core at the entrance of the network. Regarding the volatility process, this results in a more structured one-way directional flow of $h_m$ into $h_v$. As there is no overwhelming theoretical reason to constrain the volatility process to solely rely on PC-inspired inputs we modify $h_v$ in \eqref{eq:ouou} for
\begin{align*} 
 h_v^{\text{NPC}}(\boldsymbol{X}_t;[ \theta_{\mathcal{E}}^{\text{LR}},  \theta_{\mathcal{E}}^{\text{SR}},  \theta_g ,   \theta_c,\theta_v,  \theta_{\tilde{v}}]) = h_v \left( \left[ h_{\mathcal{E}}^{\text{LR}}(\boldsymbol{X}_t^{\mathcal{E}_\text{LR}}; \theta_{\mathcal{E}}^{\text{LR}}),  h_{\mathcal{E}}^{\text{SR}}(\boldsymbol{X}_t^{\mathcal{E}_\text{SR}};\theta_{\mathcal{E}}^{\text{SR}}),  h_{g}(\boldsymbol{X}_t^{g};\theta_g),  h_{c}(\boldsymbol{X}_t^{c};\theta_c) ,  h_{\tilde{v}}(\boldsymbol{X}_t;\theta_{\tilde{v}}) \right] ; \theta_v \right)
\end{align*}
where $h_{\tilde{v}}(\boldsymbol{X}_t;\theta_{\tilde{v}})$ is a subnetwork that serves in processing all variables of our high-dimensional data set to extract a time series for $h_v^{\text{NPC}}$ that carries relevant signals for the conditional variance that are not captured by the four series summing up to the conditional mean.  All subnetworks (i.e., $h_{\tilde{v}}$ and the four subnetworks of the conditional mean) precede $h_v^{\text{NPC}}$.  Figure \ref{fig:npc} summarizes the structure of HNN-NPC.

Since conditional mean hemispheres contain an unequal number of predictors,  leading to unequal a priori importance assigned to each one,  we rescale data as proposed in \cite{HNN}.  Precisely,  we divide the scaled  predictors of $\boldsymbol{X}^{i}$ by $\sqrt{\frac{{\#\text{\phantom{.}columns}(\boldsymbol{X}^{i})}}{\#\text{\phantom{.}columns}(\boldsymbol{X})}}$.  This scheme is not necessary for the $h_{\tilde{v}}$ subnetwork because it takes the whole $\boldsymbol{X}$ and its relative influence is only guided by its total number of neurons and the volatility emphasis parameter $\nu$.  Since the number of effective parameters is at least four times superior in the conditional mean subnetwork,  the $\boldsymbol{X}$'s  entering $h_{\tilde{v}}$ are multiplied by a factor of 5. Other hyperparameters are the same as described in Section \ref{sec:HNN} except that each hemisphere now consists of 200 neurons  and $\nu$ is now obtained from the blocked out-of-bag errors of \cite{HNN}'s plain NPC model.

\begin{figure}[t!]
\caption{Visualizing the Neural Phillips Curve and its Volatility}\label{NPCres}  
\begin{center} 
\vspace*{-0.8cm}
\hspace*{-0.115cm}\includegraphics[width=1.02\textwidth]{{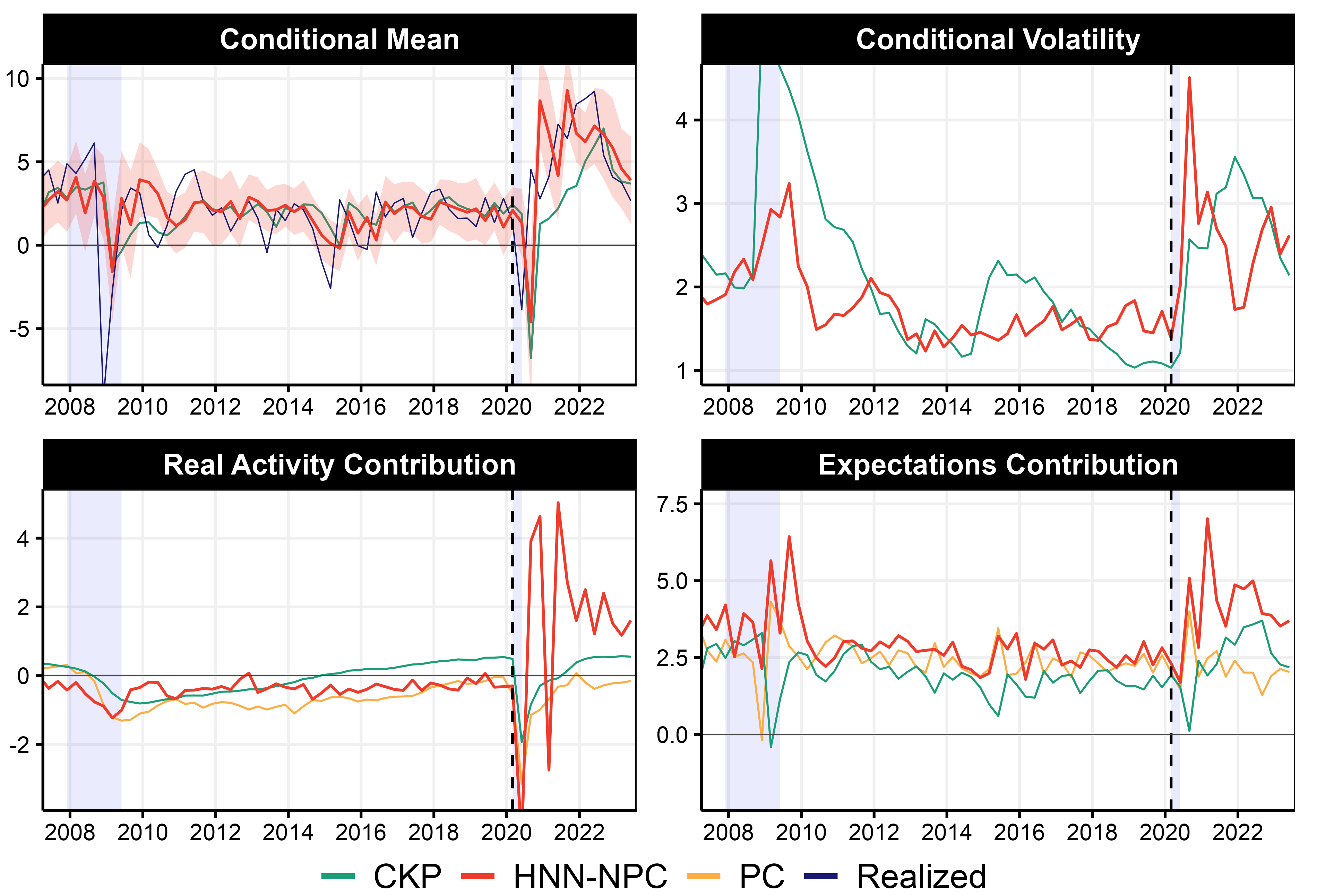}}
\end{center}
\vspace*{-0.35cm}
\scriptsize
\textit{Notes:}   HNN-NPC is the modified HNN with conditional mean structured as a Neural Phillips Curve Ã  la \cite{HNN}.   CKP refers to  \cite{chan2016}'s model of trend inflation with SV.    The upper panels show the conditional mean and the conditional variance for HNN-NPC and CKP,  which are both models providing structured inflation forecasts \textit{and} have  built-in volatility prediction.   For further comparison of contributions in the second row,  we also include PC,  which is the estimated contribution from a time-varying Phillips Curve regression using the CBO output gap as forcing variable.  In the case of expectations,  $h_{\mathcal{E},t}^{\text{LR}}+h_{\mathcal{E},t}^{\text{SR}}$ is plotted for HNN and the line for "PC" is the sum of the two lags plus the time-varying intercept.  Analogous calculations are carried out for CKP.   Up to 2019Q4 we show the in-sample results of the respective model followed by the out-of-sample results (from 2020Q1 to 2023Q2), indicated by the dotted line.  Lavender shading corresponds to NBER recessions. 
\end{figure}


First,  we can compare whether the incoming restrictions hurt or improve predictive ability both in terms of RMSE and probabilistic forecasting metrics.  Table \ref{tab:dhnn} in the appendix shows promising results with the more "specialized" HNN-NPC improving in a non-trivial fashion nearly all performance metrics including or excluding post-2020 data.  This is not completely jarring because (i) relevant restrictions will bring down variance more than they incur bias and (ii) distilling FRED-QD to include only relevant variables (according to loose macroeconomic theory) can significantly improve the performance of an otherwise dense conditional mean model.

The first row of Figure \ref{NPCres} reports the usual conditional mean and volatility,  and now compares with a small Bayesian model of similar aim.  Indeed,  \citet[][henceforth CKP]{chan2016} provide a model of trend inflation with a Phillips Curve (with unemployment as forcing variable),  drifting coefficients,  and stochastic volatility.  The first panel highlights that HNN-NPC fares particularly well during two turbulent eras by (i) avoiding the missing disinflation following the Great Recession and (ii) capturing a non-trivial part of the 2021-2023 surge in inflation.  In contrast, CKP exhibit the typical observation that traditional PC-based forecasts were not only reactive rather than proactive during recent years,  but also consistently "biased" downwards up until 2023.   Regarding volatility estimates,  HNN-NPC's proactive behavior is apparent during the Great Recession,  with $h_{v,t}^{\text{NPC}}$ rising from its bed a few quarters before the SV process embedded in CKP---the latter erupts following the 2008 oil price crash and then apparently diffuses its effects for almost three years.  Both volatility estimates spike following the initial Covid-19 crash.   The HNN-NPC spike is particularly pronounced and explains why it yields the best log score despite a highly inaccurate deflation forecast in 2020Q2: the confidence interval for that particular point in time is so large it also includes positive inflation.  As such,  the erroneous point forecast is discounted in the probabilistic performance metrics since HNN-NPC practically "knew" it was turning in a forecast of extremely low reliability.  The massive uncertainty quickly dissipates to a more reasonable level (comparable to that of post-Great Recession), then hits a low point before picking up again following the invasion of Ukraine. 

For additional comparison,  we also report contributions from a canonical PC regression in the second row of Figure \ref{NPCres}.  In the case of "PC",  those are constructed from a traditional PC specification  -- including two lags of inflation and the Congressional Budget Office (CBO) estimate of the output gap -- with time-varying coefficients obtained from \cite{GC2019}'s two-steps ridge regression approach.  As discussed in \cite{HNN},  the key statistical distinction between HNN-based inflation modeling and the two models included for reference purposes is the nonlinear processing of a rich real activity data set.  The use of neural networks serves as a convenient way to achieve that goal within a "generalized" PC environment.  



Looking at contributions in the second row,  we reaffirm some key findings from \cite{HNN}.   Among other things,  we get a rapid closing of $g_t$'s contribution to inflation post-Great Recession.  This contrasts with CKP and PC which  report more lasting downward pressures following 2008.  The next observation is the strikingly different behavior of $h_{g,t}$ starting from last quarter of 2020.  HNN-NPC sees $h_{g,t}$ contributing strongly to the highest quarterly inflation forecasts in a generation.  The two benchmarks are not nearly as agitated in 2021 and 2022, and report forecasts and contributions that fit within the popular PC-based narrative in 2021 that inflation would be transitory.   

From Figure \ref{NPCres},  we also get some additional insights from updating the data up to 2023Q2 \citep[][ends in 2021Q4]{HNN}.  First,  we see that $g_t$ is still pushing forecasts above the target range,  but its effect has massively shrunk from the highs of 2021,  mostly starting from 2022.  As of 2023Q2, the contribution of real activity to inflation as estimated by HNN-NPC is about twice as much as that from CKP, yet it is the closest they have been to agreement in the last three years.  The contribution of expectations seems to slowly decrease to pre-pandemic levels,  but is appreciably higher than that of CKP, which has already settled to rather low levels.  HNN-NPC's and CKP's point forecasts for 2023Q2 are roughly similar (with CKP gaining in performance in 2023),  and so is their latest assessment of volatility.  This agreement aligns with the observation that their predictions mainly differ during a few localized  episodes,  and 2023Q2 falls outside of that.   Usual disagreement can be traced back to their differing evaluation of economic slack and expectations,  especially when those are far from their mean.   Key disagreeing segments for those components are the missing disinflation post 2008 and non-transitory inflation surge of 2021-2023.  However, as of 2023Q2 the unemployment gap of CKP has mostly caught up with mounting real activity pressures expressed by HNN, resulting in the recent relative concordance between the two models.

\section{Concluding Remarks}\label{sec:con}

We provide a way to conduct density forecasting where both conditional mean and variance are the outputs from neural networks.  Results show that in many cases,  HNN picks up early signals of increased future volatility before the occurrence of large prediction errors.  This proactive behavior often gives a significant advantage over stochastic volatility specifications that are frequently used to close linear and nonlinear macroeconomic forecasting models.  Moreover,  its nominal coverage and overall probabilistic forecasting performance is much more consistent across targets and experiments than what we found for two leading nonlinear nonparametric machine learning alternatives.  Therefore,  HNN is an effective new tool for density forecasting in itself \textit{and}  a convenient building block for deep-learning based macroeconomic models when timely uncertainty quantification is needed.  Concerning the latter aspect,  we provided such an application by merging this paper's architecture with that of \cite{HNN}'s Neural Phillips Curve.  There are many possible extensions.  A conceptually obvious yet very relevant one  is that of a multivariate normal predictive density,  providing a MLE-based alternative for the estimation of (possible large) nonlinear/time-varying vector autoregressions.

 \pagebreak
 
 \setlength\bibsep{5pt}
    
\bibliographystyle{apalike}

\setstretch{0.75}
\bibliography{ref_gcfk_v230418}

\clearpage

\appendix
\newcounter{saveeqn}
\setcounter{saveeqn}{\value{section}}
\renewcommand{\theequation}{\mbox{\Alph{saveeqn}.\arabic{equation}}} \setcounter{saveeqn}{1}
\setcounter{equation}{0}
  
\section{Appendix}
\setstretch{1.25}
\subsection{Additional Figures and Results}\label{app:results}

\vspace{3em}
\begin{table}[h!]
\vspace{-1.8em}
\begin{threeparttable}

\centering
\scriptsize
\footnotesize
\setlength{\tabcolsep}{0.225em}
 \setlength\extrarowheight{2.5pt}
 \caption{\normalsize {Forecast Performance of Additional Quarterly Target Variables} \vspace*{-0.3cm}} \label{tab:add_targets}
 \begin{tabular}{l rrrrrrrrrrrrrrr | rrrrrrrrrrrrrrr} 
\toprule \toprule
\addlinespace[2pt]
& & \multicolumn{14}{c}{2007Q1 - 2019Q4} & &  \multicolumn{13}{c}{2007Q1 - 2022Q4} \\
\cmidrule(lr){3-15} \cmidrule(lr){18-30} \addlinespace[2pt]
& &  HNN &  & NN$_\text{\halftiny SV}$ &  & NN$_\text{\halftiny G}$ &  & {\notsotiny DeepAR} & & BART && AR$_\text{\halftiny TV}$  &  & BLR & & &
 HNN &  & NN$_\text{\halftiny SV}$ &  & NN$_\text{\halftiny G}$ &  & {\notsotiny DeepAR} & & BART && AR$_\text{\halftiny TV}$  &  & BLR   &\\
\midrule
\addlinespace[5pt] 
\rowcolor{gray!15} 
\multicolumn{30}{l}{\textbf{Unemployment Rate ($s=1$)}} &\cellcolor{gray!15} \\ \addlinespace[2pt] 

{\notsotiny RMSE} &   & \textbf{0.73} &   & 0.78 &   & 0.78 &   & 0.93 &   & 0.87 &   & 1.00 &   & 0.75 &   &   & \textbf{0.82} &   & 0.96 &   & 0.96 &   & 0.90 &   & 0.96 &   & 1.04 &   & 0.91 &   \\
$\mathcal{L}$ &   & \textbf{-0.37} &   & -0.32 &   & -0.33 &   & 0.03 &   & 1.87 &   & -0.18 &   & -0.09 &   &   & \textbf{-0.24} &   & -0.11 &   & -0.09 &   & 0.10 &   & 1.85 &   & -0.04 &   & 0.05 &   \\
$R^2_{|\varepsilon_t|}$ &   & 0.09 &   & 0.12 &   & \textbf{0.20} &   & -0.11 &   & 0.19 &   & 0.18 &   & -2.66 &   &   & -0.58 &   & -5.78 &   & -12.20 &   & \textbf{-0.09} &   & -70.33 &   & -2.02 &   & -2.40 &   \\

\addlinespace[5pt] 
\rowcolor{gray!15} 
\multicolumn{30}{l}{\textbf{Unemployment Rate ($s=4$)}} &\cellcolor{gray!15} \\ \addlinespace[2pt] 

{\notsotiny RMSE} &   & 0.74 &   & \textbf{0.69} &   & \textbf{0.69} &   & 0.85 &   & 0.75 &   & 0.97 &   & 0.82 &   &   & 0.70 &   & 2.20 &   & 2.20 &   & 0.73 &   & 0.71 &   & 0.88 &   & \textbf{0.70} &   \\
$\mathcal{L}$  &   & \textbf{-0.17} &   & 0.04 &   & 0.06 &   & 0.23 &   & 0.65 &   & 0.47 &   & 0.19 &   &   & \textbf{0.03} &   & 0.62 &   & 0.34 &   & 2.91 &   & 0.82 &   & 0.70 &   & 0.30 &   \\
$R^2_{|\varepsilon_t|}$ &   & \textbf{0.49} &   & -0.07 &   & -0.12 &   & 0.21 &   & -5.58 &   & 0.22 &   & -0.36 &   &   & -1.23 &   & -0.08 &   & -0.24 &   & \textbf{-0.02} &   & -38.15 &   & -0.24 &   & -1.12 &   \\

\addlinespace[5pt] 
\rowcolor{gray!15} 
\multicolumn{30}{l}{\textbf{Inflation ($s=4$)}} &\cellcolor{gray!15} \\ \addlinespace[2pt] 

{\notsotiny RMSE} &   & \textbf{0.93} &   & 0.95 &   & 0.95 &   & 0.99 &   & 0.99 &   & 1.08 &   & 1.00 &   &   & 0.94 &   & 0.98 &   & 0.98 &   & 0.93 &   & \textbf{0.92} &   & 1.02 &   & 1.09 &   \\
$\mathcal{L}$ &   & -3.48 &   & -3.63 &   & \textbf{-3.66} &   & -3.14 &   & -3.19 &   & -3.45 &   & -3.60 &   &   & -3.30 &   & -3.39 &   & -3.40 &   & -1.27 &   & -2.99 &   & \textbf{-3.40} &   & -3.37 &   \\
$R^2_{|\varepsilon_t|}$ &   & -0.28 &   & -0.20 &   & -0.15 &   & -0.08 &   & -2.51 &   & \textbf{0.28} &   & -0.45 &   &   & -0.09 &   & -0.03 &   & -0.04 &   & -0.04 &   & -1.55 &   & \textbf{0.36} &   & 0.00 &   \\

\addlinespace[5pt] 
\rowcolor{gray!15} 
\multicolumn{30}{l}{\textbf{S\&P 500 ($s=4$)}} &\cellcolor{gray!15} \\ \addlinespace[2pt] 

{\notsotiny RMSE} &   & 1.00 &   & 1.15 &   & 1.15 &   & 1.02 &   & \textbf{0.92} &   & 0.99 &   & 0.99 &   &   & 1.00 &   & 1.22 &   & 1.22 &   & 1.01 &   & \textbf{0.95} &   & 0.99 &   & 1.09 &   \\
$\mathcal{L}$ &   & -1.27 &   & -1.00 &   & -1.01 &   & 4.99 &   & \textbf{-1.36} &   & -1.16 &   & -1.25 &   &   & -1.27 &   & -0.96 &   & -0.99 &   & 3.80 &   & \textbf{-1.30} &   & -1.19 &   & -1.16 &   \\
$R^2_{|\varepsilon_t|}$ &   & 0.04 &   & -0.14 &   & -0.09 &   & -0.09 &   & 0.07 &   & \textbf{0.32} &   & -0.19 &   &   & 0.01 &   & -0.12 &   & -0.08 &   & -0.08 &   & 0.10 &   & \textbf{0.35} &   & -0.08 &   \\

\addlinespace[5pt] 
\rowcolor{gray!15} 
\multicolumn{30}{l}{\textbf{Housing Starts ($s=4$)}} &\cellcolor{gray!15} \\ \addlinespace[2pt] 

{\notsotiny RMSE} &   & 1.03 &   & 1.09 &   & 1.09 &   & 1.05 &   & 0.98 &   & \textbf{0.96} &   & 1.03 &   &   & 1.01 &   & 1.13 &   & 1.13 &   & 1.04 &   & 0.99 &   & \textbf{0.96} &   & 1.07 &   \\
$\mathcal{L}$ &   & -0.88 &   & -0.95 &   & -0.99 &   & 0.22 &   & \textbf{-1.10} &   & -1.05 &   & -1.02 &   &   & -0.66 &   & -0.51 &   & -0.66 &   & 0.35 &   & -0.75 &   & \textbf{-1.06} &   & -0.83 &   \\
$R^2_{|\varepsilon_t|}$ &   & 0.05 &   & -0.17 &   & -0.00 &   & -0.19 &   & -0.02 &   & \textbf{0.30} &   & -0.12 &   &   & -0.17 &   & -0.26 &   & -0.07 &   & -0.15 &   & -0.08 &   & \textbf{0.26} &   & 0.03 &   \\

\bottomrule \bottomrule
\end{tabular}
\begin{tablenotes}[para,flushleft]
  \scriptsize 
    \textit{Notes}: The table presents our forecasting evaluation metrics including the root mean square error (RMSE) relative to the AR model with constant variance, the log score ($\mathcal{L}$), and the $R^2_{|\varepsilon_t|}$ of absolute residuals. For the real activity variables (i.e., GDP and unemployment) we exclude the year 2020 from the evaluation sample. AR$_\text{\halftiny TV}$ refers to the best-performing AR specifications, which is either AR$_\text{\halftiny SV}$ or AR$_\text{\halftiny G}$.
  \end{tablenotes}
\end{threeparttable}
\end{table}

\begin{table}[h!]
\centering
            \vspace*{1em}
                  \begin{threeparttable}

\centering
\scriptsize
\footnotesize
\setlength{\tabcolsep}{0.22em}
 \setlength\extrarowheight{2.8pt}
 \caption{\normalsize {Forecast Performance of HNN vs HNN-LSTM} \vspace*{-0.3cm}} \label{tab:hrnn}
 \begin{tabular}{l ccccccccc | ccccccccc}
\toprule \toprule
\addlinespace[2pt]
& & \multicolumn{8}{c}{2007Q1 - 2019Q4} & &  \multicolumn{7}{c}{2007Q1 - 2022Q4} \\
\cmidrule(lr){2-10} \cmidrule(lr){11-19} \addlinespace[2pt]
 & & \multicolumn{4}{c}{$s=1$}  & \multicolumn{3}{c}{$s=4$}  &  &  & \multicolumn{4}{c}{$s=1$}  & \multicolumn{3}{c}{$s=4$}  & \\
 \cmidrule(lr){3-5} \cmidrule(lr){7-9} \cmidrule(lr){12-14} \cmidrule(lr){16-18} \addlinespace[2pt]
 & &  {\notsotiny RMSE} &  & $\mathcal{L}$ & &  {\notsotiny RMSE} &  & $\mathcal{L}$  & & &  {\notsotiny RMSE} &  & $\mathcal{L}$  & & {\notsotiny RMSE} &  & $\mathcal{L}$  &  \\
\midrule
\addlinespace[5pt] 
\rowcolor{gray!15}  
 \multicolumn{18}{l}{\textbf{GDP}} &\cellcolor{gray!15} \\ \addlinespace[2pt] 
 
{HNN} &   & \textbf{0.83} &   & -3.93 &   & 0.90 &   & \textbf{-3.70} &   &   & \textbf{0.85} &   & -3.87 &   & \textbf{0.85} &   & \textbf{-3.61} \\
{HNN-LSTM} &   & 0.85 &   & \textbf{-3.95} &   & \textbf{0.89} &   & -3.70 &   &   & 0.88 &   & \textbf{-3.88} &   & 0.94 &   & -3.61 \\

\addlinespace[5pt] 
\rowcolor{gray!15}  

\multicolumn{18}{l}{\textbf{Unemployment Rate}} &\cellcolor{gray!15} \\ \addlinespace[2pt] 

{HNN} &   & \textbf{0.73} &   & \textbf{-0.37} &   & \textbf{0.74} &   & \textbf{-0.17} &   &   & 0.82 &   & \textbf{-0.24} &   & \textbf{0.70} &   & \textbf{0.03} \\
{HNN-LSTM} &   & 0.76 &   & -0.31 &   & 0.78 &   & -0.08 &   &   & \textbf{0.67} &   & -0.17 &   & 0.90 &   & 0.10 \\

\addlinespace[5pt] 
\rowcolor{gray!15}  

 \multicolumn{18}{l}{\textbf{Inflation}} &\cellcolor{gray!15} \\ \addlinespace[2pt]
  
{HNN} &   & 0.94 &   & \textbf{-3.63} &   & 0.93 &   & -3.48 &   &   & 1.14 &   & \textbf{-3.41} &   & 0.94 &   & \textbf{-3.30} \\
{HNN-LSTM} &   & \textbf{0.93} &   & -3.51 &   & \textbf{0.88} &   & \textbf{-3.49} &   &   & \textbf{0.93} &   & -3.36 &   & \textbf{0.92} &   & -3.28 \\

\addlinespace[5pt] 
\rowcolor{gray!15}  
 \multicolumn{18}{l}{\textbf{S\&P 500}} &\cellcolor{gray!15} \\ \addlinespace[2pt] 
 
{HNN} &   & \textbf{0.96} &   & \textbf{-1.55} &   & \textbf{1.00} &   & -1.27 &   &   & \textbf{0.93} &   & \textbf{-1.52} &   & 1.00 &   & \textbf{-1.27} \\
{HNN-LSTM} &   & 1.02 &   & -1.49 &   & 1.00 &   & \textbf{-1.28} &   &   & 1.01 &   & -1.43 &   & \textbf{1.00} &   & -1.27 \\

\addlinespace[5pt] 
\rowcolor{gray!15}  
 \multicolumn{18}{l}{\textbf{Housing Starts}} &\cellcolor{gray!15} \\ \addlinespace[2pt] 
 
{HNN} &   & \textbf{0.99} &   & \textbf{-1.14} &   & 1.03 &   & \textbf{-0.88} &   &   & \textbf{0.86} &   & \textbf{-1.07} &   & 1.01 &   & \textbf{-0.66} \\
{HNN-LSTM} &   & 1.05 &   & -1.04 &   & \textbf{1.00} &   & -0.80 &   &   & 0.94 &   & -0.95 &   & \textbf{0.99} &   & -0.41 \\

\bottomrule \bottomrule
\end{tabular}
\begin{tablenotes}[para,flushleft]
  \scriptsize 
    \textit{Notes}: The table shows log scores ($\mathcal{L}$) and root mean square error (RMSE) for one-step and four steps ahead predictions ($s \in \{1,4\}$). We compare the performance of our proposed HNN and the LSTM extension of the model.
  \end{tablenotes}
\end{threeparttable}
\end{table}

\begin{table}[h!]
\begin{threeparttable}

\centering
\scriptsize
\footnotesize
\setlength{\tabcolsep}{0.225em}
 \setlength\extrarowheight{2.5pt}
 \caption{\normalsize {Forecast Performance of Quantile Regressions ($s=1$)} \vspace*{-0.3cm}} \label{tab:qreg_s1}
 \begin{tabular}{l rrrrrrrrrrrrrrr | rrrrrrrrrrrrrrr} 
\toprule \toprule
\addlinespace[2pt]
& & \multicolumn{14}{c}{2007Q1 - 2019Q4} & &  \multicolumn{13}{c}{2007Q1 - 2022Q4} \\
\cmidrule(lr){3-15} \cmidrule(lr){18-30} \addlinespace[2pt]
& &  HNN &  &  BART && {\notsotiny QBART} &  & BLR & & BQR & & AR$_\text{\halftiny SV}$  & & {\notsotiny BQAR} & & &
 HNN &  & BART && {\notsotiny QBART} &  & BLR & & BQR & & AR$_\text{\halftiny SV}$ && {\notsotiny BQAR} & \\
\midrule
    \addlinespace[5pt] 
\rowcolor{gray!15} 
\multicolumn{30}{l}{\textbf{GDP}} &\cellcolor{gray!15} \\ \addlinespace[2pt] 
   CRPS$_\text{\halftiny center}$   &&   0.74&&   0.76&&   0.87&&   0.84&&   \textbf{0.73}&&   0.84&&   0.85&&&   \textbf{0.77}&&   0.98&&   0.89&&   0.89&&   0.77&&   0.88&&   0.87\\
   CRPS$_\text{\halftiny left}$   &&   0.78&&   0.79&&   0.89&&   0.85&&   \textbf{0.76}&&   0.88&&   0.87&&&   0.82&&   1.06&&   0.94&&   0.92&&   \textbf{0.79}&&   0.92&&   0.91\\
   CRPS$_\text{\halftiny right}$   &&   0.62&&   0.62&&   0.91&&   0.78&&   \textbf{0.61}&&   0.74&&   0.78&&&   \textbf{0.65}&&   0.87&&   0.92&&   0.81&&   0.68&&   0.79&&   0.79\\
   $\mathcal{L}$   &&   \textbf{-3.93}&&   -3.88&&   -3.63&&   -3.69&&   -3.88&&   -3.75&&   -3.69&&&   \textbf{-3.87}&&   -3.71&&   -3.57&&   -3.63&&   -3.75&&   -3.69&&   -3.66\\
   \addlinespace[5pt] 
\rowcolor{gray!15} 
\multicolumn{30}{l}{\textbf{Unemployment Rate}} &\cellcolor{gray!15} \\ \addlinespace[2pt] 
   CRPS$_\text{\halftiny center}$   &&   \textbf{0.44}&&   0.53&&   0.56&&   0.50&&   0.46&&   0.56&&   0.57&&&   \textbf{0.50}&&   0.92&&   0.63&&   0.59&&   0.53&&   0.64&&   0.62\\
   CRPS$_\text{\halftiny left}$   &&   0.42&&   0.46&&   0.53&&   0.43&&   \textbf{0.39}&&   0.47&&   0.47&&&   \textbf{0.47}&&   0.85&&   0.58&&   0.49&&   0.53&&   0.57&&   0.60\\
   CRPS$_\text{\halftiny right}$   &&   \textbf{0.33}&&   0.50&&   0.54&&   0.46&&   0.39&&   0.50&&   0.49&&&   \textbf{0.42}&&   0.95&&   0.64&&   0.58&&   0.43&&   0.57&&   0.58\\
   $\mathcal{L}$   &&   \textbf{-0.37}&&    1.87&&    0.02&&   -0.09&&   -0.37&&   -0.16&&   -0.10&&&   \textbf{-0.24}&&    1.85&&    0.13&&    0.05&&   -0.17&&   -0.04&&    0.44\\
\addlinespace[5pt] 
\rowcolor{gray!15} 
\multicolumn{30}{l}{\textbf{Inflation}} &\cellcolor{gray!15} \\ \addlinespace[2pt] 
   CRPS$_\text{\halftiny center}$  &&   \textbf{0.99}&&   1.23&&   1.03&&   1.08&&   1.02&&   1.05&&   1.09&&&   1.07&&   1.06&&   \textbf{0.93}&&   1.09&&   1.02&&   0.97&&   0.98\\
   CRPS$_\text{\halftiny left}$   &&   \textbf{1.00}&&   1.29&&   1.10&&   1.11&&   1.01&&   1.09&&   1.06&&&   1.06&&   1.11&&   1.01&&   1.08&&   0.98&&   1.02&&   \textbf{0.97}\\
   CRPS$_\text{\halftiny right}$   &&   \textbf{0.95}&&   1.20&&   1.13&&   1.08&&   1.00&&   1.00&&   1.13&&&   1.08&&   1.08&&   0.99&&   1.17&&   1.07&&   \textbf{0.91}&&   1.01\\
   $\mathcal{L}$   &&   -3.63&&   -2.91&&   \textbf{-3.69}&&   -3.60&&   -3.21&&   -3.26&&   -2.18&&&   -3.41&&   -1.30&&   \textbf{-3.65}&&   -3.33&&   -2.38&&   -3.24&&   -2.22\\
      \addlinespace[5pt] 
\rowcolor{gray!15} 
\multicolumn{30}{l}{\textbf{S\&P 500}} &\cellcolor{gray!15} \\ \addlinespace[2pt] 
   CRPS$_\text{\halftiny center}$   &&   0.93&&   0.91&&   0.94&&   1.01&&   \textbf{0.91}&&   0.96&&   0.91&&&   0.89&&   0.87&&   0.91&&   0.96&&   \textbf{0.87}&&   0.92&&   0.89\\
   CRPS$_\text{\halftiny left}$   &&   \textbf{0.88}&&   0.89&&   0.92&&   0.99&&   0.88&&   0.95&&   0.88&&&   0.88&&   0.87&&   0.92&&   0.96&&   \textbf{0.86}&&   0.92&&   0.87\\
   CRPS$_\text{\halftiny right}$   &&   0.96&&   0.91&&   1.06&&   1.02&&   \textbf{0.91}&&   0.97&&   0.91&&&   0.91&&   \textbf{0.86}&&   1.00&&   0.96&&   0.87&&   0.94&&   0.91\\
   $\mathcal{L}$   &&   \textbf{-1.55}&&   -1.28&&   -1.40&&   -1.25&&   -1.24&&   -1.35&&   -1.40&&&   \textbf{-1.52}&&   -1.34&&   -1.38&&   -1.29&&   -1.20&&   -1.37&&   -1.41\\
   \addlinespace[5pt] 
\rowcolor{gray!15} 
\multicolumn{30}{l}{\textbf{Housing Starts}} &\cellcolor{gray!15} \\ \addlinespace[2pt] 
   CRPS$_\text{\halftiny center}$   &&   0.99&&   0.92&&   0.98&&   0.94&&   \textbf{0.85}&&   0.99&&   0.99&&&   \textbf{0.90}&&   0.91&&   0.95&&   0.98&&   0.92&&   0.99&&   0.99\\
   CRPS$_\text{\halftiny left}$   &&   0.98&&   0.97&&   1.02&&   1.01&&   \textbf{0.89}&&   0.98&&   0.98&&&   \textbf{0.91}&&   0.98&&   1.01&&   1.04&&   0.96&&   1.01&&   1.00\\
   CRPS$_\text{\halftiny right}$   &&   1.00&&   0.90&&   1.04&&   0.88&&   \textbf{0.88}&&   0.99&&   0.98&&&   0.86&&   \textbf{0.86}&&   0.95&&   0.93&&   0.87&&   0.97&&   0.96\\
   $\mathcal{L}$   &&   -1.14&&   -0.98&&   -1.07&&   -1.16&&   -0.83&&   -1.15&&   \textbf{-1.17}&&&   \textbf{-1.07}&&   -0.67&&   -1.02&&   -0.92&&   -0.48&&   -0.92&&   -0.92\\
\bottomrule \bottomrule
\end{tabular}
\begin{tablenotes}[para,flushleft]
  \scriptsize 
    \textit{Notes}: The table shows quantile-weighted continuous ranked probability score (CRPS$_{\omega}$) as well as log scores ($\mathcal{L}$). To target the left tail of the predictive distribution (downside risk, CRPS$_\text{\halftiny left}$), we set the weights to $\omega_{\tau} = (1-\tau)^2$. The right tail (upside risk, CRPS$_\text{\halftiny right}$) is targeted by setting the weights to $\omega_{\tau} = \tau^2$. We evaluate the center of the distribution (CRPS$_\text{\halftiny center}$) by using $\omega_{\tau} = \tau(1-\tau)$. We compute 19 quantiles with $\tau \in \{0.05,0.10,\dots,0.90,0.95\}$. For the real activity variables (i.e., GDP and unemployment) we exclude the year 2020 from the evaluation sample.
  \end{tablenotes}
\end{threeparttable}
\end{table}

\begin{table}[h!]
\vspace{-1.8em}
\begin{threeparttable}

\centering
\scriptsize
\footnotesize
\setlength{\tabcolsep}{0.225em}
 \setlength\extrarowheight{2.5pt}
 \caption{\normalsize {Forecast Performance of Quantile Regressions ($s=4$)} \vspace*{-0.3cm}} \label{tab:qreg_s4}
 \begin{tabular}{l rrrrrrrrrrrrrrr | rrrrrrrrrrrrrrr} 
\toprule \toprule
\addlinespace[2pt]
& & \multicolumn{14}{c}{2007Q1 - 2019Q4} & &  \multicolumn{13}{c}{2007Q1 - 2022Q4} \\
\cmidrule(lr){3-15} \cmidrule(lr){18-30} \addlinespace[2pt]
& &  HNN &  &  BART && {\notsotiny QBART} &  & BLR & & BQR & & AR$_\text{\halftiny SV}$  & & {\notsotiny BQAR} & & &
 HNN &  & BART && {\notsotiny QBART} &  & BLR & & BQR & & AR$_\text{\halftiny SV}$ && {\notsotiny BQAR} & \\
\midrule
 \addlinespace[5pt] 
\rowcolor{gray!15} 
\multicolumn{30}{l}{\textbf{GDP}} &\cellcolor{gray!15} \\ \addlinespace[2pt] 
   CRPS$_\text{\halftiny center}$   &&   0.80&&   0.80&&   \textbf{0.79}&&   0.90&&   0.80&&   0.86&&   0.89&&&   0.82&&   0.92&&   \textbf{0.80}&&   0.94&&   0.94&&   0.90&&   0.97\\
   CRPS$_\text{\halftiny left}$   &&   0.90&&   0.88&&   \textbf{0.87}&&   0.98&&   0.94&&   1.03&&   1.00&&&   0.89&&   1.01&&   \textbf{0.82}&&   0.97&&   0.95&&   0.96&&   1.03\\
   CRPS$_\text{\halftiny right}$   &&   0.66&&   0.64&&   0.70&&   0.74&&   \textbf{0.61}&&   0.67&&   0.72&&&   \textbf{0.70}&&   0.82&&   0.74&&   0.82&&   0.90&&   0.77&&   0.85\\
   $\mathcal{L}$   &&   -3.70&&   -3.70&&   \textbf{-3.72}&&   -3.59&&   -2.91&&   -3.04&&   -3.37&&&   -3.61&&   -3.55&&   \textbf{-3.65}&&   -3.51&&    0.42&&   -3.05&&   -2.95\\
   \addlinespace[5pt] 
\rowcolor{gray!15} 
\multicolumn{30}{l}{\textbf{Unemployment Rate}} &\cellcolor{gray!15} \\ \addlinespace[2pt] 
   CRPS$_\text{\halftiny center}$   &&   \textbf{0.58}&&   0.67&&   0.63&&   0.69&&   0.66&&   0.75&&   0.74&&&   0.63&&   0.90&&   \textbf{0.61}&&   0.67&&   0.73&&   0.81&&   0.85\\
   CRPS$_\text{\halftiny left}$   &&   \textbf{0.55}&&   0.69&&   0.59&&   0.59&&   0.62&&   0.65&&   0.66&&&   \textbf{0.58}&&   0.95&&   0.58&&   0.59&&   0.85&&   0.78&&   0.84\\
   CRPS$_\text{\halftiny right}$   &&   \textbf{0.51}&&   0.60&&   0.62&&   0.69&&   0.61&&   0.75&&   0.72&&&   0.62&&   0.94&&   \textbf{0.61}&&   0.70&&   0.61&&   0.80&&   0.86\\
   $\mathcal{L}$   &&   \textbf{-0.17}&&    0.65&&    0.05&&    0.19&&    0.97&&    0.97&&    0.67&&&   \textbf{0.03}&&   0.82&&   0.14&&   0.30&&   3.53&&   1.05&&   0.87\\
   \addlinespace[5pt] 
\rowcolor{gray!15} 
\multicolumn{30}{l}{\textbf{Inflation}} &\cellcolor{gray!15} \\ \addlinespace[2pt] 
   CRPS$_\text{\halftiny center}$   &&   \textbf{0.93}&&   1.11&&   0.94&&   1.02&&   1.04&&   1.08&&   0.96&&&   0.95&&   1.02&&   \textbf{0.90}&&   1.08&&   1.09&&   1.02&&   0.97\\
   CRPS$_\text{\halftiny left}$   &&   \textbf{0.97}&&   1.13&&   1.01&&   1.06&&   1.04&&   1.14&&   1.03&&&   0.98&&   1.11&&   \textbf{0.97}&&   1.08&&   1.05&&   1.08&&   0.99\\
   CRPS$_\text{\halftiny right}$   &&   \textbf{0.86}&&   1.12&&   0.95&&   0.96&&   1.03&&   0.99&&   0.91&&&   0.92&&   0.97&&   \textbf{0.87}&&   1.08&&   1.16&&   0.96&&   1.02\\
   $\mathcal{L}$   &&   -3.48&&   -3.19&&   \textbf{-3.71}&&   -3.60&&   -2.38&&   -3.45&&   -2.25&&&   -3.30&&   -2.99&&   \textbf{-3.61}&&   -3.37&&   13.55&&   -3.27&&   -0.78\\
  \addlinespace[5pt] 
\rowcolor{gray!15} 
\multicolumn{30}{l}{\textbf{S\&P 500}} &\cellcolor{gray!15} \\ \addlinespace[2pt] 
   CRPS$_\text{\halftiny center}$   &&   0.97&&   \textbf{0.91}&&   0.98&&   1.03&&   1.03&&   1.06&&   0.97&&&   0.99&&   \textbf{0.94}&&   0.99&&   1.11&&   1.04&&   1.07&&   0.99\\
   CRPS$_\text{\halftiny left}$   &&   0.96&&   \textbf{0.90}&&   0.97&&   0.97&&   1.02&&   1.04&&   0.98&&&   0.99&&   \textbf{0.94}&&   0.99&&   1.07&&   1.05&&   1.06&&   1.03\\
   CRPS$_\text{\halftiny right}$   &&   0.98&&   \textbf{0.92}&&   1.01&&   1.10&&   1.08&&   1.09&&   0.99&&&   0.99&&   \textbf{0.96}&&   1.00&&   1.19&&   1.11&&   1.11&&   1.02\\
   $\mathcal{L}$   &&   -1.27&&   \textbf{-1.36}&&   -1.25&&   -1.25&&    0.59&&   -0.59&&   -0.29&&&   -1.27&&   \textbf{-1.30}&&   -1.24&&   -1.16&&    2.21&&   -0.63&&   -0.17\\
   \addlinespace[5pt] 
\rowcolor{gray!15} 
\multicolumn{30}{l}{\textbf{Housting Start}} &\cellcolor{gray!15} \\ \addlinespace[2pt] 
   CRPS$_\text{\halftiny center}$   &&   1.06&&   0.96&&   \textbf{0.95}&&   1.05&&   1.02&&   0.97&&   0.99&&&   1.07&&   0.99&&   \textbf{0.97}&&   1.12&&   1.08&&   0.99&&   1.00\\
   CRPS$_\text{\halftiny left}$   &&   1.07&&   0.95&&   \textbf{0.94}&&   1.05&&   1.01&&   1.00&&   0.97&&&   1.06&&   1.02&&   \textbf{0.95}&&   1.12&&   1.08&&   1.02&&   0.99\\
   CRPS$_\text{\halftiny right}$   &&   1.04&&   \textbf{0.96}&&   0.96&&   1.02&&   1.10&&   0.98&&   1.02&&&   1.07&&   \textbf{0.98}&&   0.99&&   1.09&&   1.17&&   1.00&&   1.03\\
   $\mathcal{L}$   &&   -0.88&&   \textbf{-1.10}&&   -1.08&&   -1.02&&    0.30&&   -0.81&&   -0.64&&&   -0.66&&   -0.75&&   -0.78&&   \textbf{-0.83}&&    3.91&&    0.06&&   -0.04\\
\bottomrule \bottomrule
\end{tabular}
\begin{tablenotes}[para,flushleft]
  \scriptsize 
    \textit{Notes}: The table shows quantile-weighted continuous ranked probability score (CRPS$_{\omega}$) as well as log scores ($\mathcal{L}$). To target the left tail of the predictive distribution (downside risk, CRPS$_\text{\halftiny left}$), we set the weights to $\omega_{\tau} = (1-\tau)^2$. The right tail (upside risk, CRPS$_\text{\halftiny right}$) is targeted by setting the weights to $\omega_{\tau} = \tau^2$. We evaluate the center of the distribution (CRPS$_\text{\halftiny center}$) by using $\omega_{\tau} = \tau(1-\tau)$. We compute 19 quantiles with $\tau \in \{0.05,0.10,\dots,0.90,0.95\}$. For the real activity variables (i.e., GDP and unemployment) we exclude the year 2020 from the evaluation sample.
  \end{tablenotes}
\end{threeparttable}
\end{table}

\begin{table}[h!]
\vspace{0.3cm}
\centering
\begin{threeparttable}

\centering
\scriptsize
\footnotesize
\setlength{\tabcolsep}{0.225em}
 \setlength\extrarowheight{2.5pt}
\caption{\normalsize {Forecast Performance of the Neural Phillips Curve with Proactive Volatility} \vspace*{-0.3cm}} \label{tab:dhnn}
 \begin{tabular}{l ccccccccccccc | ccccccccccccc} 
\toprule \toprule
\addlinespace[2pt]
& & \multicolumn{12}{c}{2007Q1 - 2019Q4} & &  \multicolumn{11}{c}{2007Q1 - 2022Q4} \\
\cmidrule(lr){3-13} \cmidrule(lr){16-26} \addlinespace[2pt]
& &  RMSE &  & $\mathcal{L}$ &  & $R^2_{|\varepsilon_t|}$  &  & CRPS & & Cov68 && PIT-pv   & & &
 RMSE &  & $\mathcal{L}$ &  & $R^2_{|\varepsilon_t|}$ &  & CRPS & & Cov68&& PIT-pv    &\\
\midrule
\addlinespace[5pt] 
\rowcolor{gray!15} 
\multicolumn{26}{l}{\textbf{Inflation ($s=1$)}} &\cellcolor{gray!15} \\ \addlinespace[2pt] 
HNN      &        &  0.93 &        &    -3.63 &        &             -0.06 &        &    0.96 &        &  \textbf{67.3} &        &      0.41  &        &        &  1.12 &        &    -3.41 &        &             \textbf{0.17} &        &    1.05 &        &   62.5 &        &      0.61  &        \\
HNN-NPC &        & \textbf{0.88} &        &   \textbf{-3.87} &        &           \textbf{0.09} &        &    \textbf{0.91}  &        &  69.2 &        &     \textbf{0.74}  &        &        & \textbf{1.02} &        &   \textbf{-3.60} &        &            0.13 &        &   \textbf{0.98} &        &   \textbf{64.1} &        &     \textbf{0.69} &            \\
\bottomrule \bottomrule
\end{tabular}
\begin{tablenotes}[para,flushleft]
  \scriptsize 
    \textit{Notes}: The table shows our set of evaluation metrics for one-step and four steps ahead predictions ($s \in \{1,4\}$). Cov68 refers to the 68\% coverage rate and the p-value of the PIT-based auto-calibration test \citep{knuppel2022score} is given by PIT-pv.  
    We compare the performance of HNN and the extension to the Neural Phillips Curve with proactive volatility (HNN-NPC).
  \end{tablenotes}
\end{threeparttable}
\vspace{-0.3cm}
\end{table}

\clearpage

\subsection{A FRED-MD Detour}\label{app:md}
         
\noindent Since researchers and policymakers are often interested in more timely forecasts with a sampling frequency higher than quarterly, we expand our exercise to a monthly setup. This way we also gain insights in our model's behavior and performance when dealing with more noisy data. We apply our proposed model and our rich set of competitors to the FRED-MD database of \cite{mccrackenng}. Again, we explore the model's density and point forecasting performance by predicting different targets including real activity variables, monetary aggregates and inflation series in the US. In particular, we forecast nonfarm payroll, industrial production, real personal income, personal consumption expenditures, retail and food services sales, M2 money stock, and producer price inflation. We compute forecasts for one-month ahead, six months ahead and one year ahead. Similar to the quarterly case, our hold-out sample starts at the beginning of 2007 (i.e., 2007M1) and ends in 2022 (i.e., 2022M12). Results are presented in Table \ref{tab:FredMD2s1}, Table \ref{tab:FredMD2s3}, Table \ref{tab:FredMD2s6} and Table \ref{tab:FredMD2s12} for one-month, three months, six months and twelve months ahead, respectively.

In line with the findings for the quarterly application, HNN yields good forecasting results for real activity variables. Focusing on one-month ahead forecasts Table \ref{tab:FredMD2s1} shows that BLR and AR with time-varying volatility gives the lowest RMSE and log scores across many targets. However, for most of them HNN gets very close or even manages to outperform them. For example, we obtain very similar results from the best performing model and HNN for nonfarm payroll, industrial production and real personal income. For real personal consumption expenditures and retail sales, HNN outperforms all other benchmarks for the evaluation sample up to 2020 and remains highly competitive for the full sample. All nonlinear models suffer from inferior point forecasts when it comes to producer price inflation and the M2 money stock. Yet, they yield highly competitive density predictions.  Overall,  short-run monthly results reveal that alternative specifications have a hard time improving on the performance of linear benchmarks.  Yet, HNN is always near the top of the pack, highlighting its reliability even when the simplest approach comes out on top.

Nonlinearities seem to gain in importance for higher-order forecasts -- a finding that echoes to that of \cite{GCLSS2020}. In the case of three and six months ahead density forecasts (see Table \ref{tab:FredMD2s3} and Table \ref{tab:FredMD2s6}), we find that either HNN or BART yield the lowest log scores for all real activity and employment variables. Also, the nonlinear models outperform the simpler benchmarks for point forecasts when including the post-Covid periods. This holds for all cases when considering the six months ahead forecast horizon and for the most when evaluating the three months ahead horizon. A similar picture arises from forecasting our monthly target set twelve months ahead (see Table \ref{tab:FredMD2s12}). BLR often remains hard to beat but nonlinear models tend to outperform it for density predictions and the full sample.  Noteworthy,  HNN catches up with BLR when forecasting producer price inflation for twelve months ahead. It yields highly competitive point forecasts for both samples and the best density prediction for the full sample.

\begin{table}[h!]
\vspace{-1.8em}
\begin{threeparttable}

\centering
\scriptsize
\footnotesize
\setlength{\tabcolsep}{0.225em}
 \setlength\extrarowheight{2.5pt}
 \caption{\normalsize {Forecast Performance on Monthly Data ($s=1$)} \vspace*{-0.3cm}} \label{tab:FredMD2s1}
 \begin{tabular}{l rrrrrrrrrrrrrrr | rrrrrrrrrrrrrrr} 
\toprule \toprule
\addlinespace[2pt]
& & \multicolumn{14}{c}{2007M1 - 2019M12} & &  \multicolumn{13}{c}{2007M1 - 2022M12} \\
\cmidrule(lr){3-15} \cmidrule(lr){18-30} \addlinespace[2pt]
& &  HNN &  & NN$_\text{\halftiny SV}$ &  & NN$_\text{\halftiny G}$ &  & {\notsotiny DeepAR} & & BART && AR$_\text{\halftiny TV}$  &  & BLR & & &
 HNN &  & NN$_\text{\halftiny SV}$ &  & NN$_\text{\halftiny G}$ &  & {\notsotiny DeepAR} & & BART && AR$_\text{\halftiny TV}$  &  & BLR   &\\
\midrule

\addlinespace[5pt] 
\rowcolor{gray!15} 
\multicolumn{30}{l}{\textbf{Nonfarm Payroll }} &\cellcolor{gray!15} \\ \addlinespace[2pt] 

{\notsotiny RMSE} &   & 0.95 &   & 0.94 &   & 0.94 &   & 0.97 &   & 1.01 &   & 0.97 &   & \textbf{0.92} &   &   & 1.09 &   & 1.07 &   & 1.07 &   & 1.00 &   & 1.15 &   & \textbf{0.98} &   & 1.62 &   \\
$\mathcal{L}$ &   & -5.59 &   & -5.63 &   & \textbf{-5.65} &   & -5.07 &   & -5.36 &   & -5.65 &   & -5.16 &   &   & -5.38 &   & -5.51 &   & -5.40 &   & -4.99 &   & -5.25 &   & \textbf{-5.57} &   & -5.01 &   \\
$R^2_{|\varepsilon_t|}$ &   & 0.50 &   & 0.57 &   & 0.59 &   & 0.36 &   & 0.04 &   & \textbf{0.64} &   & -2.79 &   &   & \textbf{0.37} &   & 0.12 &   & -17.89 &   & 0.29 &   & 0.07 &   & -0.49 &   & -1.10 &   \\

\addlinespace[5pt] 
\rowcolor{gray!15} 
\multicolumn{30}{l}{\textbf{Industrial Production}} &\cellcolor{gray!15} \\ \addlinespace[2pt] 

{\notsotiny RMSE} &   & 0.93 &   & 0.94 &   & 0.94 &   & 1.00 &   & 0.96 &   & 0.99 &   & \textbf{0.93} &   &   & \textbf{0.92} &   & 0.97 &   & 0.97 &   & 0.97 &   & 0.95 &   & 0.98 &   & 0.97 &   \\
$\mathcal{L}$ &   & -3.71 &   & -3.62 &   & \textbf{-3.75} &   & -1.11 &   & -3.72 &   & -3.74 &   & -3.64 &   &   & -3.50 &   & -3.53 &   & \textbf{-3.62} &   & -1.19 &   & -3.49 &   & -3.59 &   & -3.53 &   \\
$R^2_{|\varepsilon_t|}$ &   & 0.07 &   & 0.08 &   & 0.13 &   & -0.10 &   & 0.12 &   & \textbf{0.20} &   & -0.51 &   &   & 0.03 &   & 0.04 &   & 0.11 &   & -0.07 &   & 0.13 &   & \textbf{0.22} &   & -0.22 &   \\
\addlinespace[5pt] 
\rowcolor{gray!15} 
\multicolumn{30}{l}{\textbf{Real Personal Income Excluding Current Transfers}} &\cellcolor{gray!15} \\ \addlinespace[2pt] 

{\notsotiny RMSE} &   & 0.93 &   & 0.95 &   & 0.95 &   & \textbf{0.93} &   & 0.96 &   & 0.99 &   & 0.94 &   &   & 0.94 &   & 0.97 &   & 0.97 &   & 0.94 &   & 0.96 &   & 0.98 &   & \textbf{0.94} &   \\
$\mathcal{L}$ &   & -3.58 &   & -3.73 &   & -3.88 &   & -1.19 &   & -3.70 &   & \textbf{-4.10} &   & -3.66 &   &   & -3.63 &   & -3.72 &   & -3.86 &   & -1.52 &   & -3.72 &   & \textbf{-4.11} &   & -3.69 &   \\
$R^2_{|\varepsilon_t|}$ &   & -0.08 &   & 0.02 &   & -0.14 &   & -0.19 &   & 0.05 &   & \textbf{0.15} &   & -0.17 &   &   & -0.08 &   & -0.02 &   & -0.16 &   & -0.27 &   & 0.04 &   & \textbf{0.13} &   & -0.21 &   \\

\addlinespace[5pt] 
\rowcolor{gray!15} 
\multicolumn{30}{l}{\textbf{Real Personal Consumption Expenditures }} &\cellcolor{gray!15} \\ \addlinespace[2pt] 

{\notsotiny RMSE}  &   & \textbf{0.92} &   & 0.98 &   & 0.98 &   & 0.96 &   & 0.97 &   & 0.97 &   & 0.94 &   &   & 1.02 &   & 1.02 &   & 1.02 &   & 1.02 &   & 1.04 &   & \textbf{1.00} &   & 1.03 &   \\
$\mathcal{L}$ &   & -4.42 &   & -4.36 &   & -4.37 &   & -4.37 &   & -4.36 &   & \textbf{-4.42} &   & -4.15 &   &   & -3.59 &   & -4.26 &   & -4.18 &   & -3.90 &   & -4.21 &   & \textbf{-4.29} &   & -4.02 &   \\
$R^2_{|\varepsilon_t|}$ &   & 0.60 &   & 0.53 &   & 0.54 &   & 0.55 &   & 0.29 &   & \textbf{0.61} &   & -0.76 &   &   & 0.28 &   & \textbf{0.29} &   & -1.21 &   & 0.28 &   & 0.02 &   & 0.08 &   & -0.46 &   \\

\addlinespace[5pt] 
\rowcolor{gray!15} 
\multicolumn{30}{l}{\textbf{Retail and Food Services Sales }} &\cellcolor{gray!15} \\ \addlinespace[2pt] 

{\notsotiny RMSE} &   & \textbf{0.85} &   & 0.97 &   & 0.97 &   & 0.91 &   & 0.87 &   & 0.98 &   & 0.89 &   &   & \textbf{0.94} &   & 0.99 &   & 0.99 &   & 0.96 &   & 0.96 &   & 0.99 &   & 0.96 &   \\
$\mathcal{L}$ &   & \textbf{-3.38} &   & -3.26 &   & -3.28 &   & -3.03 &   & -3.36 &   & -3.34 &   & -3.18 &   &   & -2.87 &   & -3.13 &   & -3.15 &   & -2.54 &   & -3.15 &   & \textbf{-3.22} &   & -3.06 &   \\
$R^2_{|\varepsilon_t|}$ &   & \textbf{0.38} &   & 0.29 &   & 0.31 &   & 0.24 &   & 0.15 &   & 0.31 &   & -0.50 &   &   & \textbf{0.18} &   & 0.14 &   & 0.14 &   & 0.10 &   & -0.00 &   & 0.18 &   & -0.27 &   \\

\addlinespace[5pt] 
\rowcolor{gray!15} 
\multicolumn{30}{l}{\textbf{M2 Nominal Money Stock}} &\cellcolor{gray!15} \\ \addlinespace[2pt] 

{\notsotiny RMSE} &   & 1.10 &   & 1.08 &   & 1.08 &   & 1.15 &   & 1.08 &   & 1.03 &   & \textbf{1.02} &   &   & 1.07 &   & 1.10 &   & 1.10 &   & 1.01 &   & 1.01 &   & \textbf{0.98} &   & 1.06 &   \\
$\mathcal{L}$ &   & -4.17 &   & -4.13 &   & -4.18 &   & -3.88 &   & -4.14 &   & \textbf{-4.24} &   & -4.20 &   &   & \textbf{-3.80} &   & -3.10 &   & -3.64 &   & -3.37 &   & -3.41 &   & -3.76 &   & -3.62 &   \\
$R^2_{|\varepsilon_t|}$ &   & -0.04 &   & -0.08 &   & -0.05 &   & -0.03 &   & 0.04 &   & \textbf{0.25} &   & -0.10 &   &   & 0.23 &   & 0.11 &   & 0.20 &   & 0.04 &   & 0.26 &   & \textbf{0.32} &   & 0.11 &   \\

\addlinespace[5pt] 
\rowcolor{gray!15} 
\multicolumn{30}{l}{\textbf{Producer Price Inflation  }} &\cellcolor{gray!15} \\ \addlinespace[2pt] 

{\notsotiny RMSE} &   & 1.07 &   & 1.07 &   & 1.07 &   & 1.05 &   & 1.00 &   & 0.97 &   & \textbf{0.96} &   &   & 1.08 &   & 1.12 &   & 1.12 &   & 1.09 &   & 1.06 &   & \textbf{0.98} &   & 1.00 &   \\
$\mathcal{L}$ &   & -3.40 &   & -3.43 &   & -3.45 &   & 1.84 &   & -3.47 &   & \textbf{-3.57} &   & -3.49 &   &   & -3.30 &   & -3.24 &   & -3.29 &   & 1.20 &   & -3.29 &   & \textbf{-3.45} &   & -3.31 &   \\
$R^2_{|\varepsilon_t|}$ &   & -0.05 &   & 0.06 &   & 0.08 &   & -0.21 &   & 0.18 &   & \textbf{0.44} &   & 0.02 &   &   & -0.00 &   & 0.02 &   & 0.00 &   & -0.05 &   & 0.19 &   & \textbf{0.39} &   & 0.09 &   \\

\bottomrule \bottomrule
\end{tabular}
\begin{tablenotes}[para,flushleft]
  \scriptsize 
    \textit{Notes}: The table presents our forecasting evaluation metrics including the root mean square error (RMSE) relative to the AR model with constant variance, the log score ($\mathcal{L}$), and the $R^2_{|\varepsilon_t|}$ of absolute residuals. For the real activity variables (i.e., IP, Consumption, Income and Retail) we exclude the year 2020 from the evaluation sample. AR$_\text{\halftiny TV}$ refers to the best-performing AR specifications, which is either AR$_\text{\halftiny SV}$ or AR$_\text{\halftiny G}$.
  \end{tablenotes}
\end{threeparttable}
\end{table}

\begin{table}[h!]
\vspace{-1.8em}
\begin{threeparttable}

\centering
\scriptsize
\footnotesize
\setlength{\tabcolsep}{0.225em}
 \setlength\extrarowheight{2.5pt}
 \caption{\normalsize {Forecast Performance on Monthly Data ($s=3$)} \vspace*{-0.3cm}} \label{tab:FredMD2s3}
 \begin{tabular}{l rrrrrrrrrrrrrrr | rrrrrrrrrrrrrrr} 
\toprule \toprule
\addlinespace[2pt]
& & \multicolumn{14}{c}{2007M1 - 2019M12} & &  \multicolumn{13}{c}{2007M1 - 2022M12} \\
\cmidrule(lr){3-15} \cmidrule(lr){18-30} \addlinespace[2pt]
& &  HNN &  & NN$_\text{\halftiny SV}$ &  & NN$_\text{\halftiny G}$ &  & {\notsotiny DeepAR} & & BART && AR$_\text{\halftiny TV}$  &  & BLR & & &
 HNN &  & NN$_\text{\halftiny SV}$ &  & NN$_\text{\halftiny G}$ &  & {\notsotiny DeepAR} & & BART && AR$_\text{\halftiny TV}$  &  & BLR   &\\
\midrule

\addlinespace[5pt] 
\rowcolor{gray!15} 
\multicolumn{30}{l}{\textbf{Nonfarm Payroll }} &\cellcolor{gray!15} \\ \addlinespace[2pt] 

{\notsotiny RMSE} &   & 0.90 &   & 0.91 &   & 0.91 &   & 0.87 &   & 0.99 &   & 0.94 &   & \textbf{0.81} &   &   & 0.61 &   & 0.60 &   & 0.60 &   & 0.53 &   & 0.59 &   & 1.08 &   & \textbf{0.48} &   \\
$\mathcal{L}$ &   & \textbf{-5.97} &   & -5.93 &   & -5.89 &   & -4.92 &   & -4.60 &   & -5.76 &   & -5.35 &   &   & -5.69 &   & \textbf{-5.76} &   & -5.66 &   & -4.76 &   & -4.67 &   & -5.62 &   & -5.23 &   \\
$R^2_{|\varepsilon_t|}$  &   & \textbf{0.54} &   & 0.38 &   & 0.35 &   & -0.57 &   & -0.05 &   & 0.52 &   & -4.43 &   &   & \textbf{0.42} &   & -1.98 &   & -15.74 &   & -0.54 &   & -0.46 &   & 0.31 &   & -9.46 &   \\

\addlinespace[5pt] 
\rowcolor{gray!15} 
\multicolumn{30}{l}{\textbf{Industrial Production}} &\cellcolor{gray!15} \\ \addlinespace[2pt] 

{\notsotiny RMSE} &   & 0.95 &   & 0.97 &   & 0.97 &   & 0.99 &   & 0.89 &   & 0.96 &   & \textbf{0.87} &   &   & 0.94 &   & 1.00 &   & 1.00 &   & 0.97 &   & \textbf{0.88} &   & 1.02 &   & 0.88 &   \\
$\mathcal{L}$ &   & \textbf{-4.16} &   & -4.10 &   & -4.14 &   & -3.81 &   & -4.08 &   & -4.06 &   & -4.02 &   &   & \textbf{-4.15} &   & -4.05 &   & -4.07 &   & -3.78 &   & -4.01 &   & -4.01 &   & -3.99 &   \\
$R^2_{|\varepsilon_t|}$ &   & 0.23 &   & 0.19 &   & 0.23 &   & 0.05 &   & 0.25 &   & \textbf{0.45} &   & -0.89 &   &   & 0.23 &   & -0.01 &   & -0.32 &   & 0.03 &   & 0.17 &   & \textbf{0.47} &   & -0.94 &   \\

\addlinespace[5pt] 
\rowcolor{gray!15} 
\multicolumn{30}{l}{\textbf{Real Personal Income Excluding Current Transfers}} &\cellcolor{gray!15} \\ \addlinespace[2pt] 

{\notsotiny RMSE} &   & \textbf{0.92} &   & 0.95 &   & 0.95 &   & 0.96 &   & 0.94 &   & 1.07 &   & 0.93 &   &   & 0.98 &   & 1.01 &   & 1.01 &   & 0.98 &   & 0.95 &   & 1.10 &   & \textbf{0.93} &   \\
$\mathcal{L}$ &   & -4.09 &   & -4.15 &   & -4.17 &   & -3.42 &   & \textbf{-4.47} &   & -4.11 &   & -4.13 &   &   & -4.05 &   & -4.11 &   & -4.12 &   & -3.45 &   & \textbf{-4.45} &   & -4.11 &   & -4.16 &   \\
$R^2_{|\varepsilon_t|}$  &   & -0.16 &   & -0.07 &   & -0.13 &   & -0.07 &   & 0.22 &   & \textbf{0.32} &   & -0.13 &   &   & -0.13 &   & -0.12 &   & -0.27 &   & -0.10 &   & 0.21 &   & \textbf{0.36} &   & -0.17 &   \\

\addlinespace[5pt] 
\rowcolor{gray!15} 
\multicolumn{30}{l}{\textbf{Real Personal Consumption Expenditures }} &\cellcolor{gray!15} \\ \addlinespace[2pt] 

{\notsotiny RMSE} &   & 1.01 &   & 1.09 &   & 1.09 &   & 1.09 &   & \textbf{0.94} &   & 0.98 &   & 1.04 &   &   & 1.06 &   & 1.08 &   & 1.08 &   & 1.07 &   & \textbf{1.06} &   & 1.24 &   & 1.07 &   \\
$\mathcal{L}$  &   & \textbf{-4.99} &   & -4.93 &   & -4.94 &   & -4.68 &   & -4.81 &   & -4.97 &   & -4.78 &   &   & -4.64 &   & -4.84 &   & -4.84 &   & -4.44 &   & -4.71 &   & \textbf{-4.86} &   & -4.68 &   \\
$R^2_{|\varepsilon_t|}$ &   & 0.36 &   & 0.35 &   & 0.36 &   & 0.39 &   & 0.16 &   & \textbf{0.53} &   & -0.86 &   &   & 0.20 &   & 0.05 &   & -0.58 &   & 0.22 &   & 0.10 &   & \textbf{0.64} &   & -0.84 &   \\

\addlinespace[5pt] 
\rowcolor{gray!15} 
\multicolumn{30}{l}{\textbf{Retail and Food Services Sales }} &\cellcolor{gray!15} \\ \addlinespace[2pt] 

{\notsotiny RMSE} &   & 0.94 &   & 1.09 &   & 1.09 &   & 0.93 &   & \textbf{0.89} &   & 0.95 &   & 0.94 &   &   & 0.99 &   & 1.07 &   & 1.07 &   & 0.96 &   & 0.97 &   & 1.05 &   & \textbf{0.95} &   \\
$\mathcal{L}$ &   & -3.71 &   & -3.46 &   & -3.57 &   & -3.03 &   & \textbf{-3.96} &   & -3.63 &   & -3.69 &   &   & -3.30 &   & -3.43 &   & -3.53 &   & -2.80 &   & -3.80 &   & -3.56 &   & \textbf{-3.63} &   \\
$R^2_{|\varepsilon_t|}$ &   & 0.12 &   & 0.05 &   & -0.16 &   & -0.04 &   & \textbf{0.36} &   & 0.30 &   & -0.25 &   &   & 0.04 &   & -0.03 &   & -0.29 &   & -0.03 &   & 0.36 &   & \textbf{0.39} &   & -0.18 &   \\

\addlinespace[5pt] 
\rowcolor{gray!15} 
\multicolumn{30}{l}{\textbf{M2 Nominal Money Stock}} &\cellcolor{gray!15} \\ \addlinespace[2pt] 

{\notsotiny RMSE} &   & 1.10 &   & 1.11 &   & 1.11 &   & 1.16 &   & 1.01 &   & 0.99 &   & \textbf{0.97} &   &   & 1.22 &   & 1.17 &   & 1.17 &   & 1.22 &   & 1.14 &   & \textbf{0.99} &   & 1.04 &   \\
$\mathcal{L}$ &   & -5.17 &   & -5.15 &   & -5.21 &   & -4.82 &   & -5.02 &   & \textbf{-5.28} &   & -5.27 &   &   & -3.60 &   & -3.59 &   & -3.54 &   & -2.47 &   & -4.11 &   & -3.04 &   & \textbf{-4.47} &   \\
$R^2_{|\varepsilon_t|}$ &   & -0.06 &   & -0.08 &   & 0.05 &   & -0.02 &   & 0.21 &   & \textbf{0.41} &   & -0.20 &   &   & 0.09 &   & 0.01 &   & 0.03 &   & 0.02 &   & \textbf{0.39} &   & 0.26 &   & 0.15 &   \\

\addlinespace[5pt] 
\rowcolor{gray!15} 
\multicolumn{30}{l}{\textbf{Producer Price Inflation  }} &\cellcolor{gray!15} \\ \addlinespace[2pt] 

{\notsotiny RMSE} &   & 1.04 &   & 1.04 &   & 1.04 &   & 1.08 &   & 1.01 &   & 0.99 &   & \textbf{0.95} &   &   & 1.04 &   & 1.10 &   & 1.10 &   & 1.09 &   & 1.00 &   & 1.00 &   & \textbf{0.97} &   \\
$\mathcal{L}$ &   & -4.34 &   & -4.42 &   & -4.48 &   & -3.44 &   & -4.33 &   & \textbf{-4.49} &   & -4.47 &   &   & -4.21 &   & -4.19 &   & -4.23 &   & -3.27 &   & -4.23 &   & -4.22 &   & \textbf{-4.31} &   \\
$R^2_{|\varepsilon_t|}$ &   & -0.08 &   & 0.03 &   & 0.09 &   & -0.03 &   & 0.19 &   & \textbf{0.42} &   & 0.01 &   &   & 0.05 &   & 0.09 &   & 0.12 &   & 0.08 &   & 0.28 &   & \textbf{0.39} &   & 0.15 &   \\

\bottomrule \bottomrule
\end{tabular}
\begin{tablenotes}[para,flushleft]
  \scriptsize 
    \textit{Notes}: The table presents our forecasting evaluation metrics including the root mean square error (RMSE) relative to the AR model with constant variance, the log score ($\mathcal{L}$), and the $R^2_{|\varepsilon_t|}$ of absolute residuals. For the real activity variables (i.e., IP, Consumption, Income and Retail) we exclude the year 2020 from the evaluation sample. AR$_\text{\halftiny TV}$ refers to the best-performing AR specifications, which is either AR$_\text{\halftiny SV}$ or AR$_\text{\halftiny G}$.
  \end{tablenotes}
\end{threeparttable}
\end{table}

\begin{table}[h!]
\vspace{-1.8em}
\begin{threeparttable}

\centering
\scriptsize
\footnotesize
\setlength{\tabcolsep}{0.225em}
 \setlength\extrarowheight{2.5pt}
 \caption{\normalsize {Forecast Performance on Monthly Data ($s=6$)} \vspace*{-0.3cm}} \label{tab:FredMD2s6}
 \begin{tabular}{l rrrrrrrrrrrrrrr | rrrrrrrrrrrrrrr} 
\toprule \toprule
\addlinespace[2pt]
& & \multicolumn{14}{c}{2007M1 - 2019M12} & &  \multicolumn{13}{c}{2007M1 - 2022M12} \\
\cmidrule(lr){3-15} \cmidrule(lr){18-30} \addlinespace[2pt]
& &  HNN &  & NN$_\text{\halftiny SV}$ &  & NN$_\text{\halftiny G}$ &  & {\notsotiny DeepAR} & & BART && AR$_\text{\halftiny TV}$  &  & BLR & & &
 HNN &  & NN$_\text{\halftiny SV}$ &  & NN$_\text{\halftiny G}$ &  & {\notsotiny DeepAR} & & BART && AR$_\text{\halftiny TV}$  &  & BLR   &\\
\midrule

\addlinespace[5pt] 
\rowcolor{gray!15} 
\multicolumn{30}{l}{\textbf{Nonfarm Payroll }} &\cellcolor{gray!15} \\ \addlinespace[2pt] 

{\notsotiny RMSE} &   & 0.82 &   & 0.88 &   & 0.88 &   & 0.97 &   & 0.88 &   & 0.91 &   & \textbf{0.77} &   &   & 0.40 &   & 0.41 &   & 0.41 &   & \textbf{0.36} &   & 0.37 &   & 1.06 &   & 0.60 &   \\
$\mathcal{L}$ &   & \textbf{-6.07} &   & -5.78 &   & -5.70 &   & -4.69 &   & -3.97 &   & -5.54 &   & -5.38 &   &   & -5.59 &   & \textbf{-5.59} &   & -5.53 &   & -3.70 &   & -4.12 &   & -5.18 &   & -5.25 &   \\
$R^2_{|\varepsilon_t|}$ &   & \textbf{0.69} &   & 0.35 &   & 0.06 &   & -0.65 &   & 0.43 &   & 0.45 &   & -2.87 &   &   & \textbf{0.54} &   & -3.46 &   & -9.68 &   & -0.64 &   & -1.20 &   & 0.38 &   & -0.17 &   \\

\addlinespace[5pt] 
\rowcolor{gray!15} 
\multicolumn{30}{l}{\textbf{Industrial Production}} &\cellcolor{gray!15} \\ \addlinespace[2pt] 

{\notsotiny RMSE} &   & 0.90 &   & 0.94 &   & 0.94 &   & 1.00 &   & 0.92 &   & 0.93 &   & \textbf{0.87} &   &   & 0.88 &   & 0.92 &   & 0.92 &   & 0.94 &   & \textbf{0.87} &   & 1.04 &   & 0.96 &    \\
$\mathcal{L}$ &   & \textbf{-4.15} &   & -3.94 &   & -4.06 &   & -3.37 &   & -4.00 &   & -4.00 &   & -4.10 &   &   & \textbf{-4.18} &   & -3.95 &   & -4.00 &   & -3.54 &   & -4.10 &   & -3.93 &   & -4.05 &   \\
$R^2_{|\varepsilon_t|}$ &   & 0.18 &   & 0.02 &   & 0.19 &   & -0.03 &   & \textbf{0.57} &   & 0.40 &   & -0.54 &   &   & 0.17 &   & -0.45 &   & -1.63 &   & -0.02 &   & 0.42 &   & \textbf{0.43} &   & -0.35 &   \\

\addlinespace[5pt] 
\rowcolor{gray!15} 
\multicolumn{30}{l}{\textbf{Real Personal Income Excluding Current Transfers}} &\cellcolor{gray!15} \\ \addlinespace[2pt] 

{\notsotiny RMSE} &   & \textbf{0.86} &   & 0.91 &   & 0.91 &   & 0.97 &   & 0.92 &   & 1.13 &   & 0.88 &   &   & 1.05 &   & 1.23 &   & 1.23 &   & 0.97 &   & \textbf{0.92} &   & 1.18 &   & 1.19 &   \\
$\mathcal{L}$ &   & -4.56 &   & -4.56 &   & -4.59 &   & -3.46 &   & \textbf{-4.68} &   & -4.33 &   & -4.57 &   &   & -4.40 &   & -4.33 &   & -4.44 &   & -3.54 &   & \textbf{-4.64} &   & -4.30 &   & -4.30 &   \\
$R^2_{|\varepsilon_t|}$ &   & -0.44 &   & -0.09 &   & -0.04 &   & -0.14 &   & 0.32 &   & \textbf{0.35} &   & -0.17 &   &   & -0.13 &   & 0.10 &   & -0.23 &   & -0.11 &   & 0.27 &   & \textbf{0.43} &   & 0.05 &   \\

\addlinespace[5pt] 
\rowcolor{gray!15} 
\multicolumn{30}{l}{\textbf{Real Personal Consumption Expenditures }} &\cellcolor{gray!15} \\ \addlinespace[2pt] 

{\notsotiny RMSE} &   & 0.91 &   & 0.92 &   & 0.92 &   & 1.16 &   & 0.95 &   & \textbf{0.89} &   & 1.01 &   &   & 0.88 &   & 0.89 &   & 0.89 &   & 0.95 &   & \textbf{0.80} &   & 1.01 &   & 1.27 &    \\
$\mathcal{L}$  &   & \textbf{-5.32} &   & -5.27 &   & -5.27 &   & -4.49 &   & -5.14 &   & -5.15 &   & -5.06 &   &   & -5.10 &   & -5.15 &   & \textbf{-5.15} &   & -4.32 &   & -5.07 &   & -5.03 &   & -4.94 &   \\
$R^2_{|\varepsilon_t|}$ &   & 0.31 &   & 0.24 &   & 0.24 &   & 0.01 &   & \textbf{0.46} &   & 0.44 &   & -0.72 &   &   & 0.29 &   & -0.25 &   & -0.73 &   & -0.01 &   & 0.08 &   & \textbf{0.33} &   & 0.16 &   \\

\addlinespace[5pt] 
\rowcolor{gray!15} 
\multicolumn{30}{l}{\textbf{Retail and Food Services Sales }} &\cellcolor{gray!15} \\ \addlinespace[2pt] 

{\notsotiny RMSE}  &   & 0.90 &   & 1.05 &   & 1.05 &   & 0.90 &   & \textbf{0.87} &   & 1.01 &   & 0.89 &   &   & 0.95 &   & 1.04 &   & 1.04 &   & 0.91 &   & \textbf{0.87} &   & 1.03 &   & 1.15 &    \\
$\mathcal{L}$ &   & -3.96 &   & -3.38 &   & -3.76 &   & -2.27 &   & -4.00 &   & -3.53 &   & \textbf{-4.03} &   &   & -3.74 &   & -3.40 &   & -3.71 &   & -2.13 &   & \textbf{-3.97} &   & -3.49 &   & -3.86 &   \\
$R^2_{|\varepsilon_t|}$ &   & 0.08 &   & 0.10 &   & 0.23 &   & -0.08 &   & \textbf{0.36} &   & 0.31 &   & -0.14 &   &   & 0.07 &   & -0.01 &   & -0.38 &   & -0.07 &   & 0.33 &   & \textbf{0.37} &   & 0.11 &   \\

\addlinespace[5pt] 
\rowcolor{gray!15} 
\multicolumn{30}{l}{\textbf{M2 Nominal Money Stock}} &\cellcolor{gray!15} \\ \addlinespace[2pt] 

{\notsotiny RMSE} &   & 1.03 &   & 1.05 &   & 1.05 &   & 1.07 &   & 0.98 &   & 1.00 &   & \textbf{0.89} &   &   & 1.10 &   & 1.07 &   & 1.07 &   & 1.11 &   & 1.01 &   & 1.03 &   & \textbf{0.93} &   \\
$\mathcal{L}$ &   & -5.84 &   & -5.78 &   & -5.81 &   & -5.58 &   & -5.57 &   & -5.83 &   & \textbf{-5.92} &   &   & -4.47 &   & -4.68 &   & -4.49 &   & -3.63 &   & -4.39 &   & -4.46 &   & \textbf{-5.19} &   \\
$R^2_{|\varepsilon_t|}$ &   & -0.01 &   & -0.08 &   & -0.09 &   & 0.02 &   & 0.04 &   & \textbf{0.29} &   & -0.28 &   &   & 0.08 &   & 0.03 &   & -0.11 &   & 0.06 &   & \textbf{0.27} &   & 0.20 &   & 0.09 &   \\

\addlinespace[5pt] 
\rowcolor{gray!15} 
\multicolumn{30}{l}{\textbf{Producer Price Inflation}} &\cellcolor{gray!15} \\ \addlinespace[2pt] 

{\notsotiny RMSE} &   & 0.98 &   & 1.00 &   & 1.00 &   & 1.09 &   & 0.93 &   & 0.99 &   & \textbf{0.88} &   &   & 0.96 &   & 0.99 &   & 0.99 &   & 1.10 &   & 0.93 &   & 1.01 &   & \textbf{0.90} &    \\
$\mathcal{L}$ &   & -4.91 &   & -4.95 &   & -4.98 &   & -2.41 &   & -5.07 &   & -4.97 &   & \textbf{-5.09} &   &   & -4.83 &   & -4.82 &   & -4.81 &   & -2.24 &   & -4.91 &   & -4.73 &   & \textbf{-4.95} &   \\
$R^2_{|\varepsilon_t|}$ &   & -0.06 &   & 0.04 &   & 0.12 &   & -0.10 &   & 0.18 &   & \textbf{0.26} &   & 0.06 &   &   & -0.00 &   & 0.07 &   & 0.13 &   & -0.04 &   & \textbf{0.26} &   & 0.24 &   & 0.16 &   \\

\bottomrule \bottomrule
\end{tabular}
\begin{tablenotes}[para,flushleft]
  \scriptsize 
    \textit{Notes}: The table presents our forecasting evaluation metrics including the root mean square error (RMSE) relative to the AR model with constant variance, the log score ($\mathcal{L}$), and the $R^2_{|\varepsilon_t|}$ of absolute residuals. For the real activity variables (i.e., IP, Consumption, Income and Retail) we exclude the year 2020 from the evaluation sample. AR$_\text{\halftiny TV}$ refers to the best-performing AR specifications, which is either AR$_\text{\halftiny SV}$ or AR$_\text{\halftiny G}$.
  \end{tablenotes}
\end{threeparttable}
\end{table}

\begin{table}[h!]
\vspace{-1.8em}
\begin{threeparttable}

\centering
\scriptsize
\footnotesize
\setlength{\tabcolsep}{0.225em}
 \setlength\extrarowheight{2.5pt}
 \caption{\normalsize {Forecast Performance on Monthly Data ($s=12$)} \vspace*{-0.3cm}} \label{tab:FredMD2s12}
 \begin{tabular}{l rrrrrrrrrrrrrrr | rrrrrrrrrrrrrrr} 
\toprule \toprule
\addlinespace[2pt]
& & \multicolumn{14}{c}{2007M1 - 2019M12} & &  \multicolumn{13}{c}{2007M1 - 2022M12} \\
\cmidrule(lr){3-15} \cmidrule(lr){18-30} \addlinespace[2pt]
& &  HNN &  & NN$_\text{\halftiny SV}$ &  & NN$_\text{\halftiny G}$ &  & {\notsotiny DeepAR} & & BART && AR$_\text{\halftiny TV}$  &  & BLR & & &
 HNN &  & NN$_\text{\halftiny SV}$ &  & NN$_\text{\halftiny G}$ &  & {\notsotiny DeepAR} & & BART && AR$_\text{\halftiny TV}$  &  & BLR   &\\
\midrule

\addlinespace[5pt] 
\rowcolor{gray!15} 
\multicolumn{30}{l}{\textbf{Nonfarm Payroll }} &\cellcolor{gray!15} \\ \addlinespace[2pt] 

{\notsotiny RMSE} &   & 0.72 &   & 0.83 &   & 0.83 &   & \textbf{0.66} &   & 0.79 &   & 0.91 &   & 0.69 &   &   & 0.58 &   & 0.58 &   & 0.58 &   & \textbf{0.36} &   & 0.41 &   & 0.97 &   & 0.75 &   \\
$\mathcal{L}$ &   & \textbf{-6.02} &   & -5.56 &   & -5.43 &   & 9.04 &   & -1.09 &   & -5.15 &   & -5.39 &   &   & -3.98 &   & -3.52 &   & -4.53 &   & 12.61 &   & -1.55 &   & -4.31 &   & \textbf{-4.95} &   \\
$R^2_{|\varepsilon_t|}$ &   & \textbf{0.73} &   & 0.06 &   & 0.08 &   & -1.02 &   & 0.42 &   & 0.29 &   & -1.79 &   &   & \textbf{0.49} &   & -0.53 &   & -0.14 &   & -0.30 &   & 0.32 &   & 0.40 &   & 0.24 &   \\

\addlinespace[5pt] 
\rowcolor{gray!15} 
\multicolumn{30}{l}{\textbf{Industrial Production}} &\cellcolor{gray!15} \\ \addlinespace[2pt] 

{\notsotiny RMSE}  &   & 0.89 &   & 1.01 &   & 1.01 &   & 0.86 &   & 0.92 &   & 1.03 &   & \textbf{0.80} &   &   & 0.94 &   & 0.99 &   & 0.99 &   & \textbf{0.81} &   & 0.82 &   & 1.02 &   & 0.97 &    \\
$\mathcal{L}$ &   & -4.10 &   & -3.76 &   & -3.64 &   & -3.06 &   & -3.78 &   & -3.69 &   & \textbf{-4.23} &   &   & \textbf{-4.07} &   & -3.68 &   & -3.63 &   & -3.21 &   & -3.89 &   & -3.64 &   & -3.97 &   \\
$R^2_{|\varepsilon_t|}$ &   & 0.17 &   & -0.28 &   & -0.26 &   & -0.03 &   & \textbf{0.43} &   & 0.35 &   & -0.35 &   &   & 0.22 &   & -0.33 &   & -0.33 &   & 0.01 &   & \textbf{0.39} &   & 0.37 &   & -0.02 &   \\

\addlinespace[5pt] 
\rowcolor{gray!15} 
\multicolumn{30}{l}{\textbf{Real Personal Income Excluding Current Transfers}} &\cellcolor{gray!15} \\ \addlinespace[2pt] 

{\notsotiny RMSE} &   & 0.84 &   & 0.88 &   & 0.88 &   & 0.87 &   & 0.88 &   & 1.06 &   & \textbf{0.77} &   &   & 1.25 &   & 1.27 &   & 1.27 &   & 0.86 &   & \textbf{0.85} &   & 1.06 &   & 1.22 &   \\
$\mathcal{L}$ &   & -4.66 &   & -4.66 &   & -4.65 &   & -4.23 &   & -4.53 &   & -4.40 &   & \textbf{-4.78} &   &   & -4.39 &   & -3.61 &   & -4.07 &   & -4.26 &   & \textbf{-4.54} &   & -4.37 &   & -4.01 &   \\
$R^2_{|\varepsilon_t|}$ &   & -0.12 &   & 0.12 &   & 0.11 &   & 0.06 &   & 0.28 &   & \textbf{0.48} &   & -0.12 &   &   & 0.16 &   & 0.09 &   & -0.07 &   & 0.13 &   & 0.31 &   & \textbf{0.50} &   & 0.09 &   \\

\addlinespace[5pt] 
\rowcolor{gray!15} 
\multicolumn{30}{l}{\textbf{Real Personal Consumption Expenditures }} &\cellcolor{gray!15} \\ \addlinespace[2pt] 

{\notsotiny RMSE} &   & 0.75 &   & \textbf{0.75} &   & \textbf{0.75} &   & 1.26 &   & 0.91 &   & 0.87 &   & 0.95 &   &   & 0.80 &   & 0.83 &   & 0.83 &   & 0.81 &   & \textbf{0.69} &   & 1.05 &   & 1.05 &   \\
$\mathcal{L}$ &   & -5.45 &   & -5.42 &   & \textbf{-5.46} &   & 2.86 &   & -4.57 &   & -5.23 &   & -5.13 &   &   & -5.12 &   & -4.78 &   & \textbf{-5.17} &   & 2.47 &   & -4.58 &   & -4.89 &   & -4.47 &   \\
$R^2_{|\varepsilon_t|}$ &   & 0.26 &   & 0.02 &   & 0.14 &   & -0.31 &   & \textbf{0.64} &   & 0.47 &   & -0.36 &   &   & 0.27 &   & -0.04 &   & -0.11 &   & -0.07 &   & \textbf{0.51} &   & 0.40 &   & 0.13 &   \\

\addlinespace[5pt] 
\rowcolor{gray!15} 
\multicolumn{30}{l}{\textbf{Retail and Food Services Sales }} &\cellcolor{gray!15} \\ \addlinespace[2pt] 

{\notsotiny RMSE} &   & 0.85 &   & 0.89 &   & 0.89 &   & \textbf{0.72} &   & 0.78 &   & 0.96 &   & 0.76 &   &   & 1.01 &   & 0.99 &   & 0.99 &   & 0.84 &   & \textbf{0.77} &   & 1.02 &   & 1.20 &    \\
$\mathcal{L}$ &   & -4.16 &   & -3.95 &   & -4.13 &   & \textbf{-4.33} &   & -3.99 &   & -3.74 &   & -4.32 &   &   & -3.59 &   & -3.35 &   & \textbf{-3.90} &   & -3.39 &   & -3.87 &   & -3.64 &   & -3.51 &   \\
$R^2_{|\varepsilon_t|}$ &   & 0.09 &   & 0.01 &   & 0.02 &   & 0.05 &   & 0.32 &   & \textbf{0.34} &   & -0.18 &   &   & 0.13 &   & 0.09 &   & 0.09 &   & 0.09 &   & 0.23 &   & \textbf{0.32} &   & 0.14 &   \\

\addlinespace[5pt] 
\rowcolor{gray!15} 
\multicolumn{30}{l}{\textbf{M2 Nominal Money Stock}} &\cellcolor{gray!15} \\ \addlinespace[2pt] 

{\notsotiny RMSE} &   & 1.01 &   & 1.07 &   & 1.07 &   & 1.21 &   & 0.90 &   & 1.09 &   & \textbf{0.87} &   &   & 1.01 &   & 1.02 &   & 1.02 &   & 1.08 &   & 0.91 &   & 1.06 &   & \textbf{0.87} &   \\
$\mathcal{L}$ &   & -6.43 &   & -6.34 &   & -6.33 &   & -5.37 &   & -6.39 &   & -6.37 &   & \textbf{-6.58} &   &   & -4.95 &   & -5.01 &   & -4.37 &   & -3.91 &   & -5.41 &   & -4.94 &   & \textbf{-5.87} &   \\
$R^2_{|\varepsilon_t|}$ &   & -0.07 &   & -0.04 &   & -0.09 &   & -0.03 &   & 0.01 &   & \textbf{0.28} &   & -0.33 &   &   & 0.06 &   & 0.06 &   & 0.07 &   & 0.06 &   & \textbf{0.31} &   & 0.14 &   & 0.12 &   \\

\addlinespace[5pt] 
\rowcolor{gray!15} 
\multicolumn{30}{l}{\textbf{Producer Price Inflation}} &\cellcolor{gray!15} \\ \addlinespace[2pt] 

{\notsotiny RMSE} &   & 0.91 &   & 0.94 &   & 0.94 &   & 0.99 &   & 0.94 &   & 1.17 &   & \textbf{0.86} &   &   & 0.94 &   & 0.95 &   & 0.95 &   & 1.01 &   & 0.93 &   & 1.14 &   & \textbf{0.92} &   \\
$\mathcal{L}$ &   & -5.76 &   & -5.74 &   & -5.71 &   & -4.29 &   & -5.63 &   & -5.63 &   & \textbf{-5.83} &   &   & \textbf{-5.66} &   & -5.57 &   & -5.57 &   & -4.19 &   & -5.51 &   & -5.50 &   & -5.65 &   \\
$R^2_{|\varepsilon_t|}$ &   & -0.10 &   & -0.03 &   & -0.06 &   & -0.03 &   & 0.19 &   & \textbf{0.22} &   & 0.01 &   &   & 0.08 &   & 0.06 &   & 0.05 &   & 0.04 &   & \textbf{0.23} &   & 0.21 &   & 0.17 &   \\

\bottomrule \bottomrule
\end{tabular}
\begin{tablenotes}[para,flushleft]
  \scriptsize 
    \textit{Notes}: The table presents our forecasting evaluation metrics including the root mean square error (RMSE) relative to the AR model with constant variance, the log score ($\mathcal{L}$), and the $R^2_{|\varepsilon_t|}$ of absolute residuals. For the real activity variables (i.e., IP, Consumption, Income and Retail) we exclude the year 2020 from the evaluation sample. AR$_\text{\halftiny TV}$ refers to the best-performing AR specifications, which is either AR$_\text{\halftiny SV}$ or AR$_\text{\halftiny G}$.
  \end{tablenotes}
\end{threeparttable}
\end{table}

\clearpage

\subsection{Results with Euro Area Data}\label{app:ead}


\noindent In this section we apply our model to euro area data. This entails a major challenge: time series for the euro area are short, most of them only dating back to the early 2000s.  The US data includes several business cycle phases and,  of special relevance recently,   high inflation periods.  There is very little if any of that for our post-2000 euro sample.  Machine learning tools -- with their edge over simpler methods depending on how much history they can learn from -- are in a difficult terrain.  Nonetheless,  due to their ability to flexibly model nonlinearities, however, several recent contributions have shown that using machine learning models for the euro area is promising \citep[see, e.g., ][]{drobetz2021empirical,barbaglia2022testing,huber2023nowcasting,lenza2023density}.  Lastly,  a more subtle problem is that the size of the overall sample limits the span of the test sample,  which, for instance,  excludes the presence of allegedly more predictable recession.  


For our exercise, we use the Euro Area Real Time Database provided by the European Central Bank \citep[see, ][]{giannone2012rtead}. The data set encompasses 165 time series covering several sectors of the real economy as well as financial market developments in the euro area. Due to missing data, we include 145 series spanning the months 2002M2 to 2022M8. Our hold-out sample runs from 2015M1 to 2022M8. We seasonally adjust all series (if applicable), transform them to stationarity (mostly corresponding to the US data set for the respective series) and standardize the data.  Our targets comprise industrial production, unemployment, inflation and the stock market index (Dow Jones Eurostoxx 50).  Forecast horizons are  one-month, three months, six months and twelve months ahead.

\vskip 0.15cm

{\noindent \sc \textbf{Results.}}  Overall, we find that AR models are hard to beat for monthly targets (see Table \ref{tab:EuroS1} to \ref{tab:EuroS12}),  particularly at shorter horizons.  Similar observations are made for the US application with monthly data (see Section \ref{app:md}).  Another common finding is that HNN yields a remarkable performance for real activity.  Point and density forecasts for industrial production rank either first or very close to the best performing model for all forecast horizons as well as both evaluation samples (i.e., including or excluding post-2020 periods).  Also, HNN explains a high share of the variation in realized volatility measured by $R^2_{|\varepsilon_t|}$, especially for higher-order forecasts. 

\begin{figure}[h!]
\vspace*{0.75em}
\caption{\textbf{Industrial Production} ($s=6$)}\label{fig:eaindpro}  
\vspace*{-0.35cm}
\begin{center} 
\hspace*{-0.1cm}\includegraphics[width=1.02\textwidth]{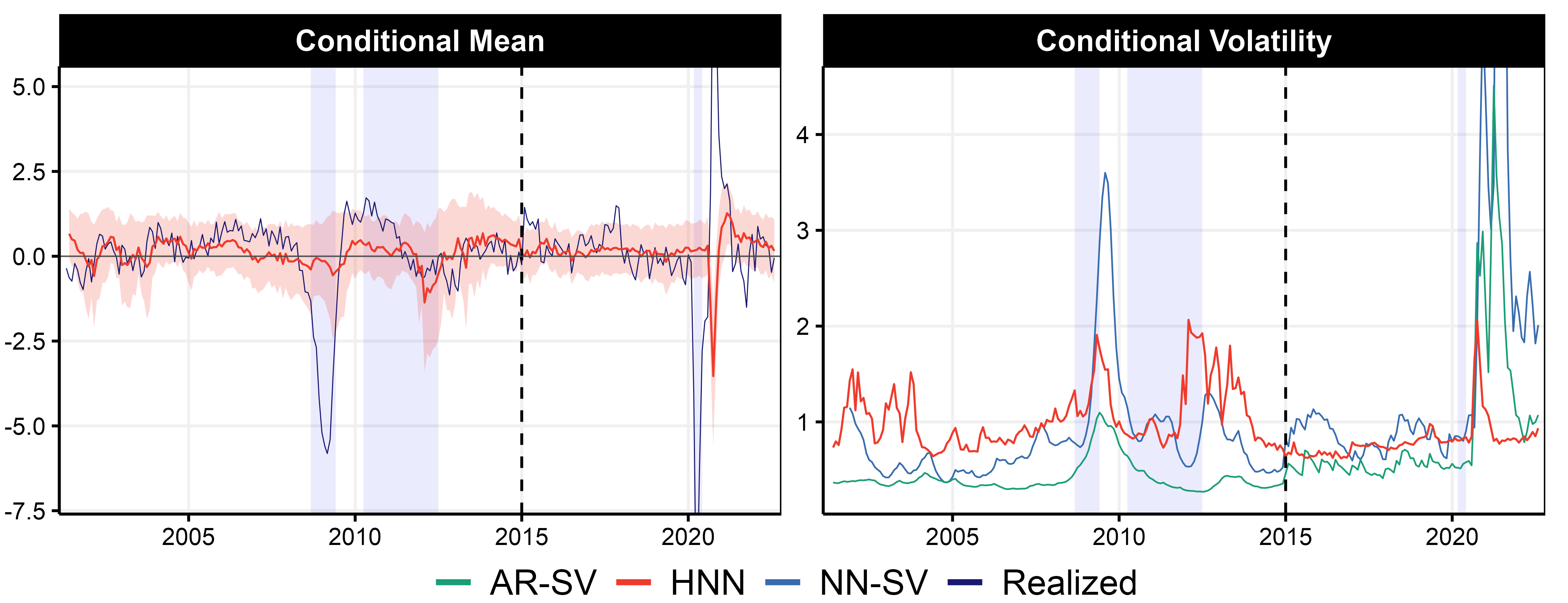}
\end{center}
\end{figure}

Visually inspecting both hemispheres gives some insights into properties of the mean and variance paths and thereby HNN's good performance. As shown in Figure \ref{fig:eaindpro} the volatility paths for HNN and selected benchmarks reflect highly uncertain as well as tranquil times of the European business cycle. Compared to its competitors, HNN marks not only the Great Recession but also the sovereign debt crisis during 2011-2013. Having overcome this long-lasting period of high fragility, the variance hemisphere shows a low and stable path until the economy was hit by the Covid-19 pandemic. While models equipped with SV show elevated uncertainty for post-2020 periods, HNN estimates a lower volatility path, which pays off as per log scores including this period in, e.g.,  Table \ref{tab:EuroS6}. 

In line with previous findings, our nonlinear competitors tend to estimate low volatility leading to inferior density forecasts. We see this pattern for multiple steps ahead forecasts  of industrial production and even more strikingly for unemployment.  BART's point forecasting performance is sometimes remarkable,  in line with the traditional wisdom on tree ensembles and small samples \citep{grinsztajn2022tree}.   However, it shows rather poor performance for density predictions---by constantly underestimating volatility as per the phenomenon described in Section \ref{sec:deepmarde}.   HNN, on the other hand, yields slightly less gains but remains competitive to AR$_\text{\halftiny SV}$ for both evaluation metrics.

\begin{figure}[h!]
\vspace*{0.75em}
\caption{\textbf{Stock Market} ($s=1$)}\label{fig:eaeurostoxx}  
\vspace*{-0.35cm}
\begin{center} 
\hspace*{-0.1cm}\includegraphics[width=1.02\textwidth]{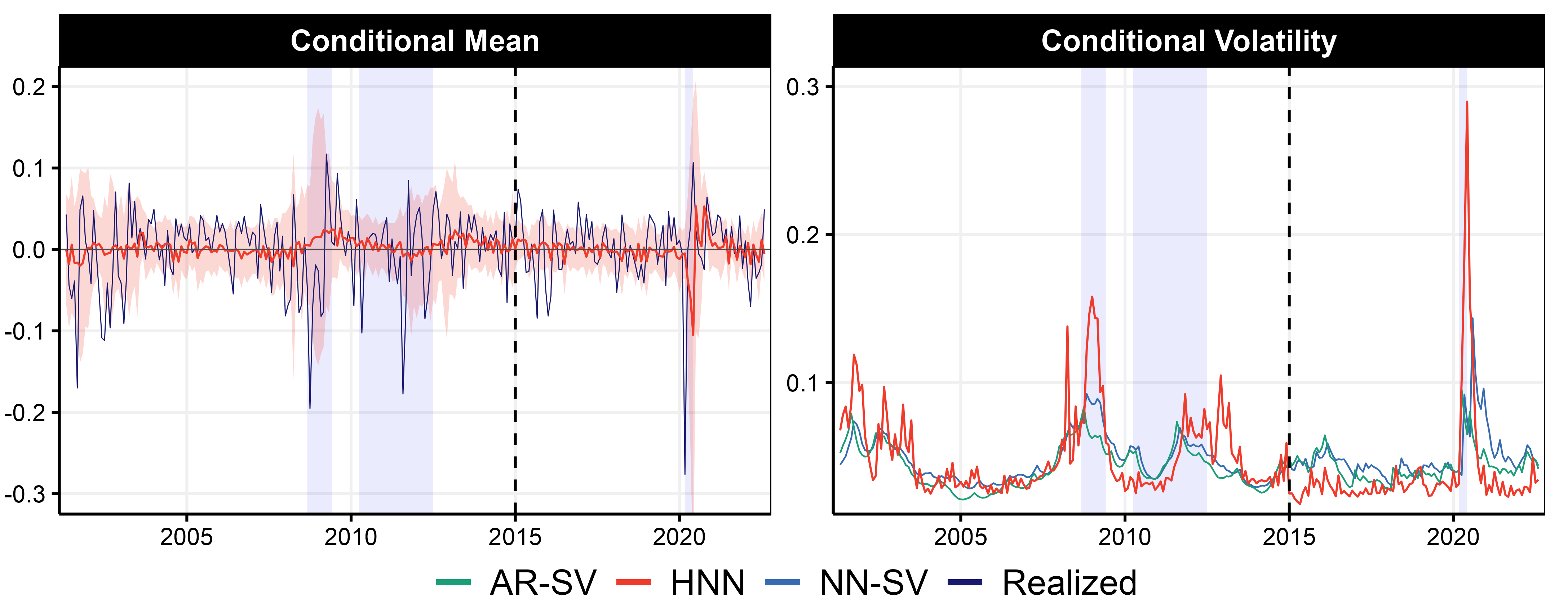}
\end{center}
\end{figure}

For the stock market index, we get good point forecasting performance of AR with time-varying volatility and BART, closely followed by HNN. For density predictions, HNN beats BART in all cases and ranks close to the AR process.  Figure \ref{fig:eaeurostoxx} reveals that the variance hemisphere proactively estimates high volatility during the Great Recession and the Covid-19 pandemic. It peaks before SV-based benchmarks and levels off rather quickly in the following periods. HNN is the only model estimating heightened uncertainty for the full duration of the sovereign debt crisis. We see variance decreasing in 2013 when financial markets gained back trust after Mario Draghi's declaration to do "whatever it takes" in order to save the euro \citep{ecb2012draghi}. The following years are characterized by stability, well captured by HNN's variance hemisphere. For the Covid-19 period, HNN shows a timely and severe peak of uncertainty already calming down in late 2020.

Similar to US inflation predictions, HNN in its unrestricted form has difficulties beating the AR model. Especially when it comes to capturing the 2021-2022 surge, neural network models suffer from rather large prediction errors. Density forecasts remain competitive showing the adaptability of HNN in terms of uncertainty and responsiveness to its own failures.  Unsurprisingly,  we also see the other nonlinear specifications struggling with the post-Covid inflation path.  Point forecasts of linear models show highest accuracy across all horizons when considering the full sample. For density forecasts we find that BART and NN$_\text{\halftiny SV}$ (similar to HNN) yield competitive results.  Even though BART and NN with time-varying volatility perform well, AR$_\text{\halftiny SV}$ often remains the best performing model.

\begin{table}[h!]
\vspace{-1.8em}
\begin{threeparttable}

\center
\scriptsize
\footnotesize
\setlength{\tabcolsep}{0.225em}
 \setlength\extrarowheight{2.5pt}
 \caption{\normalsize {Forecast Performance on Euro Area Data ($s=1$)} \vspace*{-0.3cm}} \label{tab:EuroS1}
 \begin{tabular}{l rrrrrrrrrrrrrrr | rrrrrrrrrrrrrrr} 
\toprule \toprule
\addlinespace[2pt]
& & \multicolumn{14}{c}{2015M1 - 2019M12} & &  \multicolumn{13}{c}{2015M1 - 2022M8} \\
\cmidrule(lr){3-15} \cmidrule(lr){18-30} \addlinespace[2pt]
& &  HNN &  & NN$_\text{\halftiny SV}$ &  & NN$_\text{\halftiny G}$ &  & {\notsotiny DeepAR} & & BART && AR$_\text{\halftiny TV}$  &  & BLR & & &
 HNN &  & NN$_\text{\halftiny SV}$ &  & NN$_\text{\halftiny G}$ &  & {\notsotiny DeepAR} & & BART && AR$_\text{\halftiny TV}$  &  & BLR   &\\
\midrule
\addlinespace[5pt] 
\rowcolor{gray!15} 
\multicolumn{30}{l}{\textbf{Stock Market}} &\cellcolor{gray!15} \\ \addlinespace[2pt] 

{\notsotiny RMSE} &   & 1.05 &   & 1.11 &   & 1.11 &   & 1.12 &   & 1.09 &   & \textbf{1.02} &   & 1.11 &   &   & 1.07 &   & 1.31 &   & 1.31 &   & 0.99 &   & 1.03 &   & \textbf{0.99} &   & 1.16 &   \\
$\mathcal{L}$ &   & -1.95 &   & -1.88 &   & -1.89 &   & -1.76 &   & -1.91 &   & \textbf{-1.99} &   & -1.84 &   &   & -1.53 &   & -1.45 &   & -1.38 &   & -1.61 &   & -1.23 &   & \textbf{-1.67} &   & -1.50 &   \\
$R^2_{|\varepsilon_t|}$ &   & \textbf{0.56} &   & 0.28 &   & 0.36 &   & 0.05 &   & -0.05 &   & 0.35 &   & -0.81 &   &   & \textbf{0.10} &   & 0.03 &   & -0.08 &   & 0.00 &   & -0.17 &   & 0.06 &   & -0.13 &   \\

\addlinespace[5pt] 
\rowcolor{gray!15} 
\multicolumn{30}{l}{\textbf{Industrial Production}} &\cellcolor{gray!15} \\ \addlinespace[2pt] 

{\notsotiny RMSE} &   & \textbf{0.91} &   & 0.92 &   & 0.92 &   & 0.91 &   & 0.96 &   & 0.95 &   & 0.97 &   &   & 0.95 &   & 1.03 &   & 1.03 &   & 0.95 &   & 1.08 &   & \textbf{0.93} &   & 0.99 &   \\
$\mathcal{L}$ &   & \textbf{-3.24} &   & -3.23 &   & -3.23 &   & -2.44 &   & -3.04 &   & -3.17 &   & -3.19 &   &   & -2.94 &   & -3.01 &   & -2.99 &   & -2.29 &   & -2.80 &   & \textbf{-3.04} &   & -3.02 &   \\
$R^2_{|\varepsilon_t|}$ &   & 0.08 &   & 0.01 &   & 0.03 &   & -0.11 &   & 0.06 &   & \textbf{0.28} &   & -0.07 &   &   & 0.04 &   & -0.28 &   & 0.14 &   & -0.02 &   & -0.69 &   & \textbf{0.29} &   & -0.08 &   \\

\addlinespace[5pt] 
\rowcolor{gray!15} 
\multicolumn{30}{l}{\textbf{Unemployment}} &\cellcolor{gray!15} \\ \addlinespace[2pt] 

{\notsotiny RMSE} &   & 0.99 &   & 0.99 &   & 0.99 &   & 0.94 &   & 0.94 &   & 0.99 &   & \textbf{0.92} &   &   & 1.04 &   & 1.06 &   & 1.06 &   & 1.03 &   & 0.98 &   & 1.00 &   & \textbf{0.95} &   \\
$\mathcal{L}$ &   & -1.34 &   & -1.35 &   & -1.30 &   & -1.15 &   & -0.99 &   & \textbf{-1.38} &   & -1.34 &   &   & -1.26 &   & -1.28 &   & -1.25 &   & 0.44 &   & -1.02 &   & \textbf{-1.37} &   & -1.30 &   \\
$R^2_{|\varepsilon_t|}$ &   & 0.07 &   & -0.31 &   & -0.25 &   & -0.29 &   & 0.17 &   & \textbf{0.30} &   & -1.39 &   &   & 0.03 &   & -0.37 &   & -0.44 &   & -0.41 &   & 0.04 &   & \textbf{0.27} &   & -1.15 &   \\

\addlinespace[5pt] 
\rowcolor{gray!15} 
\multicolumn{30}{l}{\textbf{Inflation}} &\cellcolor{gray!15} \\ \addlinespace[2pt] 

{\notsotiny RMSE} &   & 1.05 &   & {0.99} &   & {0.99} &   & 1.16 &   & \textbf{0.97} &   & 1.00 &   & 1.01 &   &   & 1.15 &   & 1.16 &   & 1.16 &   & 1.18 &   & 1.06 &   & \textbf{1.00} &   & 1.01 &   \\
$\mathcal{L}$ &   & -5.07 &   & -5.12 &   & \textbf{-5.14} &   & -4.53 &   & -5.13 &   & -5.13 &   & -5.12 &   &   & -3.33 &   & \textbf{-4.59} &   & -4.51 &   & 1.73 &   & -4.13 &   & -4.46 &   & -4.21 &   \\
$R^2_{|\varepsilon_t|}$ &   & -0.05 &   & -0.09 &   & 0.00 &   & -0.09 &   & 0.15 &   & \textbf{0.50} &   & -0.11 &   &   & 0.00 &   & 0.28 &   & 0.20 &   & -0.09 &   & -0.90 &   & \textbf{0.34} &   & 0.12 &   \\

\bottomrule \bottomrule
\end{tabular}
\begin{tablenotes}[para,flushleft]
  \scriptsize 
    \textit{Notes}: The table presents our forecasting evaluation metrics including the root mean square error (RMSE) relative to the AR model with constant variance, the log score ($\mathcal{L}$), and the $R^2_{|\varepsilon_t|}$ of absolute residuals. For the real activity variables (i.e., Industrial Production and Unemployment) we exclude the year 2020 from the evaluation sample. AR$_\text{\halftiny TV}$ refers to the best-performing AR specifications, which is either AR$_\text{\halftiny SV}$ or AR$_\text{\halftiny G}$.
  \end{tablenotes}
\end{threeparttable}
\end{table}

\begin{table}[h!]
\vspace{-1.8em}
\begin{threeparttable}

\center
\scriptsize
\footnotesize
\setlength{\tabcolsep}{0.225em}
 \setlength\extrarowheight{2.5pt}
 \caption{\normalsize {Forecast Performance on Euro Area Data ($s=3$)} \vspace*{-0.3cm}} \label{tab:EuroS3}
 \begin{tabular}{l rrrrrrrrrrrrrrr | rrrrrrrrrrrrrrr} 
\toprule \toprule
\addlinespace[2pt]
& & \multicolumn{14}{c}{2015M1 - 2019M12} & &  \multicolumn{13}{c}{2015M1 - 2022M8} \\
\cmidrule(lr){3-15} \cmidrule(lr){18-30} \addlinespace[2pt]
& &  HNN &  & NN$_\text{\halftiny SV}$ &  & NN$_\text{\halftiny G}$ &  & {\notsotiny DeepAR} & & BART && AR$_\text{\halftiny TV}$  &  & BLR & & &
 HNN &  & NN$_\text{\halftiny SV}$ &  & NN$_\text{\halftiny G}$ &  & {\notsotiny DeepAR} & & BART && AR$_\text{\halftiny TV}$  &  & BLR   &\\
\midrule
\addlinespace[5pt] 
\rowcolor{gray!15} 
\multicolumn{30}{l}{\textbf{Stock Market}} &\cellcolor{gray!15} \\ \addlinespace[2pt] 

{\notsotiny RMSE} &   & 1.00 &   & 1.07 &   & 1.07 &   & 1.19 &   & \textbf{0.94} &   & 1.02 &   & 1.18 &   &   & 1.02 &   & 1.19 &   & 1.19 &   & 1.03 &   & \textbf{0.92} &   & 1.01 &   & 1.22 &   \\
$\mathcal{L}$ &   & -2.35 &   & -2.27 &   & -2.26 &   & -1.78 &   & \textbf{-2.40} &   & -2.37 &   & -2.19 &   &   & -2.14 &   & -1.96 &   & -2.05 &   & -1.76 &   & \textbf{-2.21} &   & -2.11 &   & -1.92 &   \\
$R^2_{|\varepsilon_t|}$ &   & 0.51 &   & 0.23 &   & 0.24 &   & -0.19 &   & -0.29 &   & \textbf{0.55} &   & -0.51 &   &   & 0.20 &   & -0.05 &   & 0.23 &   & -0.14 &   & -0.14 &   & \textbf{0.47} &   & -0.05 &   \\

\addlinespace[5pt] 
\rowcolor{gray!15} 
\multicolumn{30}{l}{\textbf{Industrial Production}} &\cellcolor{gray!15} \\ \addlinespace[2pt] 

{\notsotiny RMSE} &   & 0.86 &   & 0.90 &   & 0.90 &   & \textbf{0.85} &   & 1.07 &   & 0.87 &   & 1.00 &   &   & 0.82 &   & 0.87 &   & 0.87 &   & 0.81 &   & 1.02 &   & \textbf{0.80} &   & 0.98 &   \\
$\mathcal{L}$ &   & -4.02 &   & -3.94 &   & -3.98 &   & -1.95 &   & -1.91 &   & \textbf{-4.03} &   & -3.81 &   &   & -3.86 &   & -3.74 &   & -3.79 &   & -1.54 &   & -2.11 &   & \textbf{-3.91} &   & -3.63 &   \\
$R^2_{|\varepsilon_t|}$ &   & 0.45 &   & 0.14 &   & 0.38 &   & 0.10 &   & -3.95 &   & \textbf{0.52} &   & -0.72 &   &   & 0.31 &   & -2.29 &   & 0.24 &   & -0.08 &   & -4.10 &   & \textbf{0.44} &   & -1.26 &   \\

\addlinespace[5pt] 
\rowcolor{gray!15} 
\multicolumn{30}{l}{\textbf{Unemployment}} &\cellcolor{gray!15} \\ \addlinespace[2pt] 

{\notsotiny RMSE} &   & 0.98 &   & 0.98 &   & 0.98 &   & 0.94 &   & 0.90 &   & 0.97 &   & \textbf{0.85} &   &   & 1.00 &   & 1.00 &   & 1.00 &   & 1.07 &   & 0.93 &   & 0.99 &   & \textbf{0.92} &   \\
$\mathcal{L}$ &   & \textbf{-1.64} &   & -1.60 &   & -1.57 &   & -1.63 &   & 4.50 &   & -1.61 &   & -1.51 &   &   & -1.52 &   & -1.51 &   & -1.47 &   & -0.35 &   & 3.02 &   & \textbf{-1.55} &   & -1.46 &   \\
$R^2_{|\varepsilon_t|}$ &   & 0.29 &   & -0.06 &   & -0.08 &   & -0.51 &   & 0.19 &   & \textbf{0.56} &   & -3.40 &   &   & 0.16 &   & -0.43 &   & -0.33 &   & -0.41 &   & -0.07 &   & \textbf{0.52} &   & -1.91 &   \\

\addlinespace[5pt] 
\rowcolor{gray!15} 
\multicolumn{30}{l}{\textbf{Inflation}} &\cellcolor{gray!15} \\ \addlinespace[2pt] 

{\notsotiny RMSE} &   & 1.12 &   & 1.00 &   & 1.00 &   & 1.26 &   & 1.03 &   & 1.05 &   & 1.03 &   &   & 1.21 &   & 1.22 &   & 1.22 &   & 1.28 &   & 1.05 &   & \textbf{0.95} &   & 0.95 &   \\
$\mathcal{L}$ &   & -5.30 &   & -5.36 &   & -5.40 &   & -4.37 &   & \textbf{-5.47} &   & -5.34 &   & -5.40 &   &   & -2.23 &   & -4.70 &   & -4.84 &   & 17.10 &   & \textbf{-5.04} &   & -4.53 &   & -4.65 &   \\
$R^2_{|\varepsilon_t|}$ &   & 0.03 &   & -0.38 &   & -0.18 &   & -0.22 &   & -0.14 &   & \textbf{0.63} &   & -0.16 &   &   & -0.18 &   & 0.26 &   & 0.39 &   & -0.08 &   & -0.38 &   & \textbf{0.45} &   & 0.18 &   \\

\bottomrule \bottomrule
\end{tabular}
\begin{tablenotes}[para,flushleft]
  \scriptsize 
    \textit{Notes}: The table presents our forecasting evaluation metrics including the root mean square error (RMSE) relative to the AR model with constant variance, the log score ($\mathcal{L}$), and the $R^2_{|\varepsilon_t|}$ of absolute residuals. For the real activity variables (i.e., Industrial Production and Unemployment) we exclude the year 2020 from the evaluation sample. AR$_\text{\halftiny TV}$ refers to the best-performing AR specifications, which is either AR$_\text{\halftiny SV}$ or AR$_\text{\halftiny G}$.
  \end{tablenotes}
\end{threeparttable}
\end{table}

\begin{table}[h!]
\vspace{-1.8em}
\begin{threeparttable}

\center
\scriptsize
\footnotesize
\setlength{\tabcolsep}{0.225em}
 \setlength\extrarowheight{2.5pt}
 \caption{\normalsize {Forecast Performance on Euro Area Data ($s=6$)} \vspace*{-0.3cm}} \label{tab:EuroS6}
 \begin{tabular}{l rrrrrrrrrrrrrrr | rrrrrrrrrrrrrrr} 
\toprule \toprule
\addlinespace[2pt]
& & \multicolumn{14}{c}{2015M1 - 2019M12} & &  \multicolumn{13}{c}{2015M1 - 2022M8} \\
\cmidrule(lr){3-15} \cmidrule(lr){18-30} \addlinespace[2pt]
& &  HNN &  & NN$_\text{\halftiny SV}$ &  & NN$_\text{\halftiny G}$ &  & {\notsotiny DeepAR} & & BART && AR$_\text{\halftiny TV}$  &  & BLR & & &
 HNN &  & NN$_\text{\halftiny SV}$ &  & NN$_\text{\halftiny G}$ &  & {\notsotiny DeepAR} & & BART && AR$_\text{\halftiny TV}$  &  & BLR   &\\
\midrule
\addlinespace[5pt] 
\rowcolor{gray!15} 
\multicolumn{30}{l}{\textbf{Stock Market}} &\cellcolor{gray!15} \\ \addlinespace[2pt] 

{\notsotiny RMSE} &   & 1.08 &   & 1.16 &   & 1.16 &   & 1.61 &   & 1.01 &   & \textbf{0.97} &   & 1.27 &   &   & 1.00 &   & 1.12 &   & 1.12 &   & 1.26 &   & \textbf{0.90} &   & 0.97 &   & 1.22 &   \\
$\mathcal{L}$ &   & -2.61 &   & -2.51 &   & -2.54 &   & -1.79 &   & -1.87 &   & \textbf{-2.72} &   & -2.47 &   &   & -2.54 &   & -2.40 &   & -2.45 &   & -1.83 &   & -2.00 &   & \textbf{-2.58} &   & -2.37 &   \\
$R^2_{|\varepsilon_t|}$ &   & 0.42 &   & 0.10 &   & 0.31 &   & 0.04 &   & 0.06 &   & \textbf{0.53} &   & -0.58 &   &   & 0.30 &   & -0.06 &   & 0.28 &   & -0.13 &   & -0.59 &   & \textbf{0.54} &   & -0.19 &   \\

\addlinespace[5pt] 
\rowcolor{gray!15} 
\multicolumn{30}{l}{\textbf{Industrial Production}} &\cellcolor{gray!15} \\ \addlinespace[2pt] 

{\notsotiny RMSE} &   & \textbf{0.80} &   & 0.81 &   & 0.81 &   & 0.84 &   & 0.87 &   & 0.94 &   & 0.94 &   &   & \textbf{0.67} &   & 0.69 &   & 0.69 &   & 0.78 &   & 0.78 &   & 0.95 &   & 1.12 &   \\
$\mathcal{L}$ &   & \textbf{-4.41} &   & -4.31 &   & -4.37 &   & -3.44 &   & >10 &   & -4.31 &   & -4.06 &   &   & \textbf{-4.31} &   & -4.04 &   & -4.25 &   & 0.55 &   & >10 &   & -4.12 &   & -3.89 &   \\
$R^2_{|\varepsilon_t|}$ &   & 0.65 &   & 0.43 &   & 0.59 &   & 0.17 &   & -0.84 &   & \textbf{0.70} &   & -2.70 &   &   & 0.59 &   & -4.96 &   & 0.34 &   & -0.31 &   & <-10 &   & \textbf{0.71} &   & -0.39 &   \\

\addlinespace[5pt] 
\rowcolor{gray!15} 
\multicolumn{30}{l}{\textbf{Unemployment}} &\cellcolor{gray!15} \\ \addlinespace[2pt] 

{\notsotiny RMSE} &   & 0.88 &   & 0.89 &   & 0.89 &   & 1.09 &   & 0.70 &   & 0.76 &   & \textbf{0.70} &   &   & 0.72 &   & 0.74 &   & 0.74 &   & 0.83 &   & \textbf{0.60} &   & 0.99 &   & 0.69 &   \\
$\mathcal{L}$ &   & -1.69 &   & -1.58 &   & -1.62 &   & 3.80 &   & >10 &   & \textbf{-1.80} &   & -1.59 &   &   & \textbf{-1.56} &   & -1.48 &   & -1.48 &   & 3.29 &   & >10 &   & -1.38 &   & -1.49 &   \\
$R^2_{|\varepsilon_t|}$ &   & 0.32 &   & -0.41 &   & -0.02 &   & -1.08 &   & -0.75 &   & \textbf{0.61} &   & -4.44 &   &   & 0.20 &   & -0.79 &   & -0.49 &   & -0.83 &   & -0.48 &   & \textbf{0.46} &   & -1.28 &   \\

\addlinespace[5pt] 
\rowcolor{gray!15} 
\multicolumn{30}{l}{\textbf{Inflation}} &\cellcolor{gray!15} \\ \addlinespace[2pt] 

{\notsotiny RMSE} &   & 1.13 &   & 1.02 &   & 1.02 &   & 1.39 &   & 1.05 &   & \textbf{0.89} &   & 0.98 &   &   & 1.12 &   & 1.14 &   & 1.14 &   & 1.11 &   & 1.04 &   & 0.94 &   &\textbf{0.89} &   \\
$\mathcal{L}$ &   & -4.69 &   & -5.46 &   & -5.46 &   & -3.21 &   & -5.44 &   & \textbf{-5.66} &   & -5.58 &   &   & -2.69 &   & -4.59 &   & -4.21 &   & >10 &   & \textbf{-5.15} &   & -4.44 &   & -4.79 &   \\
$R^2_{|\varepsilon_t|}$ &   & -0.46 &   & -0.71 &   & -0.39 &   & -0.25 &   & -1.58 &   & \textbf{0.65} &   & -0.12 &   &   & -0.11 &   & 0.12 &   & 0.13 &   & -0.04 &   & -0.57 &   & \textbf{0.43} &   & 0.27 &   \\

\bottomrule \bottomrule
\end{tabular}
\begin{tablenotes}[para,flushleft]
  \scriptsize 
    \textit{Notes}: The table presents our forecasting evaluation metrics including the root mean square error (RMSE) relative to the AR model with constant variance, the log score ($\mathcal{L}$), and the $R^2_{|\varepsilon_t|}$ of absolute residuals. For the real activity variables (i.e., Industrial Production and Unemployment) we exclude the year 2020 from the evaluation sample. AR$_\text{\halftiny TV}$ refers to the best-performing AR specifications, which is either AR$_\text{\halftiny SV}$ or AR$_\text{\halftiny G}$.
  \end{tablenotes}
\end{threeparttable}
\end{table}

\begin{table}[h!]
\vspace{-1.8em}
\begin{threeparttable}

\center
\scriptsize
\footnotesize
\setlength{\tabcolsep}{0.225em}
 \setlength\extrarowheight{2.5pt}
 \caption{\normalsize {Forecast Performance on Euro Area Data ($s=12$)} \vspace*{-0.3cm}} \label{tab:EuroS12}
 \begin{tabular}{l rrrrrrrrrrrrrrr | rrrrrrrrrrrrrrr} 
\toprule \toprule
\addlinespace[2pt]
& & \multicolumn{14}{c}{2015M1 - 2019M12} & &  \multicolumn{13}{c}{2015M1 - 2022M8} \\
\cmidrule(lr){3-15} \cmidrule(lr){18-30} \addlinespace[2pt]
& &  HNN &  & NN$_\text{\halftiny SV}$ &  & NN$_\text{\halftiny G}$ &  & {\notsotiny DeepAR} & & BART && AR$_\text{\halftiny TV}$  &  & BLR & & &
 HNN &  & NN$_\text{\halftiny SV}$ &  & NN$_\text{\halftiny G}$ &  & {\notsotiny DeepAR} & & BART && AR$_\text{\halftiny TV}$  &  & BLR   &\\
\midrule
\addlinespace[5pt] 
\rowcolor{gray!15} 
\multicolumn{30}{l}{\textbf{Stock Market}} &\cellcolor{gray!15} \\ \addlinespace[2pt] 

{\notsotiny RMSE} &   & 1.23 &   & 1.33 &   & 1.33 &   & 1.76 &   & \textbf{0.96} &   & 1.02 &   & 1.23 &   &   & 0.99 &   & 1.06 &   & 1.06 &   & 1.47 &   & \textbf{0.81} &   & 1.04 &   & 1.13 &   \\
$\mathcal{L}$ &   & -2.82 &   & -2.72 &   & -2.77 &   & -2.61 &   & -1.53 &   & \textbf{-2.97} &   & -2.83 &   &   & -2.84 &   & -2.81 &   & -2.83 &   & -2.51 &   & -1.75 &   & \textbf{-2.88} &   & -2.79 &   \\
$R^2_{|\varepsilon_t|}$ &   & 0.27 &   & -0.58 &   & -0.06 &   & 0.49 &   & -0.83 &   & \textbf{0.63} &   & -0.82 &   &   & 0.16 &   & -0.26 &   & -0.01 &   & 0.29 &   & -0.78 &   & \textbf{0.66} &   & -0.50 &   \\

\addlinespace[5pt] 
\rowcolor{gray!15} 
\multicolumn{30}{l}{\textbf{Industrial Production}} &\cellcolor{gray!15} \\ \addlinespace[2pt] 

{\notsotiny RMSE} &   & 0.79 &   & 0.77 &   & 0.77 &   & \textbf{0.67} &   & 1.00 &   & 0.76 &   & 0.93 &   &   & 0.89 &   & 0.89 &   & 0.89 &   & \textbf{0.77} &   & 0.78 &   & 0.91 &   & 1.03 &   \\
$\mathcal{L}$ &   & -4.81 &   & -4.63 &   & -4.81 &   & \textbf{-4.86} &   & >10 &   & -4.80 &   & -4.37 &   &   & \textbf{-4.60} &   & -4.16 &   & -3.88 &   & -2.42 &   & >10 &   & -4.46 &   & -3.85 &   \\
$R^2_{|\varepsilon_t|}$ &   & \textbf{0.79} &   & 0.50 &   & 0.77 &   & -0.07 &   & 0.51 &   & 0.74 &   & -4.30 &   &   & \textbf{0.59} &   & -0.67 &   & 0.25 &   & 0.05 &   & -2.38 &   & 0.39 &   & -0.04 &   \\

\addlinespace[5pt] 
\rowcolor{gray!15} 
\multicolumn{30}{l}{\textbf{Unemployment}} &\cellcolor{gray!15} \\ \addlinespace[2pt] 

{\notsotiny RMSE}  &   & 0.87 &   & 0.87 &   & 0.87 &   & 1.00 &   & \textbf{0.45} &   & 0.51 &   & 0.62 &   &   & 0.78 &   & 0.76 &   & 0.76 &   & 0.82 &   & \textbf{0.55} &   & 0.86 &   & 0.74 &   \\
$\mathcal{L}$ &   & -1.44 &   & -1.44 &   & -1.51 &   & 1.60 &   & 8.73 &   & \textbf{-2.01} &   & -1.62 &   &   & -1.26 &   & -1.16 &   & -1.28 &   & 3.92 &   & 6.33 &   & -1.23 &   & \textbf{-1.33} &   \\
$R^2_{|\varepsilon_t|}$ &   & -0.07 &   & -0.83 &   & -0.07 &   & -0.09 &   & -0.56 &   & \textbf{0.57} &   & -2.58 &   &   & 0.24 &   & -0.72 &   & -0.08 &   & -0.26 &   & -0.65 &   & \textbf{0.48} &   & -0.09 &   \\

\addlinespace[5pt] 
\rowcolor{gray!15} 
\multicolumn{30}{l}{\textbf{Inflation}} &\cellcolor{gray!15} \\ \addlinespace[2pt] 

{\notsotiny RMSE} &   & 0.87 &   & 0.98 &   & 0.98 &   & 1.17 &   & 0.94 &   & \textbf{0.58} &   & 0.79 &   &   & 0.97 &   & 1.02 &   & 1.02 &   & 1.02 &   & 0.91 &   & 0.92 &   & \textbf{0.89} &   \\
$\mathcal{L}$ &   & -4.67 &   & -5.30 &   & -5.34 &   & -0.43 &   & -4.88 &   & \textbf{-5.80} &   & -5.49 &   &   & -4.18 &   & \textbf{-4.81} &   & -3.78 &   & 5.09 &   & -4.73 &   & -3.69 &   & -4.60 &   \\
$R^2_{|\varepsilon_t|}$ &   & -0.63 &   & -0.65 &   & -0.25 &   & -0.18 &   & -1.03 &   & \textbf{0.65} &   & 0.26 &   &   & 0.12 &   & 0.11 &   & -0.04 &   & -0.12 &   & -0.10 &   & 0.34 &   & \textbf{0.35} &   \\

\bottomrule \bottomrule
\end{tabular}
\begin{tablenotes}[para,flushleft]
  \scriptsize 
    \textit{Notes}: The table presents our forecasting evaluation metrics including the root mean square error (RMSE) relative to the AR model with constant variance, the log score ($\mathcal{L}$), and the $R^2_{|\varepsilon_t|}$ of absolute residuals. For the real activity variables (i.e., Industrial Production and Unemployment) we exclude the year 2020 from the evaluation sample. AR$_\text{\halftiny TV}$ refers to the best-performing AR specifications, which is either AR$_\text{\halftiny SV}$ or AR$_\text{\halftiny G}$.
  \end{tablenotes}
\end{threeparttable}
\end{table}

\clearpage

\subsection{What are the hemispheres made of?}\label{app:VI}


To shed light on which variables drive the hemispheres in our network, we conduct a variable importance (VI) exercise similar to \cite{MRF} and \cite{HNN}. The importance of variable $k$ (for $k=1, \dots, K$) for each hemisphere $j$ (i.e., $h_m$ and $h_v$) is determined in three steps. First, variable $k$ and its lags are randomly shuffled. Second, the respective hemisphere is recomputed (but not re-estimated) with the shuffled variable $k$ all else equal. Finally, we compute the deviation of the new estimate with the transformed data ($h_j(\tilde{\bm X}_t;\theta_j)$) to the baseline result ($h_j(\bm X_t;\theta _j)$). 
The standardized $\text{VI}_k^j$,  in terms of \% of increase in MSE, is then given by
\begin{align}
 \text{VI}_k^j= 100 \times \left(\frac{\tfrac{1}{T}\sum_{t=1}^T (h _j(\tilde{\bm X}_t;\theta _j)-h _j(\bm X_t;\theta _j))^2}{Var(h _j(\bm X_t;\theta _j))}  \right).
\end{align}

\begin{figure}[h]
\caption{VI Results for \textbf{GDP} ($s=1$)}\label{fig:vi_gdps1}
  \begin{subfigure}[b]{0.5\textwidth}
  \vspace{1em}
  \caption{\footnotesize  Mean hemisphere ($h_m$)}
\hspace{-0.25cm}\includegraphics[trim={0cm 1.5cm 0cm 0cm},clip,width=0.995\textwidth]{{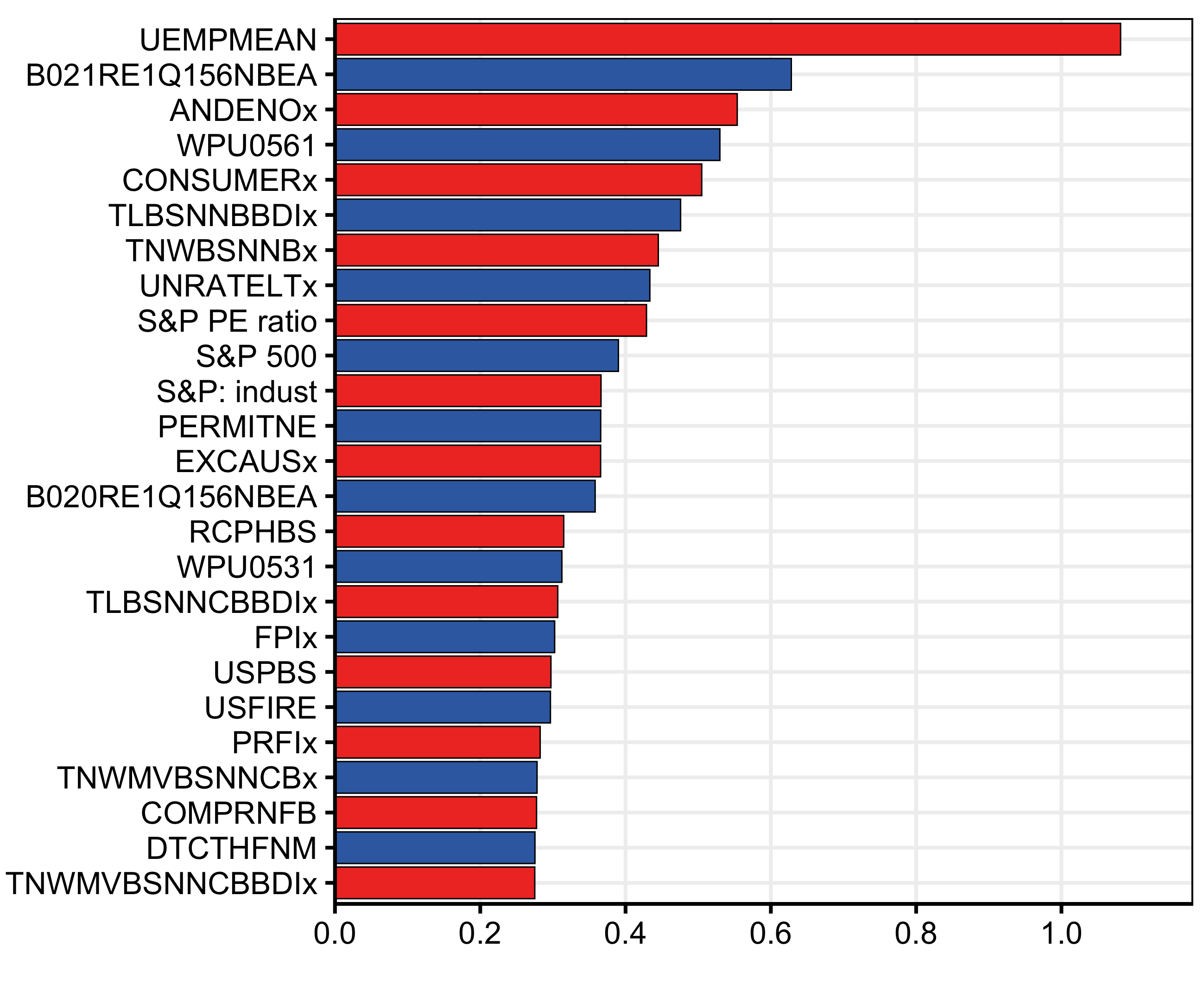}}
      \end{subfigure}
  \begin{subfigure}[b]{0.5\textwidth}
  \caption{\footnotesize Volatility hemisphere ($h_v$)}
\includegraphics[trim={0cm 1.5cm 0cm 0cm},clip,width=0.995\textwidth]{{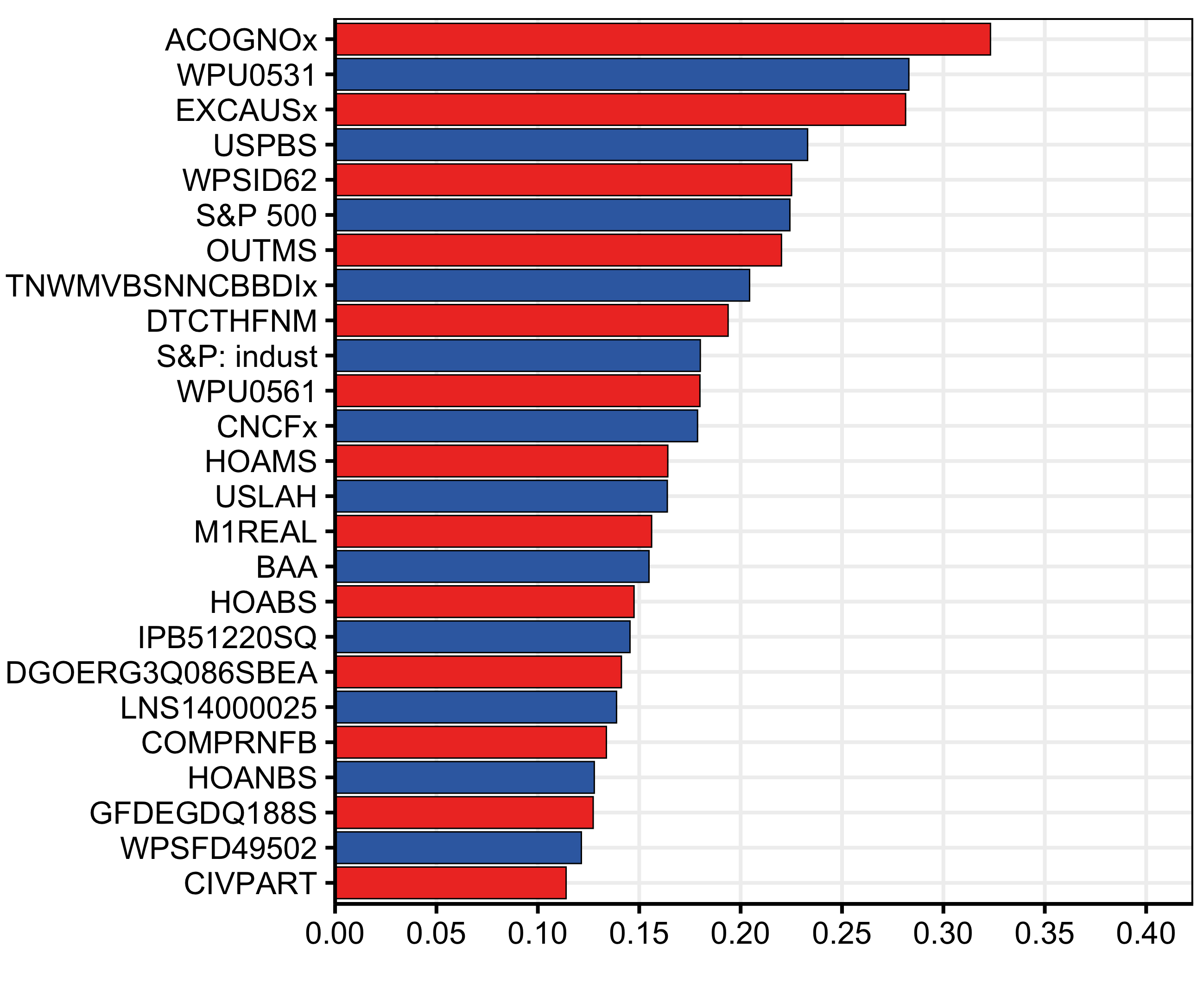}}
      \end{subfigure}
        \vspace{-2em}  

        \caption*{\scriptsize \textit{Notes:} The graph shows VI results for both hemispheres of the HNN with training ending in 2019Q4. The left panel shows the top 25 variables for the mean hemisphere and the right panel refers to the 25 most important drivers of the variance hemisphere. Mnemonics are those of FRED-QD \citep{mccracken2020fred}.}
\end{figure}

\begin{figure}[h]
\caption{VI Results for \textbf{GDP} ($s=4$)}\label{fig:vi_gdps4}
  \begin{subfigure}[b]{0.5\textwidth}
  \vspace{1em}
  \caption{\footnotesize  Mean hemisphere ($h_m$)}
\hspace{-0.25cm}\includegraphics[trim={0cm 1.5cm 0cm 0cm},clip,width=0.995\textwidth]{{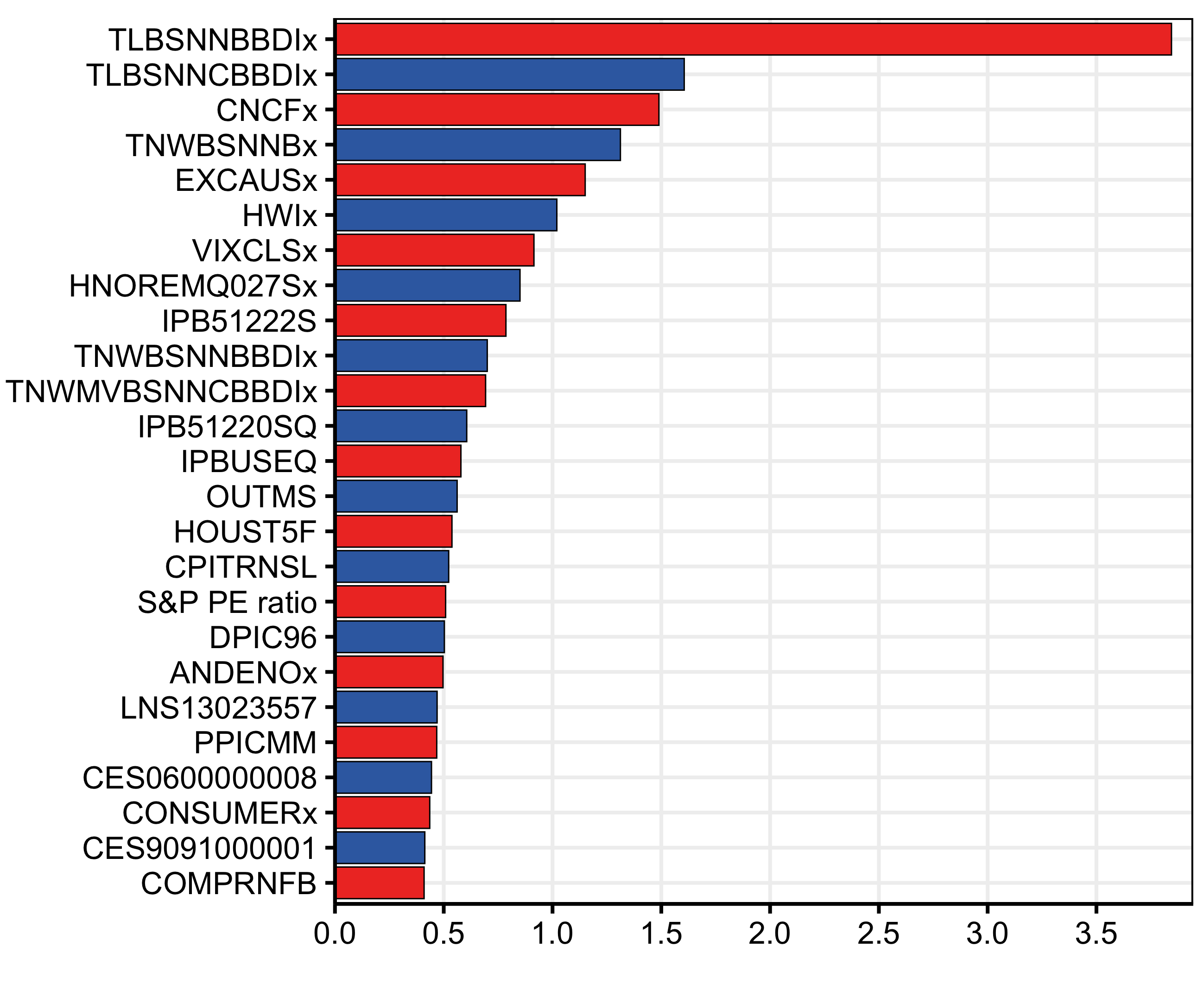}}
      \end{subfigure}
  \begin{subfigure}[b]{0.5\textwidth}
  \caption{\footnotesize Volatility hemisphere ($h_v$)}
\includegraphics[trim={0cm 1.5cm 0cm 0cm},clip,width=0.995\textwidth]{{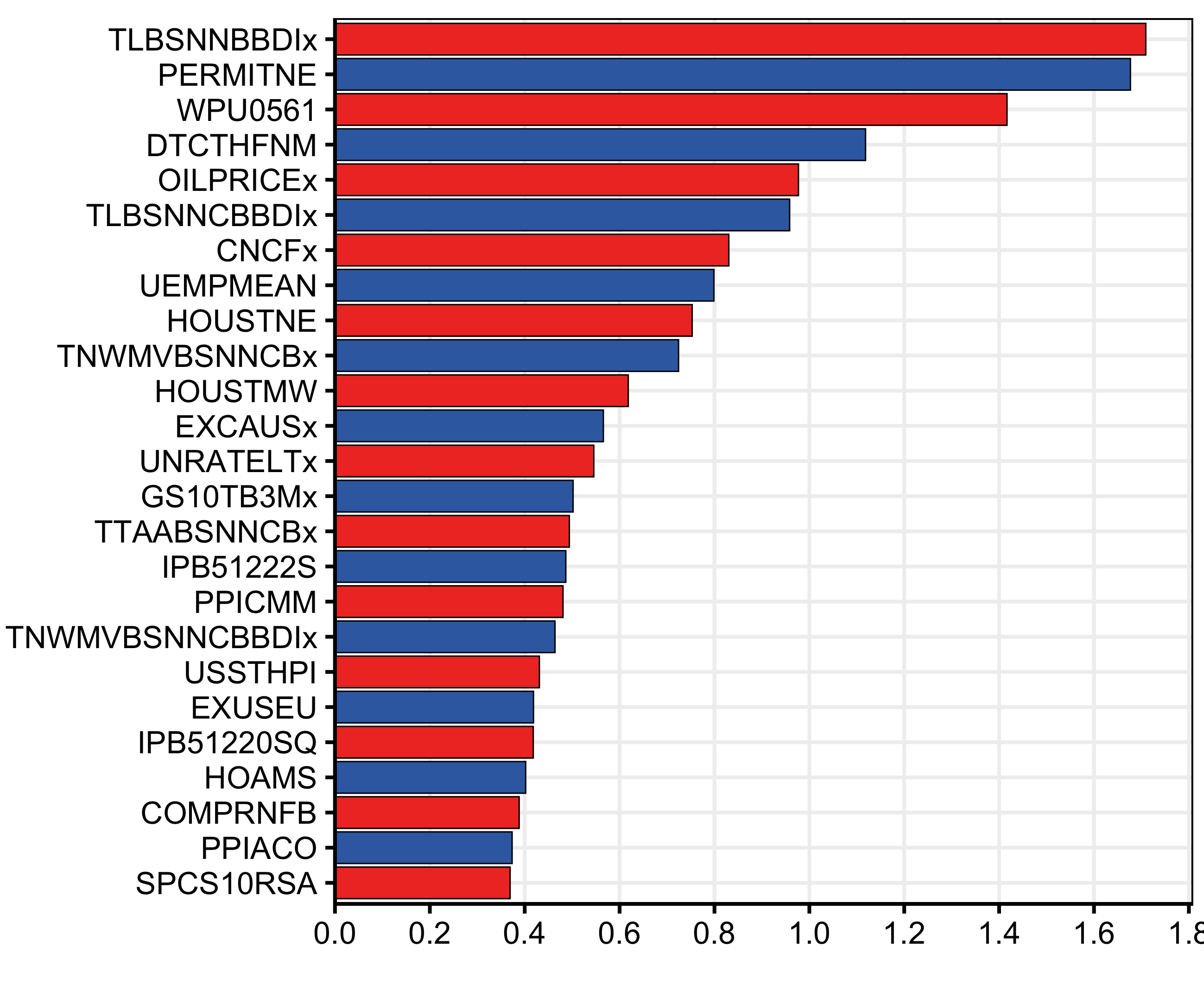}}
      \end{subfigure}
        \vspace{-2em}  

        \caption*{\scriptsize \textit{Notes:} The graph shows VI results for both hemispheres of the HNN with training ending in  2019Q4. The left panel shows the top 25 variables for the mean hemisphere and the right panel refers to the 25 most important drivers of the variance hemisphere. Mnemonics are those of FRED-QD \citep{mccracken2020fred}.}
\end{figure}

\begin{figure}[h]
\caption{VI Results for \textbf{Inflation} ($s=1$)}\label{fig:vi_cpis1}
  \begin{subfigure}[b]{0.5\textwidth}
  \vspace{1em}
  \caption{\footnotesize  Mean hemisphere ($h_m$)}
\hspace{-0.25cm}\includegraphics[trim={0cm 1.5cm 0cm 0cm},clip,width=0.995\textwidth]{{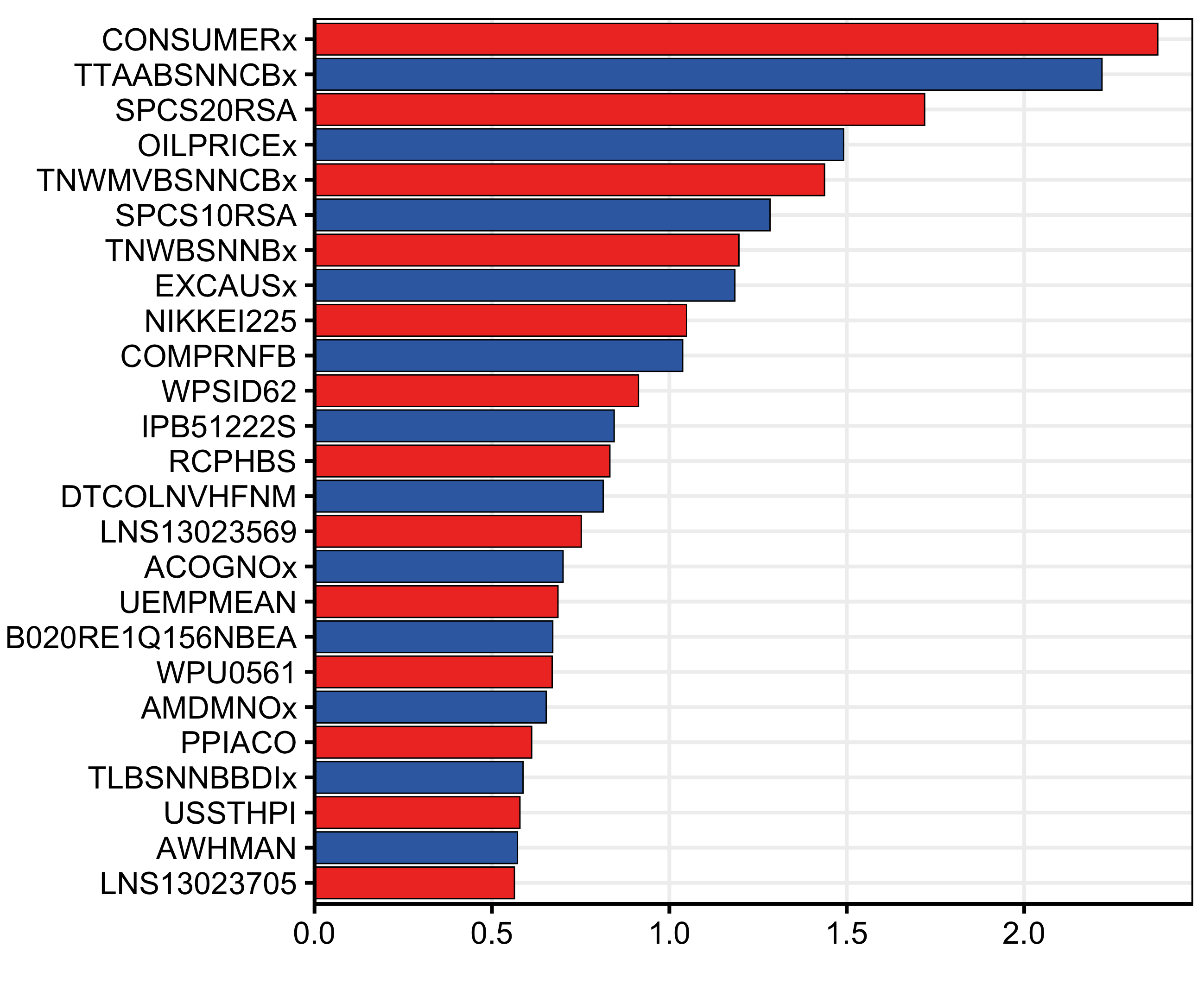}}
      \end{subfigure}
  \begin{subfigure}[b]{0.5\textwidth}
  \caption{\footnotesize Volatility hemisphere ($h_v$)}
\includegraphics[trim={0cm 1.5cm 0cm 0cm},clip,width=0.995\textwidth]{{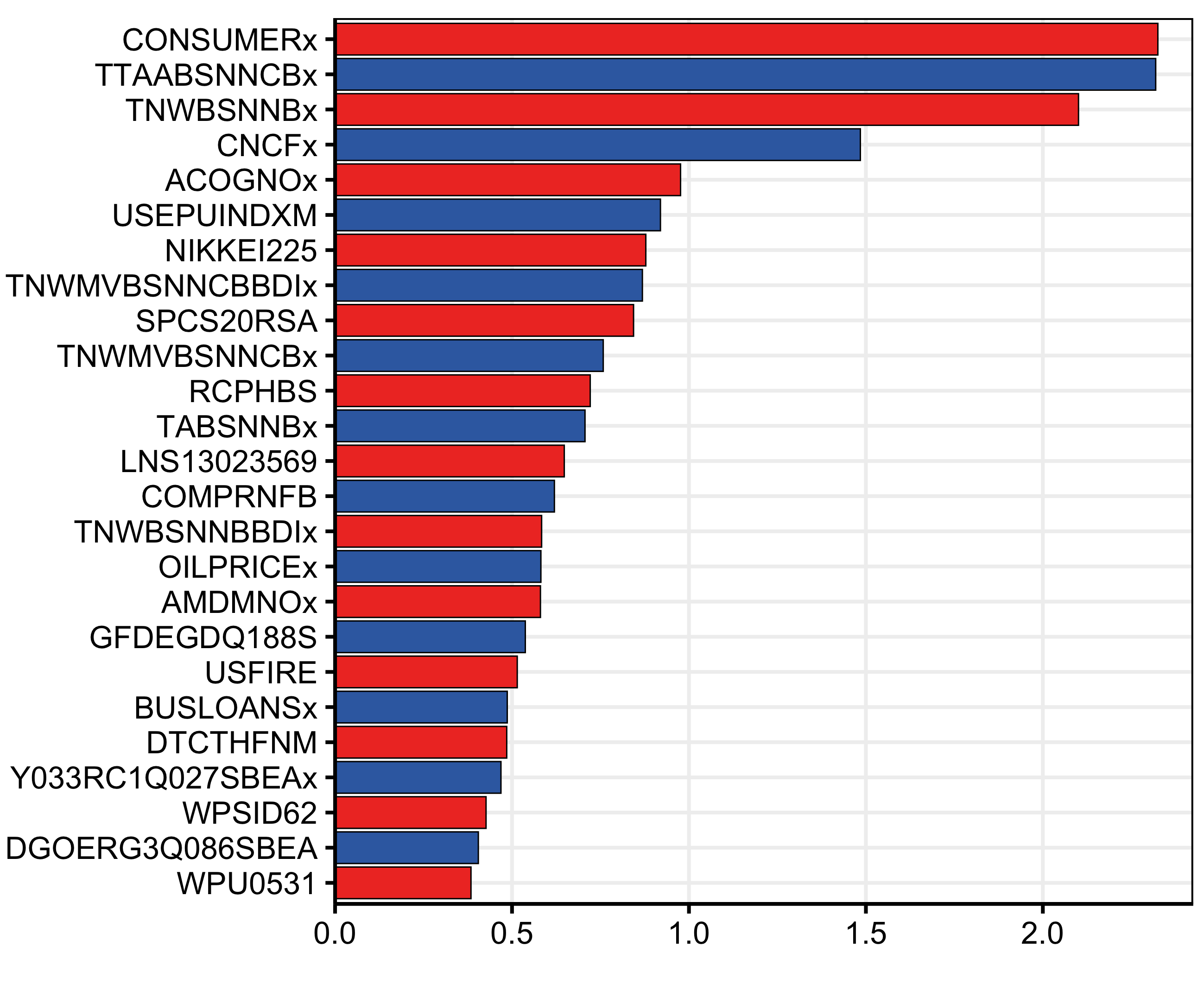}}
      \end{subfigure}
        \vspace{-2em}  

        \caption*{\scriptsize \textit{Notes:} The graph shows VI results for both hemispheres of the HNN with training ending in  2019Q4. The left panel shows the top 25 variables for the mean hemisphere and the right panel refers to the 25 most important drivers of the variance hemisphere. Mnemonics are those of FRED-QD \citep{mccracken2020fred}.}
\end{figure}

\begin{figure}[h]
\caption{VI Results for \textbf{S\&P 500} ($s=1$)}\label{fig:vi_sp500s1}
  \begin{subfigure}[b]{0.5\textwidth}
  \vspace{1em}
  \caption{\footnotesize  Mean hemisphere ($h_m$)}
\hspace{-0.25cm}\includegraphics[trim={0cm 1.5cm 0cm 0cm},clip,width=0.995\textwidth]{{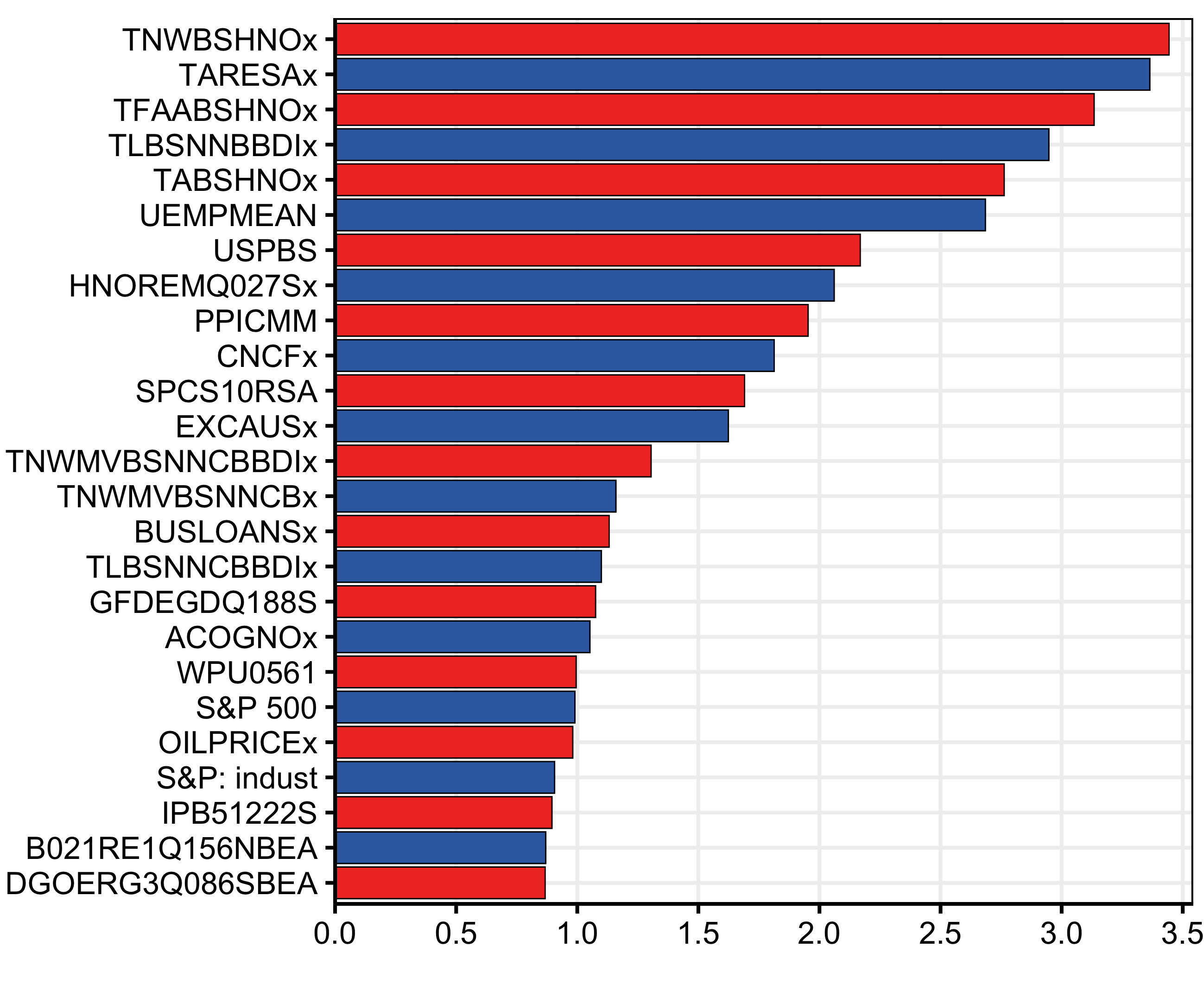}}
      \end{subfigure}
  \begin{subfigure}[b]{0.5\textwidth}
  \caption{\footnotesize Volatility hemisphere ($h_v$)}
\includegraphics[trim={0cm 1.5cm 0cm 0cm},clip,width=0.995\textwidth]{{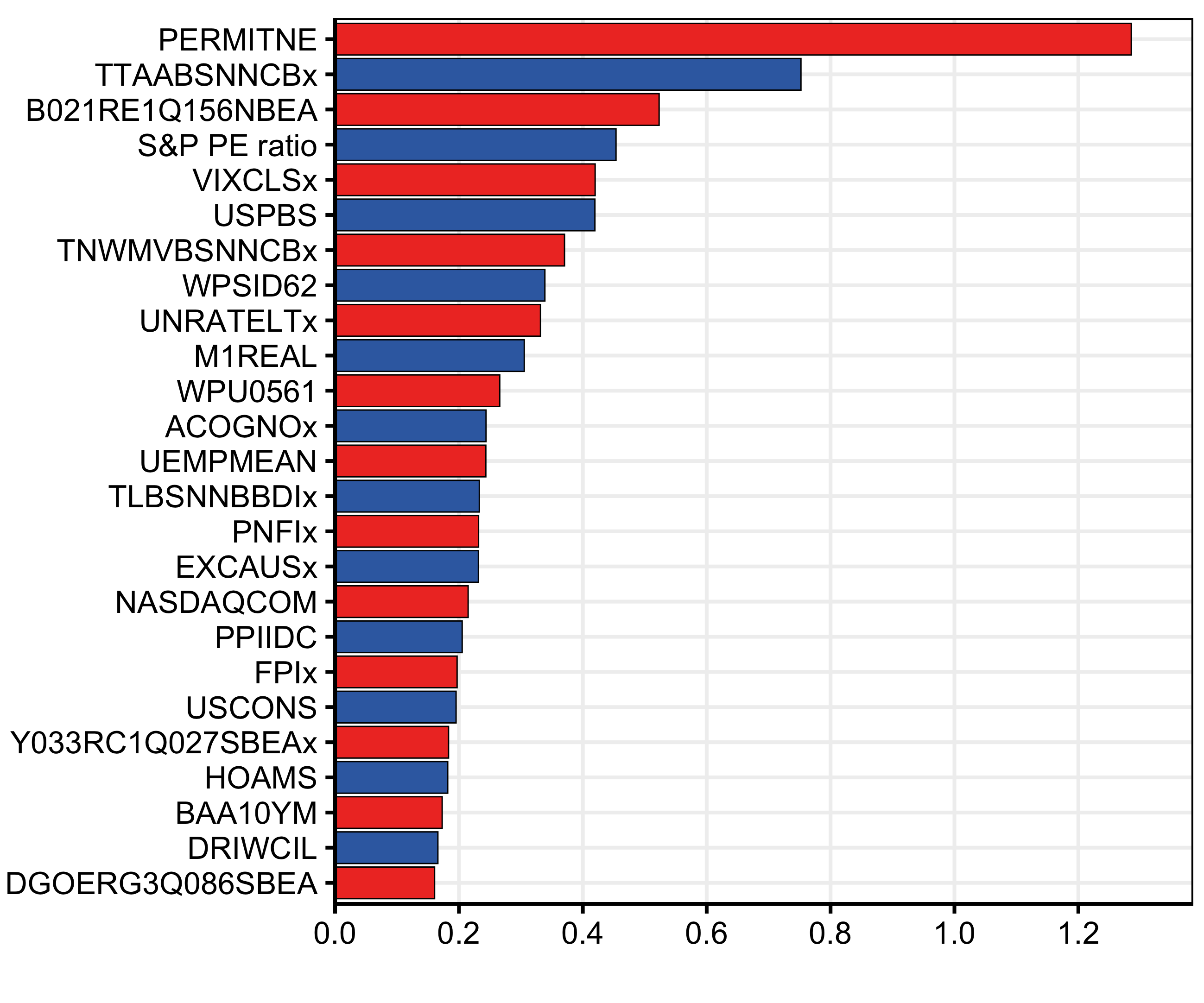}}
      \end{subfigure}
        \vspace{-2em}  

        \caption*{\scriptsize \textit{Notes:} The graph shows VI results for both hemispheres of the HNN with training ending in 2019Q4. The left panel shows the top 25 variables for the mean hemisphere and the right panel refers to the 25 most important drivers of the variance hemisphere. Mnemonics are those of FRED-QD \citep{mccracken2020fred}.}
\end{figure}

\noindent Figures \ref{fig:vi_gdps1} to \ref{fig:vi_sp500s1} report VI results for the targets discussed in Section \ref{sec:fcast_results}.

\vspace{2em}

\begin{figure}[h]
\caption{VI Results for \textbf{Housing Starts} ($s=1$)}\label{fig:vi_housts1}
  \begin{subfigure}[b]{0.5\textwidth}
  \vspace{1em}
  \caption{\footnotesize  Mean hemisphere ($h_m$)}
\hspace{-0.25cm}\includegraphics[trim={0cm 1.5cm 0cm 0cm},clip,width=0.995\textwidth]{{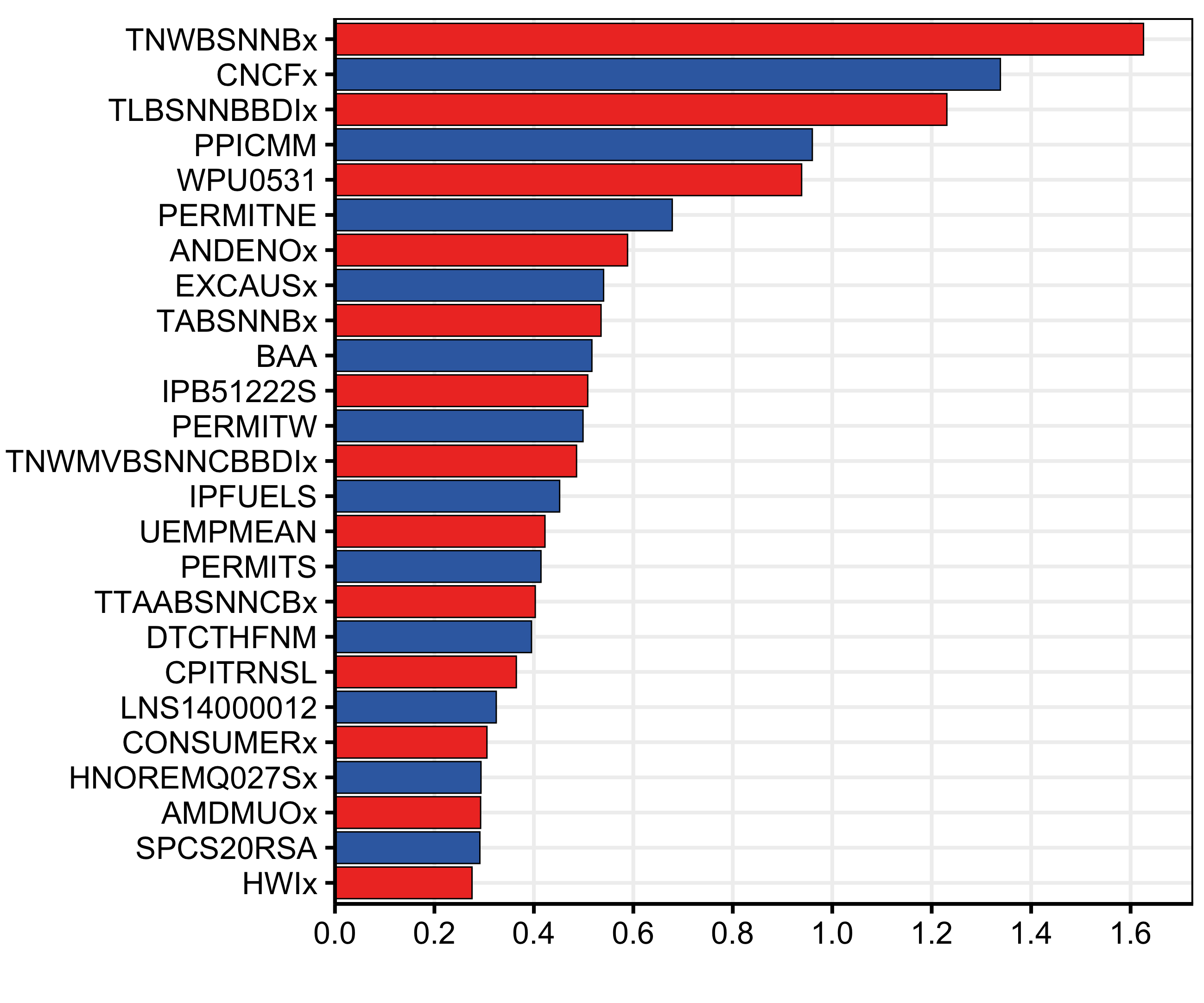}}
      \end{subfigure}
  \begin{subfigure}[b]{0.5\textwidth}
  \caption{\footnotesize Volatility hemisphere ($h_v$)}
\includegraphics[trim={0cm 1.5cm 0cm 0cm},clip,width=0.995\textwidth]{{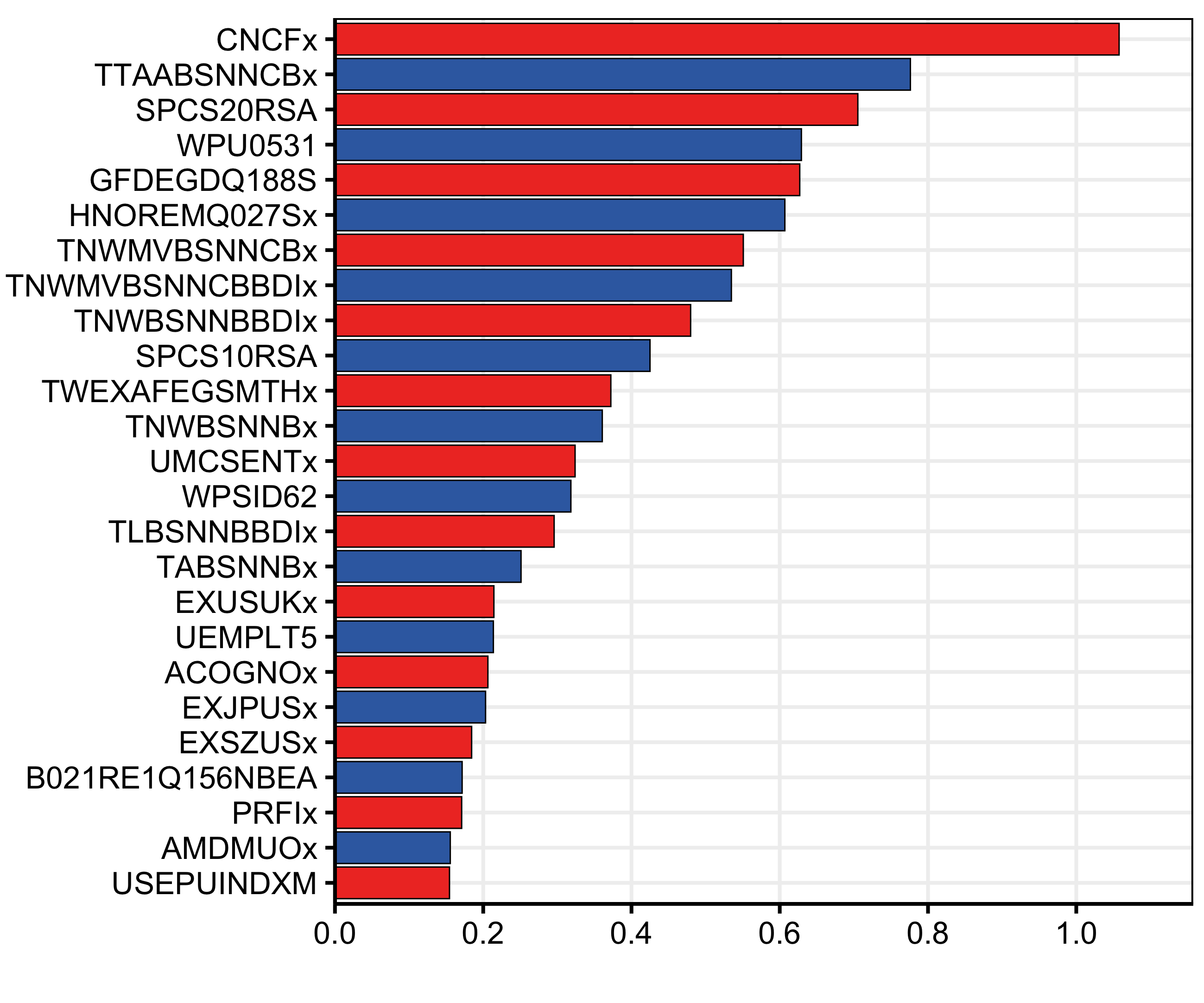}}
      \end{subfigure}
        \vspace{-2em}  

       \caption*{\scriptsize \textit{Notes:} The graph shows VI results for both hemispheres of the HNN with training ending in  2019Q4. The left panel shows the top 25 variables for the mean hemisphere and the right panel refers to the 25 most important drivers of the variance hemisphere. Mnemonics are those of FRED-QD \citep{mccracken2020fred}.}
\end{figure}

\clearpage
\subsection{Benchmark Models}\label{app:bench}

{\noindent \sc \textbf{Bayesian Linear Regression (BLR).}} The Bayesian linear regression model serves as a high-dimensional, linear benchmark in our rich set of competitors. To achieve parsimony, we implement the Normal-Gamma (NG) shrinkage prior of \cite{griffin2010inference}, which belongs to the class of hierarchical global-local shrinkage priors and as such, imposes global shrinkage common to all parameters as well as local shrinkage specific to each of them. Moreover, we estimate the model using stochastic volatility to account for time variation in the magnitudes of error terms. Formally, the model is given by
\begin{equation}
y_t = \bm X_t \bm \beta + \varepsilon_t, \quad \varepsilon_t \sim \mathcal{N}(0,\sigma^2_t).
\end{equation}
with the following prior distribution on the $k$th element of $\bm \beta$ (for $k = 1, \dots, K$):
\begin{equation}
\beta_k | \psi_k,\tilde{\lambda} \sim \mathcal{N}(0, \psi_k), \quad \psi_k|\tilde{\lambda} \sim \mathcal{G}(\vartheta,\vartheta \tilde{\lambda}/2), \quad \tilde{\lambda} \sim \mathcal{G}(e_0,e_1)
\end{equation}
The idiosyncratic scaling parameter, which ensures an individual degree of shrinkage for each element in $\bm \beta$, is denoted by $\psi_k$ whereas $\tilde{\lambda}$ gives the global shrinkage parameter. $\vartheta$ controls the tail behavior of the prior and is assumed to follow $\vartheta \sim \text{exp}(1)$. For the global shrinkage hyperparameters we assume $e_0 = e_1 = 0.01$.


To estimate the model we use a Markov chain Monte Carlo (MCMC) algorithm which iterates through the following steps. First, we draw the linear coefficients from a standard Gaussian posterior taking well-known forms. These can be found in, e.g., \cite{koop2003bayesian}. Next, we sample the additional parameters related to the NG prior. For the corresponding posteriors, we refer to \cite{griffin2010inference}. The stochastic volatilities are drawn by employing the algorithm proposed in \cite{kastner2014ancillarity}. We repeat these steps 20,000 times and discard the first 10,000 draws as burn-in.

\vskip 0.25cm

{\noindent \sc \textbf{Bayesian Additive Regression Trees (BART).}} An alternative way to approximate function $f$ is using Bayesian additive regression trees \citep{BART}. The model accomplishes this by building an ensemble model of regression trees. Let $\Lambda_d$ denote a single regression tree for $d = 1, \dots, D$ regression trees. We then take the sum over all $D$ regression trees to approximate $f$:
\begin{equation}
f(\bm X_t) \approx \sum_{d=1}^D \Lambda_d (\bm X_t | \mathcal{T}_d, \bm \rho_d).
\end{equation}
Each regression tree depends on the tree structure, $\mathcal{T}_d$, and the terminal node parameter $\bm \rho_d$. Regarding the choices on hyperparameters and priors we rely on \cite{BART}. In short, we set $D$ to 250 and use a tree-generating stochastic process for the prior on the tree structure. This process determines the probability of a given node being nonterminal, selects the variables and estimates the corresponding thresholds used in the splitting rule that spawns left and right children nodes. The priors on the terminal nodes are conjugate Gaussian prior distributions with data-based prior variances. In this setting, a certain amount of prior mass is centered on the range of the data and at the same time ensures higher degree of shrinkage with an increasing number of trees.

\vskip 0.25cm


{\noindent \sc \textbf{DeepAR.}} The DeepAR is an autoregressive neural network model based on a LSTM architecture \citep{salinas2020deepar}. It is designed for probabilistic forecasting and produces density predictions based on a user-defined distribution. In our applications, we use 2 hidden layers containing 400 LSTM cells with activation function being Hyperbolic Tangent, i.e., tanh(x) $= \frac{{e^x - e^{-x}}}{{e^x + e^{-x}}}$. Each hidden layer is subject to stochastic dropout with a rate of 0.2 during training only. We use Adam Optimizer with a learning rate of 0.001. The model is optimized according to the negative log-likelihood function over 20 epochs with a patience parameter of 5. 

\vskip 0.25cm

{\noindent \sc \textbf{Bayesian Quantile Regressions.}} Our benchmarks based on quantile regressions include a linear Bayesian quantile regression model (BQR) as well as a quantile version of BART (QBART) and the AR(2) model (QAR). In general terms, we estimate the following model for quantile $\tau \in (0,1)$:
\begin{equation}\label{eq:qreg}
y_t = f_{\tau}(\bm X_t) + u_t, \quad u_t \sim \text{AL}_{\tau}(\sigma_{\tau})
\end{equation}
To sample from the asymmetric Laplace (AL) distribution we rely on the auxiliary representation of \cite{kozumi2011gibbs} given by
$$u_t = \mu_{\tau} \upsilon_{\tau,t} + \pi_{\tau} \sqrt{\sigma_{\tau} \upsilon_{\tau,t} u_t}, \quad \mu_{\tau} = \frac{1-2\tau}{\tau(1-\tau)}, \quad \pi^2_{\tau} = \frac{2}{\tau(1-\tau)}, \quad \upsilon_{\tau,t} \sim \mathcal{E}(\sigma_{\tau})$$
This allows us to write \eqref{eq:qreg} as a conditionally Gaussian:
$$\tilde{y}_{\tau,t} = f(\tilde{\bm X}_{\tau,t}) + u_t, \quad u_t \sim \mathcal{N}(0,1)$$
with $\tilde{y}_{\tau,t} = (y_t - \mu_{\tau} \upsilon_{\tau,t})/(\pi_{\tau} \sqrt{\sigma_{\tau},\upsilon_{\tau,t}})$ and $\tilde{\bm X}_{\tau,t} = (\pi_{\tau} \sqrt{\sigma_{\tau} \upsilon_{\tau,t}} \bm I_K)^{-1} \bm X_t$. The prior on the scale parameter of the AL distribution is inverse Gamma with $\sigma_{\tau} \sim \mathcal{G}^{-1}(3,0.3)$.

In case of BQR, we define $f_{\tau}(\bm X_t) = \bm X'_t \bm \beta_{\tau}$ and estimate the large-scale model with a NG shrinkage prior. For QBART we approximate each function $f_{\tau}$ using a sum of regression trees. QAR is estimated with a weakly informative prior. For details on the posterior distributions, we refer to \cite{kozumi2011gibbs} as well as the description of BLR and BART above.

We evaluate the (tail) forecast accuracy of our quantile regression approaches using log scores and quantile-weighted CRPS (CRPS$_{\omega}$). CRPS$_{\omega}$ is computed as the sum of quantile scores (QS) over all quantiles \citep[see, e.g., ][]{giacomini2005evaluation,gneiting2007strictly}. The quantile score for quantile $\tau$ and forecast horizon $s$ is defined as:
$$\text{QS}_{\tau,t,s} = (y_{t,s} - \mathcal{Q}_{\tau,t,s}) (\tau-\mathbbm{1}\{y_{t,s} \leq \mathcal{Q}_{\tau,t,s} \}),$$
where $\mathcal{Q}_{\tau,t,s}$ is the point forecast at quantile $\tau$ and $\mathbbm{1}$ denotes an indicator function taking value 1 if the true value is at or below the quantile forecast and 0 otherwise. 
The quantile-weighted CRPS is then given by:
$$\text{CRPS}_t(\omega_{\tau}) = \int_0^1 \omega_{\tau}\text{QS}_{\tau,t} d \tau.$$
We compute quantiles $\tau \in \{0.05,0.10,\dots,0.90,0.95\}$.

\subsection{Mnemonics for HNN-NPC}\label{sec:mne}
\begin{lstlisting}[language=R]
#These are for HNN-F.  Add "trend" to the first three hemispheres to get HNN.

real.activity.hemisphere <- c("PAYEMS","USPRIV","MANEMP","SRVPRD",
	         "USGOOD" ,"DMANEMP","NDMANEMP","USCONS","USEHS",
                  "USFIRE","USINFO","USPBS","USLAH","USSERV",
                  "USMINE","USTPU","USGOVT","USTRADE",
                  "USWTRADE","CES9091000001","CES9092000001",
                  "CES9093000001","CE16OV","CIVPART",
                  "UNRATE","UNRATESTx","UNRATELTx","LNS14000012",
                  "LNS14000025","LNS14000026",
                  "UEMPLT5","UEMP5TO14","UEMP15T26","UEMP27OV",
                  "LNS13023621","LNS13023557",
                  "LNS13023705","LNS13023569","LNS12032194",
                  "HOABS","HOAMS","HOANBS","AWHMAN",
                  "AWHNONAG","AWOTMAN","HWIx","UEMPMEAN",
                  "CES0600000007", "HWIURATIOx","CLAIMSx","GDPC1",
                  "PCECC96","GPDIC1","OUTNFB","OUTBS","OUTMS",
                  "INDPRO","IPFINAL","IPCONGD","IPMAT","IPDMAT",
                  "IPNMAT","IPDCONGD","IPB51110SQ","IPNCONGD",
                  "IPBUSEQ","IPB51220SQ","TCU","CUMFNS",
                  "IPMANSICS","IPB51222S","IPFUELS") 
                  
SR.expec.hemisphere <- c("Y", "PCECTPI","PCEPILFE",
		"GDPCTPI","GPDICTPI","IPDBS", "CPILFESL","CPIAPPSL",
                  "CPITRNSL","CPIMEDSL","CUSR0000SAC","CUSR0000SAD",
                  "WPSFD49207",    "PPIACO","WPSFD49502","WPSFD4111",
                  "PPIIDC","WPSID61","WPSID62","CUSR0000SAS","CPIULFSL",
                  "CUSR0000SA0L2","CUSR0000SA0L5", "CUSR0000SEHC",
                  "spf_cpih1","spf_cpi_currentYrs","inf_mich")
                  
commodities.hemisphere <- c("WPU0531","WPU0561","OILPRICEx","PPICMM")

LR.expec.hemisphere <- c("trend")

credit.hemisphere <- c("BUSLOANSx","CONSUMERx","NONREVSLx",
                   "REALLNx","REVOLSLx","TOTALSLx","DRIWCIL",  "DTCOLNVHFNM",
                 "DTCTHFNM","INVEST","nfci","nfci_credit","nfci_nonfin")

\end{lstlisting}

\end{document}